\documentclass[aps,pra,twocolumn,superscriptaddress,nofootinbib]{revtex4-2}

\usepackage[colorlinks=True]{hyperref}
\usepackage{graphicx}
\usepackage{dcolumn}
\usepackage{bm}
\usepackage{braket}
\usepackage{amsfonts}
\usepackage{amsmath} 
\usepackage{amssymb}
\usepackage{cleveref}
\usepackage{booktabs}
\usepackage{color}
\usepackage{comment}
\newcount\ValueAtEndMainText

\newcommand{\beginsupplement}{
    \setcounter{table}{0}
    \renewcommand{\thetable}{S\arabic{table}}
    \setcounter{figure}{0}
    \renewcommand{\thefigure}{S\arabic{figure}}
    \setcounter{section}{0}
    \renewcommand{\thesection}{S\arabic{section}}
    \setcounter{equation}{0}
    \renewcommand{\theequation}{S\arabic{equation}}
}

\begin{document}
% MAIN TEXT

\title{Precision quantum simulation of magnon spectra and interactions}

\author{Google Quantum AI and Collaborators}

\date{\today}
\begin{abstract}
Quantum simulation promises to advance materials discovery by accurately simulating complex states of matter, their microscopic excitations, and macroscopic response functions~\cite{Feynman1982,Keimer2015,Altman2021}. The central challenge in resolving the underlying interacting dynamics is to combine high-fidelity evolution with the sophisticated control necessary to manipulate individual quasi-particles in quantum many-body states. Here, we report on high-precision simulation of both linear and non-linear response functions in a 2D $XY$ spin-$1/2$ magnet using an analog-digital superconducting processor of up to 97 qubits. By interleaving digital gates with analog evolution precisely characterized via Hamiltonian learning, we selectively excite magnons at tunable energy densities. Measuring first the linear magnon response -- a central probe in neutron-scattering experiments~\cite{zaliznyak2005magnetic} -- we extract temperature-dependent spectra and lifetimes. Our results reveal stark variations in magnon decay rates across the Brillouin zone, with enhancement near van Hove singularities and suppression for edge-localized modes. Next, we perform a suite of nonlinear measurements, including the study of self-scattering mechanisms, as well as pump-probe spectroscopy to directly characterize the magnon interactions. While matrix-product state simulations capture the dynamics well in either small systems or at low temperatures, their predictions become inaccurate away from these limits. This work demonstrates precise simulation of the interacting dynamics in quantum magnets, and provides key insights into quasi-particles and their microscopic scattering mechanisms.
\end{abstract}

\maketitle

Connecting microscopic atomic interactions in a material to macroscopic phenomena, such as magnetism and superconductivity, remains a foundational challenge in physics~\cite{Keimer2015}. Classical simulation methods often fail to predict the behavior of systems dominated by strong quantum fluctuations, particularly in spin-$1/2$ magnets and high-temperature superconductors~\cite{Troyer2005,Lee2006,Balents2010,Schollwock2011}. Quantum simulation~\cite{Feynman1982,Altman2021} may address this gap through two primary avenues: (i) direct validation -- simulating response functions, such as magnetic susceptibility, for comparison with experimental spectroscopy, and (ii) microscopic insight -- utilizing high-precision control to investigate quasi-particles, the emergent excitations that drive collective material behavior.

\begin{figure*}[t]
    \centering
    \includegraphics[width = \textwidth]{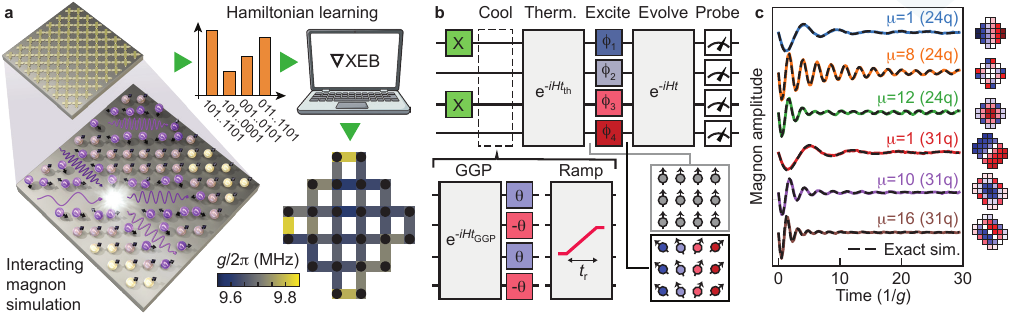}
    \caption{\textbf{High-precision magnon simulation via Hamiltonian learning and interleaved analog-digital circuits. a,} A 97-qubit superconducting Willow processor is used to simulate interacting magnons. We learn its Hamiltonian in sub-patches of 24 qubits by optimizing overlap with the bitstring distribution. Bottom right: example of learned $XY$-couplings. \textbf{b}, Circuit schematic for magnon simulations. The magnet is cooled (dashed box) using either a global gate protocol (GGP) or adiabatic ramps. Following thermalization under analog evolution, digital $Z$-gates are injected to excite specific magnon modes (spin schematics pre- and post-excitation on bottom right), before turning the analog evolution back on to propagate the perturbation and finally measure the response. \textbf{c}, Magnon response functions (spatial structures, $\varphi_i^{\mu}$, shown on right) using $A=1$ and $t_{\mathrm{th}}=0$ ns, exhibiting very good agreement with exact simulations using the learned Hamiltonian (black dashed curves).}
    \label{fig:fig1}
\end{figure*}

Due to their relatively simple microscopic models, quantum spin-$1/2$ magnets are ideal candidates for quantum simulation~\cite{Georgescu2014,Altman2021,Duan2003,Porras2004,Friedenauer2008,Simon2011}. Platforms including ultracold atoms, trapped ions, and Rydberg arrays have successfully probed magnetic ordering~\cite{BohnetMagnetismScience2016, MonroeQsimReview, MazurenkoColdAtomAFMNature2017,Ebadi2021,SchollNature2021, BrowaysXYNature2023, QuantinuumNature2026Magnetism,leclerc2026one}.  The dynamical properties of quantum magnets are governed by magnons. These quantized spin waves are not only central to our understanding of phase transitions but also serve as essential carriers for low-power information transport in spintronics~\cite{Lietal2020SpintronicsReview}. Of particular interest is precise characterization of the interactions between magnons~\cite{ZhitomirskyMagnonDecayRMP2013}, which underlie complex phenomena including Bose-Einstein condensation~\cite{DemokritovMagnonBECNature2006,GiamarchiBECNatPhys2008} and magnon hydrodynamics~\cite{xue2026MagnonHydrodynamics}. 

Quantum simulations of magnon propagation, spectroscopy, and bound states have been performed in several platforms~\cite{FukuharaNature2013, JurcevicNature2014, JurcevicPRL2015, MorvanNature2022,KranzlPRX2023,leclerc2026one,ChenBrowayesXYModelScience2026}. However, existing  approaches face significant hurdles in directly controlling magnons and probing their interactions with high fidelity. Digital processors rely on Trotterized Hamiltonian evolution, where gate errors and decoherence limit accessible evolution times and the fidelity of low-energy state preparation. Analog simulators, while offering effective adiabatic state preparation and faster dynamics less prone to noise, generally make it more challenging to precisely characterize the actual Hamiltonian under which the system evolves. Moreover, they typically lack the local control required to isolate and manipulate individual quasi-particles. Consequently, current studies often rely on bulk spectroscopy or quench experiments~\cite{VillaPRA2019, TanNatPhys2021,ChenBrowayesXYModelScience2026,sun2026experimental} that typically do not excite precise modes. Developing new high-fidelity simulation techniques to probe specific magnon modes and their interactions is therefore essential for advancing the frontier of quantum simulation of magnetism.

Here we present a high-precision quantum simulation of magnons and their interactions in a 2D magnet realized on a hybrid analog-digital Willow quantum processor of 97 superconducting transmon qubits~\cite{WillowNature2025} (Fig.~\ref{fig:fig1}a). We work in the limit where qubit nonlinearity is much larger than the photon hopping amplitude, at a filling of 1/2 photon per qubit. This naturally realizes a 2D square-lattice $XY$ model, where, in addition to the leading nearest-neighbor coupling, smaller, few-qubit terms are present~\cite{Andersen2025}. Our precision magnon spectroscopy is enabled by two key advances: (i) characterization of the Hamiltonian $H$ to a precision level of the order of $\sim 0.1\%$ of the dominant $XY$-coupling, $g/(2\pi)\sim9.7$ MHz, using learning from bitstring distributions generated in the Hamiltonian evolution (Fig.~\ref{fig:fig1}a), and (ii) introduction of the capability to interleave Hamiltonian evolution with digital gates (Fig.~\ref{fig:fig1}b). The latter allows us to initialize the system in a state with a low, tunable energy density, and to controllably excite one or more magnon modes to extract linear and non-linear response functions. 

Measuring the linear magnon response function, the
central quantity in neutron scattering experiments~\cite{zaliznyak2005magnetic}, yields temperature-dependent magnon spectra and lifetimes. These reveal strong mode dependence of the decay rate, including accelerated magnon scattering near van Hove singularities at saddle points in the dispersion, and strongly suppressed scattering for modes concentrated near edges and corners, found in a diamond sample geometry. Next, we move beyond the linear-response regime and perform controlled non-linear spectroscopy measurements, yielding a mode-resolved understanding of scattering mechanisms inaccessible in the linear regime. Finally, to fully isolate individual scattering pathways, we introduce pump-probe and coherent two-mode spectroscopy techniques, in which particular target modes are collided to reveal their scattering products.

The analog-digital protocol leveraged in our experiment is illustrated in Fig.~\ref{fig:fig1}b. The first key step is to prepare an initial state $|\psi_0\rangle$ with tunable energy density $\varepsilon$. Starting with a staggered product state in the $Z$-basis, we utilize two complementary approaches: (i) a slow ramp protocol across a quantum phase transition between a paramagnetic phase and an $XY$-ordered antiferromagnetic (AFM) phase, tuned by a staggered $Z$-field that is controlled by the qubit frequencies, and (ii) a global gate protocol (GGP), comprising a short (6 ns) $XY$-coupling pulse, followed by spatially alternating $Z$-rotations,  $\exp(-i \theta\sum_i (-1)^i Z_i)$ (see SI). Both protocols allow for controlling the energy density in the range $\varepsilon\equiv\langle X_iX_{i+1}+Y_iY_{i+1}\rangle/2 \geq -0.4$ for GGP, and $\varepsilon \geq -0.53$ for ramps, compared to the ground state $\varepsilon_0\approx -0.554$ (from matrix product state, MPS, calculations), with energy here and below given in units of $g$. In both cases, Hamiltonian evolution over time $t_{\rm th}$ can be used to let the system equilibrate after state preparation~\cite{Andersen2025}. 
 
We perform Hamiltonian learning and characterize fidelity using cross-entropy benchmarking (XEB) at $T=\infty$~\cite{AruteNature19,neill2018blueprint,ShawNature2024}, while the magnons are probed at lower temperatures. A key feature of the GGP is that it allows for switching between these conditions via $\theta$ without other major changes to the circuit, thus enabling directly connecting the fidelity and magnon measurement precision. The Hamiltonian parameters, including higher-order terms, are learned in sub-patches of 24 qubits by collecting bitstrings after evolution times $6\leq gt\leq180$ and optimizing the XEB with classical simulations (see SI). Despite the large Hilbert space, this is made possible by an accurate initial guess from precise device modeling~\cite{Andersen2025}, the fast repetition rate of our platform, and an effective backpropagation algorithm. Upon learning, we find XEB errors relative to ideal evolution under the learned Hamiltonian of $4.7\cdotp 10^{-4}$ per qubit per cycle, which only changes marginally to $6.2\cdotp 10^{-4}$ after patching the subsystems together. The high fidelity is supported by the low decoherence rate in the Willow architecture of $\sim 1.9\cdotp 10^{-4}$ per cycle during analog operation, after postselection to mitigate T1 errors. The resultant effective heating rate is less than $5\cdotp 10^{-4}g$, which allows for resolving intrinsic magnon lifetimes with minimal effects of decoherence (see SI).

We next turn to magnon excitations at lower temperature ($\theta=\pi/8$, $\varepsilon\approx-0.4$). Following state preparation, we excite the $\mu$-th magnon mode with spin distribution $\varphi_{i}^{\mu}$ ($\mu\in [1,N_Q-1]$; found from linearized Holstein-Primakoff representation, see SI) by employing an interleaved analog-digital scheme. Specifically, we rapidly (2 ns) turn off the couplers, detune the qubits for 6 ns to impart spatially patterned $Z$-phases, $\exp(\pm i A\sum_i \varphi^{\mu}_i Z_i/2)$, and turn the couplers back on to evolve the magnon (see SI for calibration procedure). By measuring the time-dependent overlap, $\sum_i \varphi_i^{\mu} \langle Z_i(t)\rangle$, in two experiments with opposite excitation phases and subtracting the results~\cite{knap2013probing}, we directly probe the retarded Greens function, $\chi_{\mu}(t)=-iA\bra{\psi_0} [\hat{b}_{\mu}(t),\hat{b}^{\dagger}_{\mu}(0)] \ket{\psi_0}$ (see SI). In contrast to more commonly employed methods where the excitation is created either before cooling the state~\cite{roushan2017spectroscopic,roberts2024manybody}, via a quench~\cite{TanNatPhys2021,ChenBrowayesXYModelScience2026} or by globally driving the system at a particular frequency~\cite{sun2026experimental}, applying the digital excitation gates within the analog evolution allows for probing the actual response function of $\psi_0$ at the temperature of interest rather than more qualitative proxies of this quantity. 

Fig.~\ref{fig:fig1}c shows the measured $\chi_{\mu}(t)$ following GGP for various modes in small (24- and 31-qubit) systems to facilitate comparison with exact simulations. We observe underdamped magnon oscillations that are in excellent quantitative agreement with numerical simulations (dashed black). Across all measurements, the median relative precision in the frequency $\omega$ and amplitude decay rate $\Gamma$ is found to be $7\cdotp10^{-3}$ and $7.8\cdotp 10^{-2}$, respectively (see SI).

\begin{figure}[t]
    \centering
    \includegraphics[width=\columnwidth]{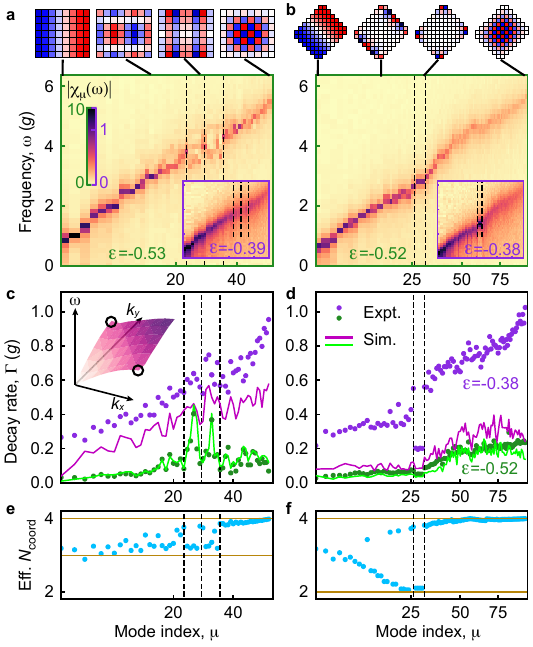}
    \caption{\textbf{Effects of mode distribution and van Hove singularity on magnon decay. a,b,} Magnon dispersions in 56-qubit rectangle (\textbf{a}) and 97-qubit diamond (\textbf{b}) geometries. The main and inset show results at low ($\varepsilon=-0.53,-0.52$) and elevated ($\varepsilon=-0.39,-0.38$) energy densities, prepared using adiabatic ramp and GGP, respectively, including thermalization for $t_{\mathrm{th}}=100$ ns in both cases. Examples of phase patterns used for excitation ($\varphi_i^{\mu}$) are shown in the top row. For efficiency, we excite $N_{\mathrm{q}}/8$ uniformly distributed modes in eight batches, using a small amplitude $A=0.5$ to minimize interaction effects.
    \textbf{c,d}, Extracted decay rates from \textbf{a} and \textbf{b}. The rectangular geometry (\textbf{c}) shows increased decay rates for particular modes, marked by dashed vertical lines, which we attribute to van Hove singularities at the saddle points in the dispersion (inset). In the diamond geometry, we instead find that the modes concentrated on the corners have a strongly suppressed decay rate, particularly at higher temperature (purple). Solid curves show MPS simulations (bond dimension, $\chi=512$), with substantially worse agreement at the elevated temperature. \textbf{e,f,} Effective coordination number of the magnon modes, weighted by $|\varphi_i^{\mu}|^2$. Both geometries show two branches, corresponding to modes that are concentrated at the edge and in the bulk. The branching behavior matches well with the decay rates observed in \textbf{c,d}.}
    \label{fig:fig2}
\end{figure}

Having established the precision of our technique, we next turn to the full mode dependence of the magnon response in the frequency domain, shown for a $7\times8$ qubit rectangle and a 97-qubit diamond geometry in Fig.~\ref{fig:fig2}a,b. The main and inset show low and high energy densities, respectively, where the former is near the Kosterlitz-Thouless transition around $\varepsilon=-0.53$~\cite{Andersen2025}. Since momentum is not well-defined in the diamond geometry, we plot the dependence on mode index, $\mu\sim k^2$, and find a near-linear dependence, $\omega \propto\sqrt{\mu}\propto k$, as expected for magnons in the 2D $XY$-model. As can be seen from both the insets and the extracted decay rate in Figs.~\ref{fig:fig2}c,d, increasing the temperature induces line broadening, consistent with increased thermal populations and enhanced scattering. Moreover, in both geometries, we observe that the decay rate generally increases with the energy of the mode, as expected from both enhanced scattering matrix elements and increased phase space for scattering. 
\begin{figure*}
    \centering
    \includegraphics[width=\textwidth]{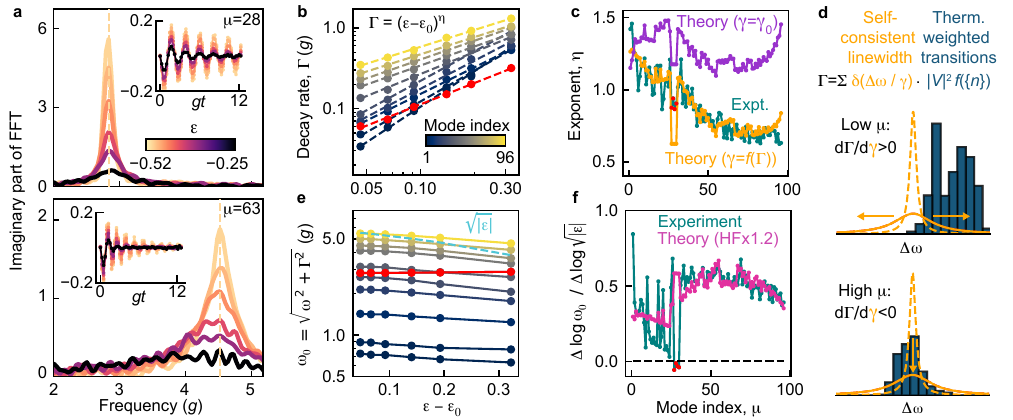}
    \caption{\textbf{Temperature dependence of magnon decay rate and resonance frequency. a,} Magnon response for modes 28 (top) and 63 (bottom) in Fourier (main) and time-domain (inset). Mode 63 shows red-shift with increasing $\varepsilon$. \textbf{b}, Dependence of decay rate on energy elevation above ground state. Dashed lines show power-law fits. The data is collected using the same parallel excitation scheme as in Fig.~\ref{fig:fig2}. Red indicates corner mode. \textbf{c}, Power-law exponent, $\eta$, extracted from \textbf{b} (teal). The decrease below unity is not captured theoretically when assuming constant linewidths, $\gamma=\gamma_0$ (purple), pointing to the important effect of self-consistent broadening effects (yellow). \textbf{d}, The temperature dependence of $\Gamma$ arises from changes in not only thermally available transitions (dark blue), but also broadening of linewidths (yellow). Transitions at low mode index (top; $\mu=6$) are relatively non-resonant and thus accelerated by broadening ($d\Gamma/d\gamma>0$). Decay of higher index modes (bottom; $\mu=91$) is more resonant, and thus decelerated by broadening ($d\Gamma/d\gamma<0$), causing $\eta<1$ in \textbf{c}. \textbf{e}, Undamped magnon resonance frequency, $\omega_0$, grows with energy density, except for in corner modes (red). Solid lines are guides to the eye. Clear discrepancy from classical spring prediction, $\omega_0\propto\sqrt{|\varepsilon|}$ (turquoise) is observed. \textbf{f}, Extracted relative change, $\Delta \log \omega_0$ normalized by $\Delta \log\sqrt{|\varepsilon|}$ shows correlation with the effective coordination number (Fig.~\ref{fig:fig2}f), indicating stronger renormalization in edge than bulk modes. Hartree-Fock (blue) approximately agrees with high-$\mu$ behavior, but requires rescaling with empirical factor of 1.2, and does not capture higher-order effects governing lower-$\mu$-modes.}
    \label{fig:fig3}
\end{figure*}

Beyond this general trend, however, both geometries display additional mode structure: in the rectangular geometry, certain modes (between dashed vertical lines) are found to have substantially broader linewidths. We attribute this to van Hove singularities at the saddle points of the Brillouin zone, where a high density of states enables rapid scattering and magnon decay. For the small amplitude ($A=0.5$) used here, the average number of magnons created in each excited mode is only $A^2/\pi^2=0.025$ (see SI). Therefore, decay is expected to be dominated by $2\rightarrow 2$ magnon scattering in which the magnon scatters with a thermally populated mode. Since mainly low-energy modes are populated at the low temperature, only a small amount of energy can be interchanged, hence making a high density of states accelerate the decay. In the diamond geometry, on the other hand, we instead find that four of the modes have substantially lower decay rates, particularly evident at higher temperatures. These are modes that are densely concentrated at the corners. Both of these features are part of a broader division into two branches of decay rates, which follow the effective coordination number of the modes (weighted by $|\varphi_i^{\mu}|^2$; Fig.~\ref{fig:fig2}e,f). While one branch corresponds to modes concentrated in the bulk ($N_{\mathrm{coord}}\sim 4$), the other is due to modes with higher weight on the edges.

As the data in Fig.~\ref{fig:fig2} are beyond the reach of exact simulation, we compare to approximate simulations using MPS calculations. These agree well with the experimental data at the low temperature (see additional comparisons in SI), consistent with the high experimental precision observed in Fig.~\ref{fig:fig1}c. At higher temperatures, on the other hand, where the entanglement entropy is higher, much larger deviations are observed. Extensive analysis of MPS truncation errors, as well as implementations of alternative methods such as projected entangled pair states (PEPS), support the idea that tensor network simulation of the frequency and decay rate for sufficiently long-lived low-frequency modes requires infeasibly large resources to reach the experimental precision level (see SI).

Following the observed temperature-induced broadening in Fig.~\ref{fig:fig2}, we next leverage our technique to further understand how the interacting magnon dynamics are impacted by increasing temperature. Fig.~\ref{fig:fig3}a shows the response for modes 28 and 63 as we vary the energy density via the ramp time, revealing not only broadening, but also a systematic red-shift with increasing temperature for mode 63. To discern the two effects, we first consider the decay rates (Fig.~\ref{fig:fig3}b), which are well captured by a power-law dependence on the energy elevation above the ground state, $\Gamma_i\propto(\varepsilon-\varepsilon_0)^\eta=\delta E^{\eta}$, using $\varepsilon_0\sim-0.57\pm0.01$ directly estimated from the data (see SI). Importantly, the exponent $\eta$ is found to be strongly mode-dependent, decreasing from values well above unity to around $\sim 0.7$ as one moves from low- to high-energy modes (Fig.~\ref{fig:fig3}c). 

To understand this behavior, we model the decay rate using the scattering vertex $|V_{ij,kl}|^2$ obtained from Holstein-Primakoff representation (SI). Since it is computed for the pure $XY$-model and is a mean-field approximation, we do not expect exact quantitative agreement, but rather a theoretical guide to better understand our observations. The decay rate of the amplitude of mode $i$ is found from summing over thermally populated scattering partners, $j$, and the outgoing modes, $k$ and $l$:

\begin{equation}
\Gamma_i=4\pi\sum_{j,k,l}|V_{ij,kl}|^2 f(n_j,n_k,n_l)\delta(\Delta\omega_{ijkl}/\gamma_{ijkl})
\label{eqn:meanfield}
\end{equation}
where $f(n_j,n_k,n_l)=[n_j(1+n_k)(1+n_l)-(n_j+1)n_kn_l]$ accounts for the Bose-Einstein statistics of the involved modes with populations $n$, and $\delta((\omega_k+\omega_l-\omega_i-\omega_j)/\gamma_{ijkl})$ ensures energy conservation within a Lorentzian of linewidth $\gamma_{ijkl}$. 

\begin{figure*}
    \centering
    \includegraphics[width = \textwidth]{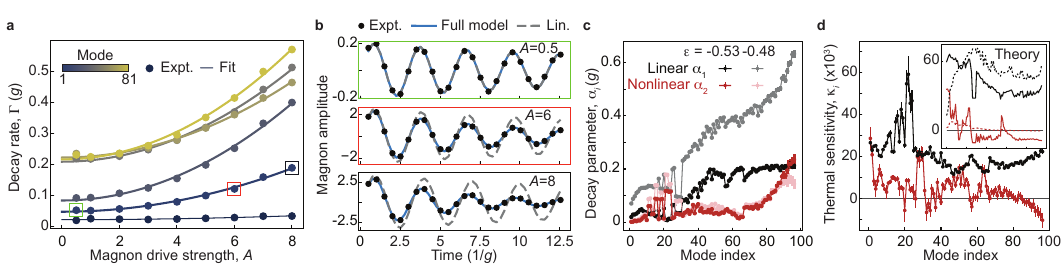}
    \caption{\textbf{Nonlinear magnon decay pathways. a,} The decay rate scales roughly as $A^2$, consistent with a nonlinear self-scattering process. Points indicate experimental data, and lines are quadratic fits. Every sixteenth mode in a 97-qubit system is shown. \textbf{b,} Examples of fits to the nonlinear decay equation (Eq. \ref{eqn:nonlinear}) for mode 17 (corresponding points boxed in \textbf{a}). We collectively fit (solid lines) the data (markers) at all amplitudes. The linear decay component (dashed lines) describes the low-amplitude data well, but underestimates the decay for higher $A$. For clarity, every other time point is displayed. \textbf{c,} Fit values for the linear and nonlinear decay parameters, $\alpha_1$ and $\alpha_2$. Dark and light lines indicate values at $\varepsilon_0 = -0.53$ and $\varepsilon_1 = -0.48$, respectively. Error bars represent the fit uncertainty. \textbf{d,} We quantify temperature sensitivity of the decay channels with $\kappa_i=(\log \alpha_i(\varepsilon_1)-\log \alpha_i(\varepsilon_0))/(\varepsilon_1-\varepsilon_0)$. Error bars are propagated from \textbf{c}. $\kappa_2$ (red) is typically several times lower than $\kappa_1$ (black), and can also be negative. Inset: these qualitative observations are consistent with a mean-field calculation of the $XY$-model. Solid black/red lines are the calculated $\kappa_{1/2}$, including contributions from both thermal occupancy changes and from line broadening, while dashed lines only include thermal occupancy changes. While $\kappa_1$ has comparable contributions from the two effects, $\kappa_2$ as a self-scattering process is dominated by line broadening.}
    \label{fig:fig4}
\end{figure*}
Assuming a constant $\gamma_{ijkl}$ in Eq.~\ref{eqn:meanfield} predicts $\eta>1$ for all modes (purple curve in Fig.~\ref{fig:fig3}c), in contrast to the observed sublinear behavior of higher-energy modes. 
However, when computing the broadening self-consistently, by setting $\gamma_{ijkl}=\Gamma_j+\Gamma_k+\Gamma_l$, we find good agreement in $\eta$ across the spectrum. This shows that $\Gamma(\varepsilon)$ is not only governed by the number of thermally populated scattering partners in $f(\{n\})$, but also strong feedback effects from broadening of $\gamma_{ijkl}$. While line broadening accelerates the relatively non-resonant collisions of low-energy modes, it instead slows down the near-resonant collisions at higher energies (Fig.~\ref{fig:fig3}d). The latter causes a sublinear dependence of $\Gamma$ on $\delta E$ for high-energy modes.

We next turn to the undamped magnon frequency $\omega_0=\sqrt{\omega^2+\Gamma^2}$, which generally decreases with temperature, as shown in Fig.~\ref{fig:fig3}e. Intuitively, as we move away from the ground state, the (absolute) $XY$-correlations decrease, reducing the effective torsional spring constant $K\propto|\varepsilon|$ of our system and renormalizing the magnon frequency as $\omega_0\propto \sqrt{|\varepsilon|}$. However, this classical spring model (dashed turquoise curve) fails to capture the observed behavior. The deviation is particularly stark for corner modes (red), whose frequency instead increases with rising temperature. To further understand the mode-dependent renormalization, we next plot the relative change, $\Delta \log\omega_0$, normalized by $\Delta \log\sqrt{\varepsilon}$ such that unity corresponds to the classical spring picture. We find that this ratio splits into two branches, similar to the effective coordination numbers considered in Fig.~\ref{fig:fig2}f, demonstrating that the edge modes are more prone to renormalization. On the one hand, the bulk modes exhibit behavior that is at least qualitatively captured by Hartree-Fock calculation for $\mu\gtrsim30$, although with a substantial rescaling factor of 1.2. On the other hand, the lower energy modes --- particularly in the edge branch -- deviate more strongly, suggesting higher-order renormalization effects. 

The magnon properties explored so far belong to the low-amplitude limit, with 0.1 magnons or fewer created on average in each driven mode. As a result, the excited magnons scatter primarily with thermally populated modes. However, our excitation technique also allows us to precisely tune the magnon amplitude and characterize higher-order magnon processes in the nonlinear regime. Tuning the magnon drive strength $A$ from 0.5 to 8, we observe that the decay rate increases roughly as $A^2$ (Fig.~\ref{fig:fig4}a). This indicates the opening of an additional, nonlinear, decay channel, in which the magnon mode scatters with itself. However, the quadratic dependence by itself does not allow for distinguishing the specific scattering channel. To characterize the nonlinear decay further, we fit the magnon decay curves to a simple ansatz combining both linear and nonlinear decay (see SI for details):
\begin{equation}
\frac{dn(t)}{2dt} = -\alpha_1 n(t) -\alpha_2 n^2(t) -c\left(n(0)-n(t)\right)n(t).
\label{eqn:nonlinear}
\end{equation}
The rate $\alpha_1$ captures the linear decay, and corresponds to $\Gamma$ in the low-amplitude limit. The rate $\alpha_2$ describes the nonlinear decay, while $c$ represents the additional thermal scattering that occurs as the initially populated magnon mode decays and raises the overall temperature of the system. For each mode, we simultaneously fit measurements for multiple amplitudes to this model, with example results shown in Fig.~\ref{fig:fig4}b. While this simplified model is not expected to fully capture the intricate magnon decay dynamics, the good agreement over a range of amplitudes, times, and modes suggests that it is reliably extracting the dominant components of the nonlinear decay process.

Fit values for the linear and nonlinear couplings, $\alpha_1$ and $\alpha_2$, are presented in Fig.~\ref{fig:fig4}c at two energy densities. While $\alpha_1$ closely tracks the decay rate at low amplitude (compare Fig.~\ref{fig:fig2}d), $\alpha_2$ displays significantly different variations with mode index, including sharp jumps that correspond to qualitative changes in mode spatial structure and an upturn at high mode index. In Fig.~\ref{fig:fig4}d we present the thermal sensitivity for both parameters, calculated as $\kappa_i=(\log \alpha_i(\varepsilon_1)-\log \alpha_i(\varepsilon_0))/(\varepsilon_1-\varepsilon_0)$. The temperature sensitivity is substantially smaller for the nonlinear coefficient, identifying it as arising from a $2M \rightarrow B+C$ process, rather than alternative mechanisms where an incoming thermal mode is involved (e.g. $2M+A \rightarrow B$). We also find that $\kappa_2$ becomes negative for the highest-energy modes, consistent with the broadening effects observed in Fig.~\ref{fig:fig3}: since the nonlinear process does not require an incoming thermal mode, its thermal sensitivity is dominated by line broadening contributions rather than thermal populations, causing a change in not just the exponent, but even the sign of $\kappa$. Self-consistent Born approximation calculations using Eq.~\ref{eqn:meanfield} support this intuition, although with substantial quantitative differences (Fig.~\ref{fig:fig4}d inset).

\begin{figure}
    \centering
    \includegraphics[width=\columnwidth]{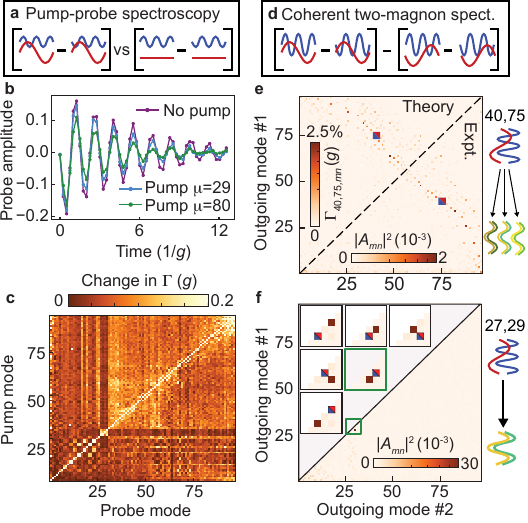}
    \caption{\textbf{Directly probing magnon scattering via collision experiments. a,} Equal-time pump-probe measurements show how a high-amplitude pump mode (red; $A=4$) impacts the decay rate of low-$A$ probe modes (blue; $A=0.5$). We use four measurements, including alternating signs of the probe, and comparison with and without pumping. For efficiency, we excite 24 probe modes in parallel (more than in other experiments because we focus on the differential signal with and without pumping, which is more robust). \textbf{b}, Response of mode 50 without pumping (purple), compared with pumping of corner mode 29 (blue) and bulk mode 80 (green). The latter has substantially stronger effect. \textbf{c}, Pump-induced change in the probe decay rate for all pairs of pump and probe modes. The dark ``cross" emanating from a bright square indicates that the corner modes are decoupled from other modes, but interact strongly with each other. \textbf{d}, Coherent two-mode spectroscopy enables studying the outgoing mode pairs, by combining four measurements with alternating signs of the two incoming modes ($|A|=2$.) \textbf{e}, Outgoing mode pair distribution from colliding modes 40 and 75 (blue and red squares). Upper and lower triangles show, respectively, the theoretically predicted scattering rate and experimentally observed pair amplitudes (see text for details). The distribution is concentrated along the anti-diagonal, where energy is conserved. \textbf{f}, Same as in \textbf{e}, but for incoming corner modes 27 and 29. Contrasting the wide range of outgoing pairs found in \textbf{e}, scattering is now almost purely into a single pair of two (other) corner modes. Insets: zoomed-in results for all six combinations of incoming corner mode pairs (marked by blue and red squares).}
    \label{fig:fig5}
\end{figure}

The measurements described so far gave valuable information about both thermal scattering and self-scattering. Next, we introduce more sophisticated experimental techniques to perform precision 
quantum simulations of the magnon scattering processes~\cite{KarpovPRR2022, farrell2025digitalquantumsimulationsscattering} even more directly. Specifically, our analog-digital method allows for exciting two modes simultaneously, before turning the analog evolution back on to let the modes interact. We first employ this magnon collider setup to study which modes interact with each other in an experiment inspired by pump-probe spectroscopy (Fig.~\ref{fig:fig5}a). Specifically, we excite one ``pump" mode with amplitude $A=4$ and study how it affects the lifetime of weakly excited ($A=0.5$) probe modes. An example of this is shown in Fig.~\ref{fig:fig5}b, which displays the decay of mode 50 both in the absence of pumping (purple), and with separate pumping of a corner mode (mode index 29; blue) and a bulk mode (80; green). Interestingly, we find that pumping mode 80 induces a large increase in the decay rate, while pumping the corner mode has almost no effect. To understand this better, we next plot the pump-induced change in decay rate for all pairs of pump and probe modes in Fig.~\ref{fig:fig5}c. The data reveals very non-monotonic structure, suggesting that we are resolving interactions between the pump and probe modes, rather than simply heating effects, although we note that the latter is also expected to be present. We observe a ``cross" of low interaction intensity, centered on a square of strong interactions corresponding to the corner modes. This indicates that the corner modes are highly decoupled from the other modes, but interact strongly with each other, thus explaining the very low decay rates observed for these modes in Fig.~\ref{fig:fig2}.

Besides resolving which incoming modes scatter with each other, our collider method also allows for studying which modes they scatter into. This is achieved by instead performing four measurements with alternating signs of the two incoming modes ($A=\pm 2$), combined together as shown in Fig.~\ref{fig:fig5}d to isolate the non-trivial scattering products. We then determine the amplitude of outgoing mode pair ($m,n$) by computing the overlap with the two-point correlator, $A_{mn}(t)=\sum_{m,n}\varphi_i^m \varphi_j^n (\langle Z_i(t)Z_j(t) \rangle-\langle Z_i(t)\rangle \langle Z_j(t) \rangle$). Fig.~\ref{fig:fig5}e shows the outgoing mode pairs (quantified by the standard deviation in $A_{mn}$ over time) arising from colliding bulk modes 40 and 75. We observe a wide range of outgoing mode pairs, clustered around the anti-diagonal, showing how energy conservation governs the allowed pathways. Comparing with the theoretically computed rates from the scattering vertex, $|V_{40,75,m,n}|^2$, we find very good qualitative agreement on which outgoing mode pairs dominate, although we note that the mean-field treatment is insufficient to quantitatively describe the dynamics.

When instead colliding two corner modes (Fig.~\ref{fig:fig5}f), we observe very different behavior: in contrast to the wide range of outgoing pairs observed in Fig.~\ref{fig:fig5}e, we now observe scattering into only a single pair, namely the two other corner modes. By exciting the six possible combinations of the four corner modes, we observe that this is general: every pair of corner modes scatters into another pair. The combination of pump-probe and this more coherent two-mode spectroscopy technique has therefore shown that the corner modes not only tend to scatter \textit{with} each other, but also \textit{into} each other. 

Through this suite of high-precision magnon measurements enabled by our interleaved analog-digital protocol and Hamiltonian learning, we have directly probed individual quasi-particle properties, as well as their interactions and dependence on temperature, mode profile, and excitation amplitude. We find that considerable efforts using classical methods (see SI) fail to provide equally precise simulations of the magnon decay rate, particularly at intermediate temperatures. Notably, the decay rate is also the central quantity of interest in our study from a physical perspective, containing the key information about magnon scattering mechanisms. Our work thus represents an important step towards the broader use of quantum processors to simulate materials and quasi-particle dynamics that are beyond the reach of classical computers. Looking ahead, we expect that high-resolution quantum simulation of response functions and quasi-particles, as demonstrated here, will provide a powerful route to uncovering microscopic mechanisms underlying magnetic and superconducting order in correlated materials, thereby advancing their predictive modeling, experimental validation, and ultimately guiding materials discovery.

{\bf Acknowledgements.--} We thank Eugene Demler, Thierry Giamarchi, Sid Parameswaran, Louk Rademaker, and Peter Zoller and for insightful discussions.

{\bf Competing interests.--} Provisional patent applications have been filed for the global gate protocol and calibration of gates interleaved with analog evolution.

\textit{Note added.--} During the preparation of this manuscript, we became aware of linear magnon response studies in a 1D digital superconducting processor~\cite{lee2026benchmarking}.

\newpage
\onecolumngrid
\vspace{5mm}

\begin{flushleft}

{\hypertarget{authorlist}{${}^\dagger$}  \small Google Quantum AI and Collaborators}

\bigskip

    \renewcommand{\author}[2]{#1\textsuperscript{\textrm{\scriptsize #2}}}
    \renewcommand{\affiliation}[2]{\textsuperscript{\textrm{\scriptsize #1} #2} \\}
    \newcommand{\corrauthora}[2]{#1$^{\textrm{\scriptsize #2}, \hyperlink{corra}{\ddagger}}$}
    \newcommand{\corrauthorb}[2]{#1$^{\textrm{\scriptsize #2}, \hyperlink{corrb}{\mathsection}}$}

\begin{footnotesize}

\newcommand{\xGoogle}{\affiliation{1}{Google Research, Mountain View, CA, USA}}

\newcommand{\xPrincetonECE}{\affiliation{2}{Department of Electrical and Computer Engineering,
Princeton University, Princeton, NJ, USA}}

\newcommand{\xGenevaB}{\affiliation{3}{Department of Quantum Matter Physics, University of Geneva, Geneva, Switzerland}}

\newcommand{\xGeneva}{\affiliation{4}{Department of Theoretical Physics, University of Geneva, Geneva, Switzerland}}

\newcommand{\xHarvard}{\affiliation{5}{Department of Physics, Harvard University, Cambridge, MA, USA}}

\newcommand{\xUMD}{\affiliation{6}{Joint Quantum Institute and Joint Center for Quantum
Information and Computer Science, NIST/University of Maryland, College Park, MD, USA}}

\newcommand{\xEPFL}{\affiliation{7}{Center for Computation Quantum Physics, École Polytechnique Fédérale de Lausanne, Lausanne, Switzerland}}

\newcommand{\xUMass}{\affiliation{8}{Department of Electrical and Computer Engineering, University of Massachusetts, Amherst, MA, USA}}

\newcommand{\xUConnStorrs}{\affiliation{9}{Department of Physics, University of Connecticut, Storrs, CT, USA}}

\newcommand{\xAuburnECE}{\affiliation{10}{Department of Electrical and Computer Engineering, Auburn University, Auburn, AL, USA}}

\newcommand{\xUCSB}{\affiliation{11}{Department of Physics, University of California, Santa Barbara, CA, USA}}

\newcommand{\xPrinceton}{\affiliation{12}{Department of Physics,
Princeton University, Princeton, NJ, USA}}

\newcommand{\Google}{1}
\newcommand{\PrincetonECE}{2}
\newcommand{\GenevaB}{3}
\newcommand{\Geneva}{4}
\newcommand{\Harvard}{5}
\newcommand{\UMD}{6}
\newcommand{\EPFL}{7}
\newcommand{\UMass}{8}
\newcommand{\UConnStorrs}{9}
\newcommand{\AuburnECE}{10}
\newcommand{\UCSB}{11}
\newcommand{\Princeton}{12}

\corrauthorb{T. I.~Andersen}{\Google},
\corrauthora{N. Astrakhantsev}{\Google},
\corrauthora{J. Martinez}{\Google,\! \PrincetonECE},
\corrauthora{W. Morong}{\Google},
\author{J. Motruk}{\GenevaB},
\author{D. Rossi}{\Geneva},
\author{B. Ware}{\Google},
\author{B. Kobrin}{\Google},
\author{W. Wu}{\Google,\! \Harvard},
\author{E. Bennewitz}{\Google,\! \UMD},
\author{M. Rudolph}{\Google,\! \EPFL},
\author{T. Westerhout}{\Google},
\author{A. Abbas}{\Google},
\author{R. Acharya}{\Google},
% ...
\corrauthorb{X. Mi}{\Google},
\corrauthorb{D. Abanin}{\Google,\! \Princeton}

\bigskip

\xGoogle
\xPrincetonECE
\xGenevaB
\xGeneva
\xHarvard
\xUMD
\xEPFL
\xUMass
\xUConnStorrs
\xAuburnECE
\xUCSB
\xPrinceton

\vspace{2mm}

{\hypertarget{corrb}{${}^\mathsection$} Corresponding authors: trondiandersen@google.com, mixiao@google.com,
abanin@google.com}\\
{\hypertarget{corra}{${}^\ddagger$} These authors contributed equally to this work.}\\

\end{footnotesize}
\end{flushleft}

\twocolumngrid

\let\oldaddcontentsline\addcontentsline
\renewcommand{\addcontentsline}[3]{}

\let\addcontentsline\oldaddcontentsline

% SUPPLEMENTARY INFORMATION

\clearpage

\beginsupplement

\onecolumngrid

\begin{center}

    \vspace*{-0.5cm} 

    {\large\bfseries\MakeUppercase{Supplementary information}} \\[0.5cm]

\end{center}

\twocolumngrid

\tableofcontents 

\section{Experimental details}

\subsection{Global gate protocol (GGP)} \label{sec:ggp}
We here provide additional details about the global gate protocol presented in the main text, which allows for preparing quantum states with precise temperature control in a very short amount of time. The protocol proceeds as follows:
\newline
1) Prepare qubits in alternating $\ket{1}$ and $\ket{0}$ states
\newline
2) Turn on uniform and time-independent $XY$-Hamiltonian for a time duration $t_{\mathrm{GGP}}$
\newline
3) Rotate qubits by alternating phases, $\pm\theta$, around the $Z$-axis.

The key idea behind the technique is that it allows for tuning correlations in two stages: in step (2), when the couplers are turned on, spin density moves from sites with $\ket{1}$ to sites with $\ket{0}$. This spin current is equivalent to $\langle XY-YX\rangle$-correlations; the spins become increasingly correlated in the $XY$-plane, but they point in perpendicular directions. In step (3), we rotate neighboring spins in opposite directions, so that the correlations are turned from $XY-YX$ (perpendicular) to $XX+YY$ (parallel).

\begin{figure}[h]
    \centering
    \includegraphics[width=\columnwidth]{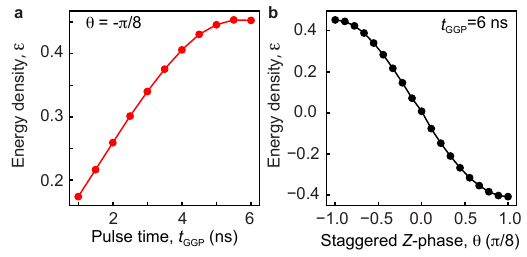}
    \caption{\textbf{Global gate protocol (GGP). a,} Dependence of final energy density on pulse time (step 2; \textbf{a}) and staggered $Z$-phases (step 3; \textbf{b}) in a system of 24 qubits. The energy is minimized by using a pulse time of $\sim 0.4/g$ and staggered $Z$-phases of $\pm \pi/8$.}
    \label{fig:BAM_SM}
\end{figure}

Crucially, the protocol allows for targeting a wide range of temperatures, tuned via both the time $t_{\mathrm{GGP}}$ and the rotation angle $\theta$ (see Fig.~\ref{fig:BAM_SM}). The former controls the strength of the perpendicular correlations developed in step 2, while the latter sets the extent to which these are rotated to become parallel in step 3. Specifically, infinite temperature (zero parallel correlations) is achieved by either setting $t_{\mathrm{GGP}}=0$ such that no perpendicular correlations are developed in step 2, or $\theta=0$, such that the correlations remain purely perpendicular in step 3. The strongest achievable correlations (lowest temperature) is typically in the range of about 70$\%$ to 80$\%$ of the ground state value (smaller system sizes allow for somewhat larger correlations), and are reached by setting $t_{\mathrm{GGP}}\sim 0.4/g$ and $\theta=\pi/8$. Notably, only one of the parameters needs to be tuned to reach temperatures across the full available range. Tuning via $\theta$ is often preferred, since this leaves the duration of the protocol unchanged and thus minimizes confounding effects.
Importantly, the duration of the protocol is very short, requiring only 6 ns for step 2 for a typical coupling of $g=2\pi\cdotp10$ MHz, and only about 5-10 ns for step 3. This allows for minimizing the accumulation of noise effects and control errors.
\begin{figure}
    \centering
    \includegraphics[width=\columnwidth]{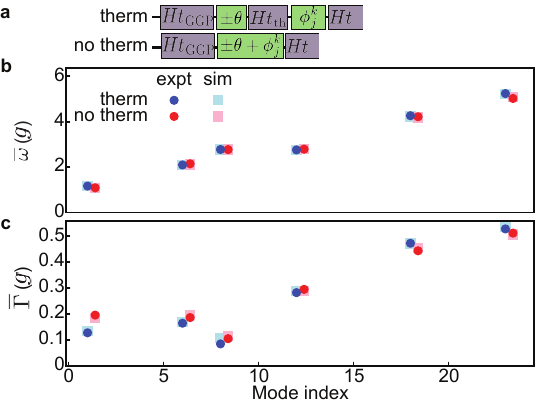}
    \caption{\textbf{Comparison of GGP measurements with and without intermediate thermalization.} \textbf{a}, Comparison of pulse sequences, which differ in whether $Z$ rotations to create the background thermal state and to excite magnons are applied together or separately. Typical times are $t_{\text{GGP}}=0.4/g$, $t_{\text{th}}=6.3/g$, and $t=0$ to $38/g$. \textbf{b,c,} Extracted magnon frequencies (\textbf{b}) and decay rates (\textbf{c}) for select modes in $N_{\mathrm{q}}=24$. Red points are offset in the horizontal direction for clarity. All observables are similar for both preparation approaches but show some variations on the few-percent level that are well-captured by exact simulations. Magnon frequency and decay rate are quantified with the model-free measures $\overline{\omega}$ and $\overline{\Gamma}$ as defined in Sec. \ref{sec:sm_accuracy}.}
    \label{fig:sm_bam_compare}
\end{figure}

We use two variations of this preparation. The first (``therm") allows time for thermalization before magnons are created, and is more appropriate for studies of magnon physics and for comparison with our ramped state preparation (which always includes a thermalization period after the ramp). The second (``no therm") is optimized for speed, and compiles the GGP phases and magnon phases together, so that the magnons are created on an initially non-thermalized background. This is advantageous for high-fidelity experiments, since the overall protocol is as quick as possible to minimize decoherence. However, it also presents the most feasible protocol for numerics, and therefore is a useful classical benchmark for the computational complexity of accurate magnon simulations. We use this approach primarily in the complexity and precision studies in SM sections \ref{sec:sm_accuracy} and \ref{sec:complexity}, as well as in main text Fig. 1c. In Fig. \ref{fig:sm_bam_compare} we compare the observables resulting from these two approaches. In general they both show the same trends, but have small systematic offsets which are consistent with numerical simulations. We find that the median differences between the magnon frequency and decay rate measured without thermalization as compared to with thermalization are, respectively, 2\% and 10\%.

\subsection{Phase calibration for injected digital $Z$-gates}

\begin{figure}[h]
    \centering
    \includegraphics[width=\columnwidth]{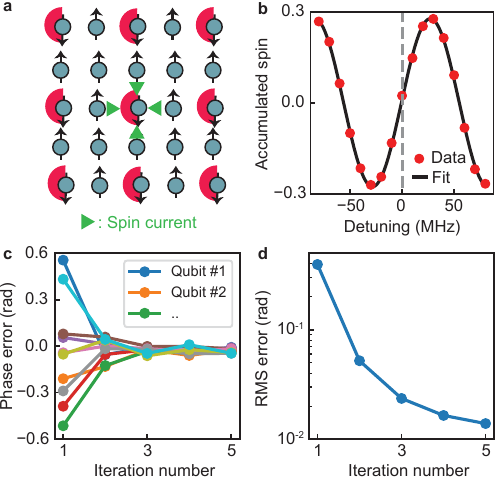}
    \caption{\textbf{Phase calibration for injected $Z$-gates. a,} Protocol for calibration: the qubit phases are swept (via detuning) in four parallel layers, giving rise to non-zero spin-current (green arrows). \textbf{b}, By measuring the accumulated spin current a short time ($t<10$ ns) after the $Z$-pulse as a function of the qubit detuning, we extract the effective phase accumulation time and the phase required to acquire zero phase (gray dashed line). \textbf{c,d}, The protocol is iterated to minimize the phase errors, typically achieving root-mean-square (RMS) errors on the scale of 10 mrad.}
    \label{fig:SI_phase_cal}
\end{figure}

In our experiments, we inject digital $Z$-gates by first rapidly (2 ns) turning off the couplers, detuning the qubits by pre-calibrated frequency shifts $\{\Delta\omega_i\}$ for $\tau=6$ ns to accumulate the desired phases $\{\phi_i\}$, and then turning the couplers back on. Since the $Z$-fields do not commute with the $XY$-coupling, turning off the couplers during the gate is crucial to achieve the correct perturbation. In realistic experimental conditions, one typically finds that $\phi_i=(\Delta\omega_i+\delta\omega_i)(\tau+\delta\tau)$, where $\delta\omega_i$ arises from additional frequency shifts, e.\,g. due to dispersive shifts from the nearest neighbor couplers, and $\delta\tau$ is the effective additional time due to phase accumulation during ramps. While approximate estimates of $\{\Delta\omega_i\}$ can be obtained from either modeling or measurements of isolated systems of one or two qubits, these typically do not provide the necessary accuracy for our high-fidelity experiments: methods purely based on modeling typically suffer from inaccuracies in the qubit and coupling trajectories (causing errors in $\delta\tau$), and neither technique captures the full set of dispersive shifts present in full-scale experiments. Hence, it is necessary to devise calibration protocols that are performed in the actual full-scale context of the quantum simulation.

In our scheme, we achieve this by taking advantage of the spin currents arising from the accumulated phases. In particular, non-zero phases cause rotation of the existing $\langle X_iX_{i+1}\rangle$ in a low-temperature state into non-zero spin current, $\langle X_iY_{i+1}\rangle$, the accumulation of which can be easily measured in the $Z$-basis. In an iterative protocol, we therefore sweep the detuning applied to each qubit centered around the current best guess $\Delta\omega_i=-\delta\omega_i$, performed in a parallelized fashion across 4 qubit subsets, while keeping the remaining qubits at $\Delta\omega_i=-\delta\omega_i$ (Fig.~\ref{fig:SI_phase_cal}a). We then measure the accumulated spin current a short time after turning the couplers back on, and fit its dependence on detuning with a sinusoid (Fig.~\ref{fig:SI_phase_cal}b) to extract $\tau+\delta\tau$ from its period. When it comes to the phase, the extracted offset $\delta\phi_i$ of each fitted sinusoid is not the error relative to the \textit{correct} phase, but rather relative to the current attempt at setting neighboring phases to zero in that iteration. Hence, we find the actual errors $\{\delta\tilde{\phi}_i\}$ by solving the set of equations given by:
$\delta\phi_i=\tilde{\delta\phi_i}-\sum_{j\in\mathrm{NN}}\tilde{\delta\phi_j}/N_{\mathrm{NN}}$, where the sum is over the $N_{\mathrm{NN}}$ nearest neighbors of qubit $i$. We then shift the current guess of $\delta\omega_i$ by $\tilde{\delta\phi_i}/(\tau+\delta\tau)$ and repeat the iterative process until the parameters have converged (typically 3-4 iterations; see Figs.~\ref{fig:SI_phase_cal}c,d)). 

\subsection{Benchmarking of injected $Z$-gates}
\label{sec:zerr}
\begin{figure}[h]
    \centering
    \includegraphics[width=\columnwidth]{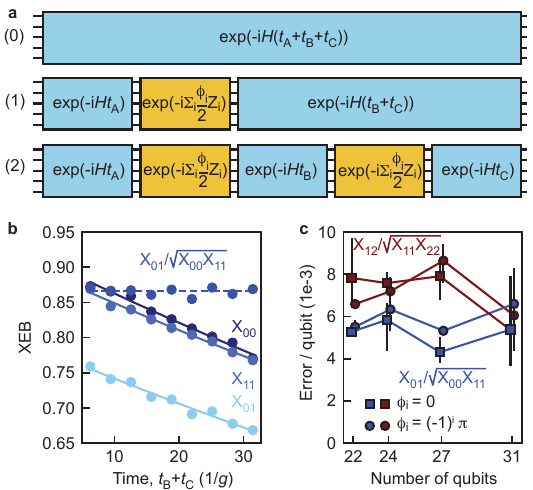}
    \caption{\textbf{Benchmarking of injected digital $Z$-gates. a,} Circuit schematics used in benchmarking, where the analog evolution (blue) is interrupted 0, 1 or 2 times when applying digital $Z$-rotations (yellow). $t_{A}$ is set to 6 ns to mimic GGP as well as possible. \textbf{b,} Cross-entropy between experiment 0 and 1, $X_{01}$, is normalized by the self-XEB of the individual experiments, $X_{00}$ and $X_{11}$, to benchmark the fidelity of the $Z$-gates (see text for details). \textbf{c,} Error per qubit measured in three patches of four different system sizes, for injected phases of $\phi_i=0$ (squares) and $\phi_i=(-1)^i\pi$ (circles). Error bars extend from lowest to highest measured error, and markers indicate the median. The markers are separated slightly horizontally for improved visibility. For the first pulse (blue), the error is about $\sim 5\cdotp 10^{-3}$ and $\sim 6\cdotp 10^{-3}$ for the two phase sets. For the second pulse (red), the error is $\sim 7\cdotp 10^{-3}$ for both phase sets.}
    \label{fig:Zpulsebench}
\end{figure}
To characterize the fidelity with which we turn off the analog Hamiltonian and perform digital $Z$-rotations, we perform cross-entropy benchmarking between experimental datasets (as opposed to comparing with simulation). This is advantageous, since it allows for determining the fidelity without making assumptions about the Hamiltonian during evolution. Specifically, we perform three different experiments (Fig.~\ref{fig:Zpulsebench}a) that involve evolution under the analog Hamiltonian with 0, 1 or 2 sets of injected $Z$-gates with rotations that are ideally equivalent to the identity operator ($\phi_i=0$ or $\phi_i=\pm\pi$). We then compute the XEB overlap between measurements with $n-1$ and $n$ pulses ($n=1,2$), given by $X_{n-1,n}=D\sum_x p^{n-1}_xp^{n}_x-1$, where $p^n_x$ is the measured probability of bitstring $x$ from the experiment with $n$ injected sets of $Z$-rotations, and $D$ is the Hilbert space dimension. Importantly, $X_{n-1,n}$ includes not only fidelity loss due to the $Z$-gate, but also from other effects, which we remove by normalizing by the geometric mean of the self-XEB of the two individual measurements, $X_{i,i}=D\sum_x(p^i_x)^2-1$. Fig.~\ref{fig:Zpulsebench}b shows an example of these quantities for $n=1$ in 24 qubits, plotted against the duration of the second analog evolution interval, using phases of $\pm \pi$ which maximizes the frequency excursion during the gate. $X_{00}$, $X_{11}$ and $X_{01}$ decay with very similar rates, and the combined fidelity of all the $Z$-gates, $X_{01}/\sqrt{X_{00} X_{11}}$ is found to be around 87\%, independent of the time. In Fig.~\ref{fig:Zpulsebench}c, we plot the extracted gate injection errors from 4 different system sizes, measured across three patches and for both $\phi_i=0$ (squares) and $\phi_i=(-1)^i\pi$ (circles). For the first pulse, the error of the injected gate is found to be around $\sim5\cdotp10^{-3}$ per qubit for $\phi_i=0$ and approximately $\sim6\cdotp10^{-3}$ per qubit for $\phi_i=(-1)^i\pi$. For the second pulse, we find a somewhat higher error of $\sim7\cdotp10^{-3}$ for both cases. This difference is likely related to the fact that while the first pulse is calibrated directly after an analog evolution pulse, the second is instead calibrated following low-temperature state preparation via GGP. While this is done to keep the calibration circuits the same as in the actual magnon experiment, the differences from the circuits used in benchmarking is expected to induce some error. Seeing as the two phase sets ($\phi_i=0$ and $\phi_i=(-1)^i\pi$) involve very differently sized frequency excursions, their similar performance indicates that the error is more dominated by effects of the changing coupling or the calibrated phase offsets of the qubits (see dashed gray line in Fig.~\ref{fig:SI_phase_cal}b), rather than the frequency excursion itself.

\subsection{Isolated vs parallel measurements of modes}

\begin{figure}
    \centering
    \includegraphics[width=\columnwidth]{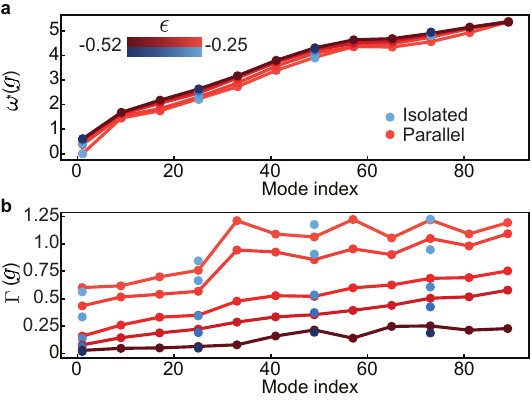}
    \caption{\textbf{Characterization of parallel vs isolated magnon mode measurements.} Magnon frequency $\omega$ (\textbf{a}) and decay rate $\Gamma$ (\textbf{b}) are presented for twelve modes measured in parallel (red) and a subset of four modes measured in isolation (blue), for various energy densities and in a 97 qubit system. The values of $\omega$ and $\Gamma$ extracted from parallel measurements typically differ by a few percent from the isolated measurements, but show the same general trends.}
    \label{fig:sm_parallel}
\end{figure}

In order to scalably characterize many magnon modes in a large system, as in Figs. 2 and 3 of the main text, we excite multiple modes simultaneously and project the resulting dynamics onto the different mode profiles. In these parallel measurements, we use a very low amplitude ($A=0.5$) to minimize interactions between the modes (note that the larger-amplitude, nonlinear measurements in Fig. 4 are measured in isolation). In Fig.~\ref{fig:sm_parallel} we compare the magnon frequency and decay rate of modes extracted in parallel and in isolation. The two measurements show similar behavior, with somewhat larger differences in the decay rates than in the magnon frequencies. We note that numerical simulations shown with experiments are always run using the same pulse sequence.

\subsection{Additional decay rate data}\label{sec:adrd}

\begin{figure*}
    \centering
    \includegraphics[width=1.9\columnwidth]{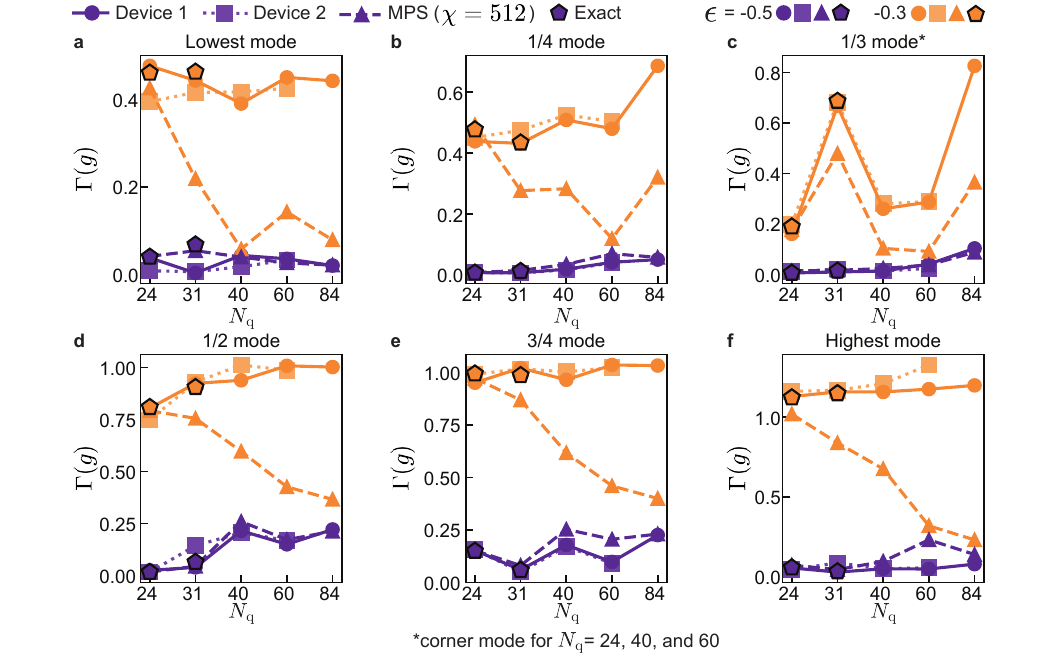}
    \caption{\textbf{Additional decay rate data.} Measurements of $\Gamma$ for system sizes ranging from $N_{\mathrm{q}}=24$ to 84, six modes across the spectrum at each size, varied temperatures, and for an additional Willow device (Device 2), which has a maximum patch size of $N_{\mathrm{q}}=60$. While the comparison between devices is complicated by variations in the exact Hamiltonian realized, the measured $\Gamma$ is largely consistent as the system is scaled up and as temperature is varied, indicating the reliability of experimental measurements. Numerical simulations of Device 1 with MPS track the experiment and exact numerics closely for small sizes and/or low temperatures, but strongly underestimate the decay rate as temperature is raised in a large system.}
    \label{fig:sm_mps_compare}
\end{figure*}

In Fig. \ref{fig:sm_mps_compare} we present additional magnon decay rate data surveying a range of system sizes, points in the magnon spectrum, temperatures, and experimental devices. We generally see a strong consistency between the two devices. The differences that are observed are largely due to differences in the exact Hamiltonian realized, as each device shows good agreement with individual exact simulations at small sizes (Sec. \ref{sec:sm_accuracy}). Notably, we generally do not observe any increase in disagreement between the two devices as the system size is increased. MPS numerics of device 1 are approximately converged at the smallest size, and agree well with both the experiment and with exact numerics. For large sizes, this agreement remains relatively good for low energies, but at high energies the decay rate is dramatically underestimated, consistent with large entanglement entropy and thus higher difficulty in classical simulations. In Sec. \ref{sec:complexity} the computational complexity of classical simulation of the magnon decay is analyzed, and the origin of these trends is discussed in detail.

\subsection{Ground state energy estimation from decay rate data}

\begin{figure}[h!]
    \centering
    \includegraphics[width=0.9\columnwidth]{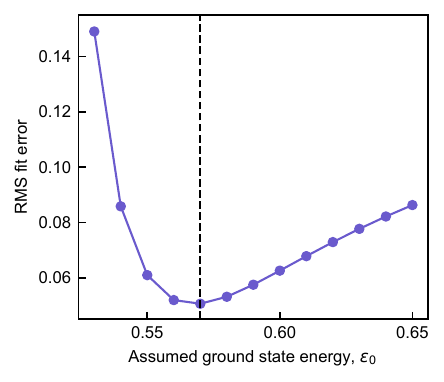}
    \caption{\textbf{Estimating $\epsilon_0$ from decay rate data.} We estimate the ground state energy, $\epsilon_0$, by computing how the root-mean-square (RMS) error in fitting $\Gamma(\epsilon)$ (see Fig. 3 in the main text) depends on the assumed $\epsilon_0$. The lowest error is found for $\epsilon_0=-0.57\pm0.01$.}
    \label{fig:Egs}
\end{figure}

In Fig. 3 in the main text, we consider the energy density dependence of the decay rate, $\Gamma(\epsilon)\propto (\epsilon-\epsilon_0)^\eta$. From this data, we estimate the ground state energy, $\epsilon_0$, by evaluating which assumed value of $\epsilon_0$ gives the lowest fitting error (Fig.~\ref{fig:Egs}). We estimate $\epsilon_0=-0.57\pm 0.01$, which is in good agreement with the value predicted by MPS calculations (-0.554).

\subsection{Nonlinear magnon analysis}
\label{sec:sm-nonlinear}

\begin{figure}
    \centering
    \includegraphics[width=\columnwidth]{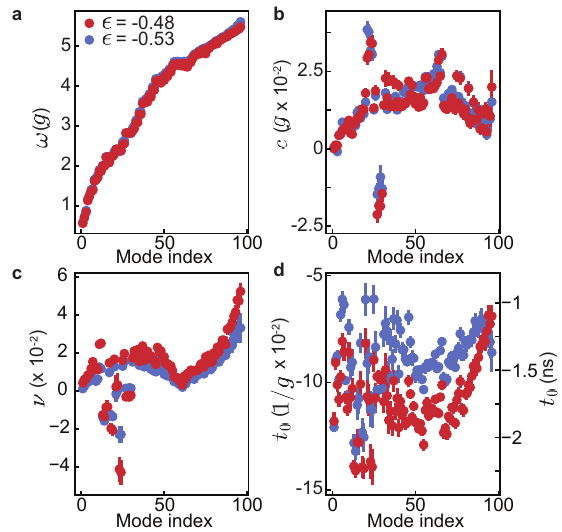}
    \caption{\textbf{Values for nonlinear fit parameters.} Values for other shared parameters in the fits to the nonlinear decay model shown in main text Fig. 4. We see anomalous values for some fit parameters for the edge and corner modes, such as the negative values of $c$ for the corner modes. This indicates that our simple model (Eq.~\eqref{eq:nlinfit}), which assumes decay to a featureless continuum of states, is imperfect for these states. We also see a clear dependence of the time offset $t_0$ on energy density, which is attributed to the finite settling rate of the tunable couplers that mediate interactions following ramps of different speeds.}
    \label{fig:sm_fits}
\end{figure}

In Figure 4 of the main text, we jointly fit multiple measurements at different magnon drive strengths to the following model:
\begin{align}
& a(t) = \sqrt{n(t)}\sin\left(\omega (1-n(t) \nu )(t+t_0)\right)+y_0, \\
& n(t) = \frac{\alpha' a_0e^{-\alpha'(t+t_0)}}{\alpha'+(2\alpha_2-2c) a_0(1-e^{-\alpha'(t+t_0)})}, \label{eq:nlinfit} \\
&\alpha' = 2(\alpha_1+c a_0) \nonumber
\end{align}
where $n(t)$ is the solution to Eq. 2 of the main text. In addition to the parameters in $n(t)$, we include $\nu$ to capture the nonlinear shifts of the magnon mode, and $y_0$ and $t_0$ to capture experimental non-idealities. Of the fit parameters, $a_0$, $f$, $\alpha_1$, $\alpha_2$, $c$, $\nu$, $t_0$, and $y_0$, all except $a_0$ are fit simultaneously to all the magnon drive strength curves for a given mode and energy density. As each fit for data in main text Fig. 4c,d is performed using data at four amplitudes and 51 time points, this results in the fitting of 11 parameters to 204 points in each case. Fig. \ref{fig:sm_fits} shows the fit values for all shared parameters other than $\alpha_1$ and $\alpha_2$ (which are shown in main text Fig. 4c) and $y_0$ (whose magnitude never exceeds 0.01). Of these fit values, the four corner modes (mode indices 27-30) show a negative value of $c$, indicating that they are not fully described by our model. This is no surprise, as the model assumes decay to a featureless bath, while our results in Fig. 5 of the main text show that these states are primarily coupled to each other. Anomalies for these modes are also seen in other fit parameters such as $\nu$, and in $d\omega/dt$ as shown in main text Fig. 3f. In general, while they may contain additional information, all of these terms are highly decoupled from both $\alpha_1$ and $\alpha_2$, the focus of the main text. $\omega$, $\nu$, and $t_0$ primarily control the oscillations, not the decay envelope, and $c$ has a very small impact on the dynamics as it is suppressed both at short and long times by the factor of $n(t)(n(t)-n(0))$ (see Eq. 2 of the main text).

\subsection{Corner modes in pump-probe experiments}

\begin{figure}[h]
    \centering
    \includegraphics[width=\columnwidth]{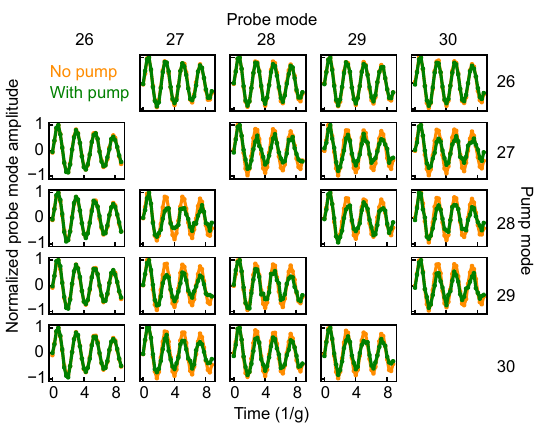}
    \caption{\textbf{Pump-probe signal of corner modes.} Amplitude in time domain for probe modes 26-30, shown with (orange) and without (green) pump. While mode 26 shows very little interactions with the corner modes (27-30), the corner modes are strongly coupled with each other, causing coherent swapping behavior rather than purely exponential decay.}
    \label{fig:corner_PP}
\end{figure}

In Fig. 5 in the main text, we show the change in decay rate induced by the pump mode. Since the corner modes are strongly coupled, resonant, and form a small manifold of only 4 modes, their interactions in the presence of the large-amplitude ($A=4$) pump give rise to non-monotonic effects in certain cases, which we attribute to coherent swapping. In Fig.~\ref{fig:corner_PP}, we show this behavior for all pairs of corner modes, as well as for mode 26 for comparison.

\subsection{Decay dynamics of modes near van Hove singularity}

\begin{figure}[h]
    \centering
    \includegraphics[width=\columnwidth]{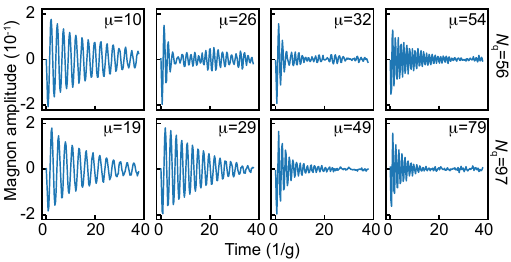}
    \caption{\textbf{Decay of modes near van Hove singularity.} Top: Magnon amplitude in time domain for $\mu=$10, 26, 32 and 54 in 56 qubits. The middle two modes are near the van Hove singularity, and show deviation from exponential behavior at long times, which we attribute to scattering into near-resonant modes. Modes 10 and 54 are shown for comparison. Bottom: Magnon dynamics in $N_{\mathrm{q}}=97$ shown for further comparison.}
    \label{fig:56q_decay}
\end{figure}

In the 56-qubit geometry, we identify magnon modes with enhanced scattering due to the the high density of states near van Hove singularities at saddle points in the dispersion. While the non-exponential decay behavior observed at small system sizes in e.\,g. Fig.~\ref{fig:GGPbench} generally diminishes significantly in larger systems, some modes display similar features in our 56-qubit system, particularly those near the van Hove singularities. This is shown for mode indices $\mu=$26 and 32 in Fig.~\ref{fig:56q_decay} and is suggestive of swapping with near-resonant modes present at the saddle points. For comparison, we also display the decay for other modes far from the singularities ($\mu=$10 and 54), as well as four modes in the 97-qubit geometry.  

 \subsection{Correlation and energy measurements}
\label{sec:correlate}
Throughout this manuscript, energy density is defined as the average nearest-neighbor $XY$ correlations:
\begin{equation}
\epsilon = \sum_{\langle i,j \rangle}^{N_b} \frac{\langle X_iX_j + Y_iY_j \rangle}{2N_{\text{b}}},
\end{equation}\label{eq:energy}
where $\langle X_iX_j\rangle=\langle Y_iY_j\rangle$ in our $U(1)$ symmetric system (confirmed explicitly in Ref.\,\cite{Andersen2025}).
We typically measure nearest-neighbor correlations by applying a layer of digital single- and two-qubit gates to project into the Bell basis, which allows for simultaneously measuring the relative parity of neighboring qubits and their total excitation number~\cite{Andersen2025} and thus performing the same postselection on total excitation number that is applied to our magnon measurements. Notably, the phases of the single-qubit rotations are defined by the frequencies of the microwaves used, which are not equal to the common frequency of the resonant analog dynamics and also differ across qubits. If not compensated for, this causes the measured correlators between two qubits to precess from $\langle X_iX_j\rangle=\langle Y_iY_j\rangle$ into $\langle X_iY_j\rangle=-\langle Y_iX_j\rangle$ at a rate equal to the difference of their microwave frequencies. To mitigate this effect, we take advantage of the fact that the precession preserves the norm, $\sqrt{(|\langle{X_i X_j}\rangle|^2 + |\langle{X_i Y_j}\rangle|^2 + |\langle{Y_i X_j}\rangle|^2 + |\langle{Y_i Y_j}\rangle|^2)/2}=\sqrt{(|\langle{X_i X_j}\rangle|^2 + |\langle{X_i Y_j}\rangle|^2}$, meaning that measurements of this quantity are unaffected by the phase accumulation. Beyond the thermalization time of a few $1/g$ (see Ref.~\cite{Andersen2025}), spin currents ($\langle X_iY_j-Y_iX_j\rangle$) are zero, and one finds that $\sqrt{(|\langle{X_i X_j}\rangle|^2 + |\langle{X_i Y_j}\rangle|^2}=\langle{X_i X_j}\rangle=\langle{Y_i Y_j}\rangle$. In other words, in this regime, measurements of the norm allow for determining the energy density, $\epsilon$.  

For longer-range correlators, conversion to the Bell basis requires more layers of two-qubit gates. Hence, in that case, we instead measure the norm from single-qubit $X$-and $Y$-measurements. This technique does not allow for postselection, and therefore has additional errors from excitation decays, but allows for resolution of the full correlations across the sample. It is used only in the following section \ref{sec:corr}.

\subsubsection{Experimental heating rate}
\label{sec:heating}

\begin{figure}
    \centering
    \includegraphics[width=\columnwidth]{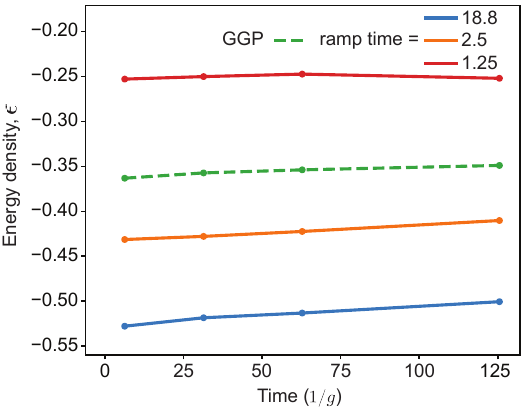}
    \caption{\textbf{Experimental heating rate.} Energy over hold time for ramp times that cover the experimentally used energy densities, and for the GGP (for $N_{\mathrm{q}}=97$). The ramps and GGP prepare a broad range of initial energies, and all have slow heating with a time constant of $2100/g$ (34 $\mu$s) or above. Only points after $6.3/g$ (100 ns) are shown, to avoid correlations due to transient non-equilibrium spin currents caused by the preparation (see \ref{sec:correlate}).}
    \label{fig:sm_heating}
\end{figure}

While ideal Hamiltonian evolution conserves energy, noise in a realistic experimental setting causes the system to approach infinite temperature ($\epsilon=0$) with time. Here, we show that the rate of this heating process is very slow compared to the coupling rate in our system. Figure \ref{fig:sm_heating} shows the experimentally measured heating rate from a variety of prepared states. Across ramps to prepare various thermal states, and GGP state preparation, the fastest heating time observed is $2100/g$ (34 $\mu$s).

\subsubsection{Experimental correlation length}
\label{sec:corr}

\begin{figure}
    \centering
    \includegraphics[width=\columnwidth]{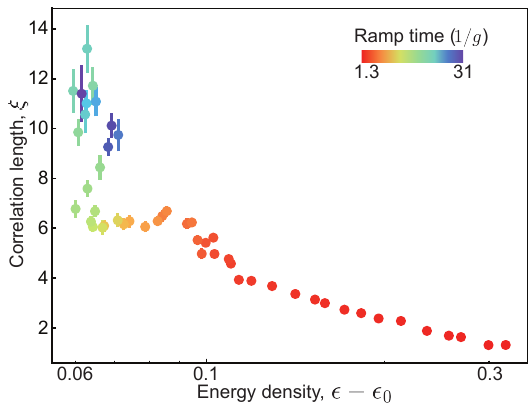}
    \caption{\textbf{Experimental correlation length vs energy.} Correlation length $\xi$ versus energy density and ramp duration (for $N_{\mathrm{q}}=97$). As ramp duration increases, we initially see a steady growth in correlation length, followed by non-monotonic behavior and increasing deviations from exponential decay that signify the onset of the Kosterlitz-Thouless transition regime \cite{Andersen2025}. The largest distance between any two qubits is $\sqrt{148}\approx12.1$ sites.}
    \label{fig:sm_corr}
\end{figure}

In Fig. \ref{fig:sm_corr}, we track the growth of the correlation length with system temperature, controlled by the speed of the ramp from the non-interacting initial state to the final state.  The correlation length is extracted from a linear fit to the logarithm of the mean correlation value vs distance. Note that unlike measurements of nearest-neighbor correlations, this does not use Bell state mapping or postselection, either for $\xi$ or $\epsilon$ (see \ref{sec:correlate}).

\subsection{Postselection fidelity}
To reduce the effect of decoherence on the measured dynamics, we postselect on shots having the correct number of measured qubits in the 1 state (described here as ``photons''). The effect of this mitigation method can be impacted by readout errors, although the latter are beneficially biased towards errors that reduce photon number and thus avoid erroneously canceling out T1 errors. Specifically, the probability of measuring 0 on a site given a state of 1 is $P_{01}=0.01$, while the probability of measuring 1 on a site given a state of 0 is $P_{10}=0.003$ (values of $P_{01}$ and $P_{10}$ are taken from means of measurements taken simultaneously on all qubits). To evaluate the mitigation efficacy, we consider a sample of our magnon data prepared at the lowest temperature, which has the lowest observed postselection rates as a consequence of the long ramp times needed to prepare the initial state. Measuring the probability of observing $[50, 49, \cdots 46]$ photons, where $N_\text{ph}=49$ is the target number of excitations for half-filling, we measure a distribution of values (values throughout this section shown to two significant digits) of $f_\text{{ph,Obs}} = [0.03, 0.22, 0.32, 0.24, 0.12, 0.04]$ at $t=0$ of the usual magnon experimental sequence. We then construct a simple excitation number confusion matrix using the measured mean readout errors for the 0 and 1 state, approximating independent errors on each site:
\begin{align}
M_{ij} = &\sum_{N_1=0}^{N_\mathrm{q}}\binom{j}{N_1}\binom{N_\mathrm{q}-j}{i-j+N_1}(1-P_{10})^{N_\mathrm{q}-i-N_1} \nonumber\\
&(P_{10})^{i-j+N_1}(P_{01})^{N_1}(1-P_{01})^{j-N_1}
\end{align}
in which $M_{ij}$ is the probability of measuring $i$ photons given an actual value of $j$ photons ($i,j \geq 0$) and $N_1$ is the number of $P_{01}$ errors.  Approximately solving the matrix equation $M f_{\text{ph}} = f_{\text{ph,Obs}}$, we find that $f_{\text{ph}} = [0.00,  0.34, 0.38, 0.20, 0.08, 0.00]$. Of the 0.22 observed probability of $N_\text{ph}=49$, the relative contribution of each number to this is $[0.00, 0.85, 0.14, 0.01, 0.00, 0.00]$. Thus, in this worst case scenario of the longest ramps, we expect readout errors to cause an admixture of roughly 0.85 at the intended excitation number, 0.14 at one excitation less, and 0.01 at all other photon numbers. 

\section{Hamiltonian learning}\label{sec:hamlearning}
Knowing the Hamiltonian under which the quantum system evolves is an essential challenge in analog quantum simulation. In Ref.\,\cite{Andersen2025}, we demonstrated a scalable protocol for computing the effective spin Hamiltonian (projected to two-level systems only), based on a series of single- and two-qubit calibration measurements. Using these measurements and known coupling efficiencies of the chip, the protocol numerically finds localized dressed qubit states and an effective spin Hamiltonian as matrix elements between them.

To further improve the effective spin model, we utilize Hamiltonian learning based on optimizing cross-entropy benchmarking (XEB) for a high-temperature state evolved to late times ($gt \sim 100$). In this regime, under ergodic dynamics, XEB provides an estimator of the wave function fidelity~\cite{ShawNature2024}. Because Hamiltonian errors cause fidelity loss to accumulate over time, XEB becomes increasingly sensitive to individual Hamiltonian terms as the system evolves~\cite{AruteNature19,neill2018blueprint}. Optimizing XEB at long evolution times therefore yields high-precision Hamiltonian learning with relatively few experimental shots, especially compared to learning schemes based on local observables~\cite{Bairey_2019,Huang_2020}. While prior work has demonstrated XEB optimization to learn a few unknown parameters\,\cite{BenchmarkingSoonwonTheoryPRL2023,Andersen2025,neill2018blueprint}, its application to learning an extensive number of parameters in large-scale systems has thus far been unexplored. 

The system size used in such XEB optimization should be large enough for two reasons. (i) First, the number of parameters in such a Hamiltonian scales polynomially with the qubit count $N_{\text{q}}$, while the total number of possible measurements (e.\,g. XEB speckle patterns) grows exponentially; therefore, it is necessary to consider large $N_{\text{q}}$ to avoid overfitting. (ii) Second, the effective Hamiltonian may include couplings beyond nearest-neighbor, e.\,g.~arising from higher-order mediated processes.
Correctly learning such couplings requires a subsystem large enough to accommodate them.

Optimization of a cost function such as XEB in a long time-evolution process with many qubits, is a computationally heavy task. The complexity of the time-evolution itself scales as $\mathcal{O}(D t W)$, where $D$ is the Hilbert space size, $t$ is the total evolution time and $W$ is the Hamiltonian bandwidth~\cite{Andersen2025}. In Refs.\,\cite{Wiebe_2014,Granade_2012}, various gradient-free approaches were used. Direct application of such approaches introduces an extra $\mathcal{O}(N_{\text{params}})$ runtime cost factor. More recently, with the advent of automatic differentiation libraries, it has become possible to efficiently compute gradients with respect to the Hamiltonian parameters using the ML-inspired forward/backward pass approach\,\cite{broughton2021tensorflowquantumsoftwareframework}. This approach, however, by default stores all intermediate ``hidden layers''. Applied naively to a quantum system, this would result in prohibitive $\mathcal{O}(N_{\text{steps}} \times D)$ memory cost, scaling with the total number of time evolution steps $N_{\text{steps}}$. 

The ML--inspired approach can be improved due to the invertible nature of quantum mechanics. Having only the wave function at the final time point, we can recover wave functions at any earlier moments of time. This {\it adjoint path} method has been applied in Ref.\,\cite{jones2020efficientcalculationgradientsclassical} to large-scale circuits, and in Ref.\,\cite{Leung_2017} to small-scale Hamiltonian dynamics. In this appendix, we show how combining this method with the {\it non-branching} Hamiltonian representation\,\cite{Wallerberger_2022}, Lin tables for $U(1)$ basis indexing~\cite{lin_tables} and \texttt{JAX}--powered GPU allowed us to perform $\mathcal{O}(10^2)$ gradient descent (or ADAM) steps in up to $\sim 30$-qubit patches with hundreds of optimization parameters within a day of simulation. 

\subsection{Hamiltonian terms}
As discussed in Ref.\,\cite{Andersen2025}, the effective Hamiltonian within the qubit subspace is generated by tracing out the coupler excited states and non-computational qubit states. In the perturbative regime --- set by the large qubit-coupler detuning $\Delta \sim 100\,g$ and anharmonicity $\eta \sim 20\,g$ --- the magnitude of terms in the effective spin Hamiltonian are exponentially suppressed by the number of photons they act on and their spatial extent.  
In this work, we include the terms large enough that they could be reliably learned at available experiment times:
\begin{align} \label{eq:H_full}
    \hat{H} &= \sum\limits_i \omega_i \hat{n}_i + \sum\limits_{d_{ij} \leqslant 3} g_{ij}^{XX} \hat{H}_{ij} + \sum\limits_{d_{\alpha \beta} = 1} g^{nn}_{\alpha \beta} \hat n_\alpha \hat n_\beta  \\ &+
     \sum_{\substack{d_{i\alpha}=1, \\ d_{\alpha j} = 1, 2}} g^{XnX}_{i\alpha j}\hat{H}_{ij} \hat{n}_{\alpha} 
     + \nonumber \sum_{\substack{d_{i \alpha} = d_{\alpha \beta} = d_{\beta j} = 1}} g^{XnnX}_{i\alpha \beta j}\hat{H}_{ij} \hat{n}_{\alpha} \hat{n}_{\beta}, 
\end{align}
where $d_{ij}$ is the Manhattan distance between sites $i$ and $j$ on the 2D grid, and $\hat{H}_{ij} = (\hat{X}_i \hat{X}_j + \hat{Y}_i \hat{Y}_j) / 2$. The first two terms with $d_{ij} = 1$ represent the standard 2D $XY$-model, which arises at leading order in $g$.
The remaining terms include direct couplings with $2 \leqslant d_{ij} \leqslant 3$ and perturbative processes arising at order $\mathcal{O}(g^2 / \eta)$,  $\mathcal{O}(g^2 / \Delta)$, and $\mathcal{O}(g^3 / \eta^2)$; for $g = 2\pi \cdot 10$ MHz, the latter processes are characterized by a typical magnitude of $\sim 0.1 g$, $\sim 0.02 g$, and $\sim 0.01\,g$, respectively.   
Potential higher-order terms with magnitudes $\lesssim 0.001\,g$, are truncated due to their negligible effect on XEB fidelity. In total, the above Hamiltonian in $24$ qubits contains 900 terms (as compared to $\sim100$ in the simplest nearest-neighbor XXZ Hamiltonian with $Z$-disorder), of which $\sim60\%$ are smaller than $0.01g$ and $\sim80\%$ are $<0.05g$.

\subsection{Scalable gradient computation}
In this section, we formulate the adjoint path method for continuous time-dynamics. Let us consider optimization of a scalar function of the form
\begin{gather}
    \label{eq:functional}
    \mathcal{C} = \langle r | \prod\limits_{i=0}^{N_t - 1} \hat{U}_i(\vec g) |\psi_0\rangle,
\end{gather}
where $\hat{U}_i(\vec g) = e^{-i \hat{H}(\vec{g}, t_i) \delta t}$ is a small time-evolution operator at step $i$, $|\psi_0\rangle$ is some initial state, $|r\rangle$ is the so-called {\it adjoint} vector, and $\vec{g}$ is a vector of parameters. The derivative reads
\begin{gather}
    \label{eq:derivative}
    \nabla_{\vec g} \mathcal{C} = \sum\limits_k \langle r | \prod\limits_{i = 0}^{k - 1} \hat{U}_i(\vec g) \left[\nabla_{\vec g} \hat{U}_k(\vec g)\right] \prod\limits_{j = k + 1}^{N_t - 1} \hat{U}_j(\vec g) |\psi_0\rangle,
\end{gather}
which suggests a computation algorithm: starting with two vectors
\begin{gather}
    |L\rangle = |r\rangle, \;
    |R\rangle = \prod\limits_i^{N_t - 1}\hat{U}_i(\vec g) |\psi_0\rangle,
\end{gather}
where $|R\rangle$ is the final wave function, initialize an array of derivatives $\nabla_{\vec g} \mathcal{C}$, filled with zeros, set $j = N_t - 1$ and repeat $N_t$ times:
\begin{gather}
    |R\rangle \leftarrow \hat{U}^{\dagger}_j(\vec g) |R\rangle, \\
    \nabla_{\vec g} \mathcal{C} \leftarrow \nabla_{\vec g} \mathcal{C} + \langle L | \nabla_{\vec g} \hat{U}_j(\vec g) | R\rangle, \\
    |L\rangle \leftarrow \hat{U}^{\dagger}_j(\vec g) |L\rangle, \\
    j \leftarrow j - 1.
\end{gather}

Applying inverse operations to $|R\rangle$ and $|L\rangle$ carries twice the cost of the forward pass. Without the loss of generality, we assume that $\hat{U}_j(\vec g)$ is approximated as a polynomial of Hamiltonians evaluated at different time points around the time $t_j$. Such a polynomial can be written down for any integration procedure, such as Runge-Kutta\,\cite{Dormand1980}. 

Hence, with the chain rule, derivative computation reduces to evaluating $\left[\nabla_{\vec g} \hat{H}(\vec g, t)\right] |R\rangle$. Each Hamiltonian term has its own optimized amplitude, so each component of $\nabla_{\vec g} \hat{H}(\vec g, t)$ has only a few matrix elements, computed using the {\it non-branching term} representation $(x, s, v)$\,\cite{Wallerberger_2022}. Therefore, computing $\left[\nabla_{\vec g} \hat{H}(\vec g, t)\right] |R\rangle$ is as costly as a single action $\hat{H}(\vec g, t)|R\rangle$, $\mathcal{O}(D N_{\text{params}})$, the same as the cost of the $N_{\text{params}}$ subsequent dot products with $|L\rangle$. Hence, the whole gradient requires only four times the forward pass computation and double the memory.

\subsection{Details of XEB optimization}
The cross-entropy benchmark (XEB) reads
\begin{gather}
    \text{XEB}(\vec g) = \frac{D}{M}\sum_x p_{\text{sim}}(x, \vec g) - 1,
\end{gather}
where $x$ is one of the $M$ bitstrings sampled from a quantum computer. Expanding the simulated probability, we obtain
\begin{gather}
    \text{XEB}(\vec g) = \frac{D}{M}\sum_x \vert\langle x|\prod\limits_i \hat{U}_i(\vec g)|\psi_0\rangle\vert^2 - 1.
\end{gather}

While, unlike Eq.~\eqref{eq:functional}, this cost function is non-linear in the time-evolution operator, its derivative can be written in the form of Eq.\,\eqref{eq:derivative}: 
\begin{gather}
    \nabla_{\vec g} \text{XEB}(\vec g) = 2 \text{Re}\left(\frac{D}{M}\sum_x \left[\nabla_{\vec g} \langle x|\prod\limits_i \hat{U}_i(\vec g)|\psi_0\rangle\right] \times \right. \\ \nonumber \left. \langle x|\prod\limits_i \hat{U}_i(\vec g)|\psi_0\rangle^* \right).
\end{gather}

Now, introducing:
\begin{gather}
    |r(\vec g)\rangle= 2 \frac{D}{M}\sum_x \langle x |\prod\limits_i \hat{U}_i(\vec g)|\psi_0\rangle |x\rangle,
\end{gather}
we obtain 
\begin{gather}
\nabla_{\vec g} \text{XEB}(\vec g) = \text{Re} \left[\langle r(\vec g) | \nabla_{\vec g} \left( \prod\limits_i \hat{U}_i(\vec g) \right)|\psi_0\rangle\right],
\end{gather}
matching the Eq.\,\eqref{eq:derivative} representation. Note that $\ket{r(\vec g)}$ has to be recomputed every time the parameters $\vec g$ are updated. 

\subsection{Learning results}

\begin{figure}[t!]
    \centering
    \includegraphics[width=\columnwidth]{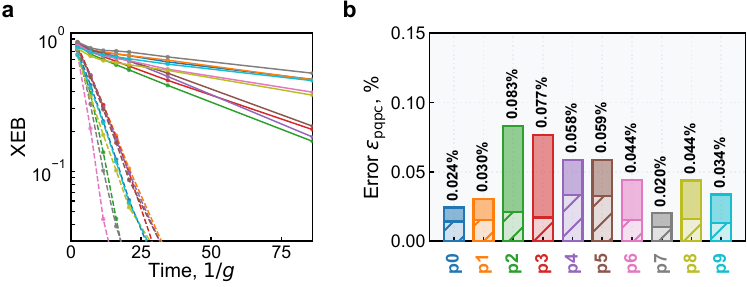}
    \caption{\textbf{XEB improvement from Hamiltonian learning in $24$ qubits.} 
    {\bf a,} XEB curves before (dashed) and after (solid) learning in ten representative 24-qubit systems. The optimization is performed in a multi-stage fashion at gradually increasing times of 17, 34, 86 and 170 cycles. At 170 cycles, the post-selected shot count is $\sim 10^4$ per patch. {\bf b,} The resulting post-learning coherent (opaque) and incoherent (transparent) errors of 24-qubit patches.}
    \label{fig:xeb_impr}
\end{figure}

In this work, we use the ADAM optimizer and learn each term in Eq.\,\eqref{eq:H_full}\,\cite{kingma2015adam}. We first optimize at short times to ensure a sufficient fidelity for the initially unoptimized Hamiltonian, and then progressively move to longer times to increase the sensitivity to Hamiltonian errors, performing the last step at $gt \sim 170$ evolution cycles.

A typical example of the learning procedure is shown in Fig.\,\ref{fig:xeb_impr}, where we find post-training per-qubit-per-cycle errors of only $4.7 \times 10^{-4}$ in 24-qubit optimizations with an average incoherent error of $1.9 \times 10^{-4}$, leaving a remaining control error of $2.8 \times 10^{-4}$. Importantly, training at a particular time increases the XEB at \emph{all} times. This validates that we are not overfitting the XEB signal at a single time slice but are genuinely improving the Hamiltonian description of the device. Furthermore, due to the large Hilbert space and long evolution times, the XEB speckle patterns caused by incoherent errors are near-orthogonal~\cite{manole2025learnquantumrandomcircuit} with $\mathcal{O}(1 / \sqrt{D})$ corrections. This allows us to independently learn the Hamiltonian parameters in an unbiased way without taking into account incoherent processes.

\subsection{Hamiltonian patching protocol}
When performing Hamiltonian learning in subsystems, there is a trade-off between the accuracy of capturing longer-range effects and the speed of learning. On the one hand, the subsystem needs to be sufficiently large that it captures non-local hybridization effects and thus leads to Hamiltonian terms that accurately match a larger system\footnote{As we shall see from the patching algorithm below, we only need each term in the larger system to be accurately represented in one of the smaller patches, not the other way around.}. On the other hand, the computational time for optimization increases exponentially with system size and in practice is limited to $\sim 30$ qubits. We find that patches of 24 qubits is appropriate for learning, resulting in accurate results (see below) within a few hours\footnote{Actual numbers depend on the evolution time we train our XEB at, and the hardware used. With $\mathcal{O}(10^3)$ terms, a single optimization step at $\sim 90$ cycles with an H100 NVIDIA GPU takes about 15 minutes.}. 

When patching together the Hamiltonian of a larger system, $P$, we evaluate the usefulness of each term in a learned patch, $p$, according to the following ``neighborhood similarity'' metric:
\begin{equation}
M=\sum_{q\in P\cup p} 10^{-d(q)} -\sum_{q\in p\setminus P} 10^{-d(q)}.
\end{equation}
Here, $d(q)$ is the Manhattan distance from the qubit $q$ to the closest qubit in the Hamiltonian term support. The exponential decay in $d(q)$ is motivated by how the hybridization effects fall off with distance, and the base 10 was chosen to optimize performance. Intuitively, the metric rewards patch $p$ for any qubit it contains that is also in $P$, and penalizes it for any non-matching qubits, weighted by the expected effect of the qubit on the strength of the term. For each term in the Hamiltonian of $P$, we use the learned value in the patch with highest neighborhood similarity. 

\subsection{Fidelity benchmarking after patching}

\begin{figure}[t!]
    \centering
    \includegraphics[width=\columnwidth]{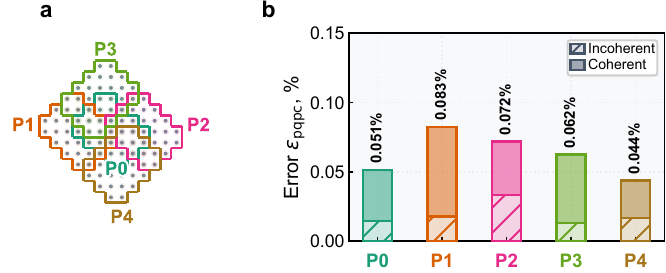}
    \caption{\textbf{XEB in the $24 \to 31$ patching.} 
    {\bf a,} 31-qubit systems covering device 1. {\bf b,} The resulting coherent and incoherent errors obtained after patching the respective 31-qubit systems.
    }
    \label{fig:xeb_patching}
\end{figure}

In Fig.\,\ref{fig:xeb_patching}, we benchmark the fidelity after learning and patching, by evaluating the XEB fidelity in 31-qubit systems that were patched using the algorithm described above. The coherent error increases by only $1.5 \times 10^{-4}$ per qubit per cycle (pqpc) after patching, from $2.8 \times 10^{-4}$ in the 24-qubit subsystems used for learning, to $4.3\times10^{-4}$ pqpc after patching the 31-qubit systems. The average incoherent error is found to be $1.9 \times 10^{-4}$, resulting in a total error of $6.2 \times 10^{-4}$. The total error rates were obtained using direct state-vector simulation and XEB benchmarking, while incoherent rates were estimated from the ratio between the experimental and simulated (ideal) self-XEB~\cite{Boixo_2018,Andersen2025}:
\begin{gather}
    F_{\text{incoherent}} = \sqrt{\frac{\text{self-XEB}_{\text{exp.}}}{\text{self-XEB}_{\text{sim.}}}}.
\end{gather}

The experimental self-XEB in 31-qubit patches was obtained using an unbiased estimator~\cite{Andersen2025} that has low variance for shot counts $M \gg \sqrt{D}$, which is made possible in our platform by its high repetition rate.

The corresponding fidelity after 10 cycles of evolution (typically sufficient to determine magnon frequency and decay rate) is 55\,\% for 97 qubits, and 34\,\% after including fidelity loss due to one layer of injected $Z$-gates (see Sec.~\ref{sec:zerr}). In the case where an additional 100\,ns of thermalization is included and magnon excitation is performed in a separate $Z$-gate layer from GGP, the total fidelity is estimated to be 12\,\%.

\subsection{Accuracy improvement after Hamiltonian learning}
\begin{figure}[t!]
    \centering
    \includegraphics[width=\columnwidth]{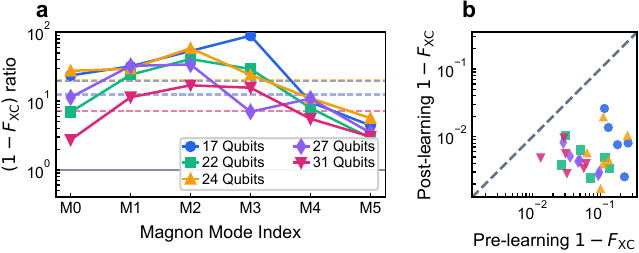}
    \caption{\textbf{Accuracy improvement from Hamiltonian learning.} 
    {\bf a,} Ratio of $1 - F_{\text{XC}}$ (see definition in Eq.\,\eqref{eq:xc}) between post- and pre-optimized Hamiltonians for various system sizes. {\bf b,} Pre- and post-optimization $1 - F_{\text{XC}}$.
    }
    \label{fig:snr_impr}
\end{figure}

Having verified that Hamiltonian learning significantly improves control errors and scales favorably with patching, we verify that this also leads to better agreement in a local observable. As a measure of agreement, we use the $F_{\text{XC}}$ metric (see Eq.\,\eqref{eq:xc}) being the normalized correlation between signals maximized over possible time shifts between signals, and $0 \leqslant F_{\text{XC}} \leqslant 1$. In Fig.\,\ref{fig:snr_impr}, we show that the deviation of this metric $1 - F_{\text{XC}}$ from the perfect signal decreases with learning by approximately an order of magnitude to values consistently under $10^{-2}$.

\section{Magnon theory}\label{sec:magnons_construction}
The 2D $XY$-model can be studied analytically by means of the Holstein-Primakoff (HP) transformation~\cite{holstein_field_1940}. This mean-field technique allows us to get insights about the excitations of the system at a low computational cost. The transformation is based on a high spin $S$ expansion around the classical magnetization of the system. For this analysis let us consider the spin Hamiltonian
\begin{equation}    \label{eq:initial_spin_Hamiltonian}
    \mathcal{H}=J\sum_{\langle i,j\rangle}\left(S^x_iS^x_j+S^y_iS^y_j+\Delta S^z_iS^z_j\right)
\end{equation}
where we also introduce the spin anisotropy $\Delta$. In comparison with the main text Hamiltonian, we have $J=g/(2S^2)$.

The Holstein-Primakoff transformation involves expanding the spin operators in terms of bosonic particles (spin waves or magnons) along a quantization axis. This axis of quantization has to be chosen along the classical magnetization of the system in order to end up with a quadratic Hamiltonian at first order, the so-called linear spin wave (LSW) Hamiltonian. In the pure $XY$ phase the classical magnetization pattern is antiferromagnetic in the $x$-direction. Hence, we take $\pm \hat{x}$ as quantization axis on the two sublattices.

We now define a rotated coordinate system for convenience. Using $S^{\hat{\alpha}}$ for this rotated frame we get the Hamiltonian
\begin{align}    \label{eq:parametric_rotation}
    \mathcal{H}=&-J \sum_{\langle i,j\rangle}\Big[\Delta S^{\hat{x}}_iS^{\hat{x}}_j- S^{\hat{y}}_iS^{\hat{y}}_j+S^{\hat{z}}_iS^{\hat{z}}_j\Big]. \nonumber\\
\end{align}

We apply a spin-wave expansion using the Holstein-Primakoff (HP) transformation \cite{holstein_field_1940}
and perform the expansion for high $S$ and low $n_i$\footnote{Note that the smallness parameter in this expansion is $\langle \hat{n}_i \rangle / (2 S)$, and due to the strong magnetic ordering, the magnon density in the ground state $\langle \hat{n}_i \rangle \sim 0.197$~\cite{anderson1952approximate}.}
\begin{subequations}
\begin{align}
S^{\hat{x}}_i&=\sqrt{\frac{S}{2}}(a_i+a^\dagger_i) + \mathcal{O}(a^3)\\
S^{\hat{y}}_i&=-i\sqrt{\frac{S}{2}}(a_i-a^\dagger_i) + \mathcal{O}(a^3)\\
S^{\hat{z}}_i&=S-a^\dagger_ia_i
\end{align}
\end{subequations}

We can now perform the Holstein-Primakoff transformation and separate the terms linear, quadratic and at higher order in the bosons $a_i$. With the correct choice of quantization axis, the linear terms vanish, leaving as first order approximation the LSW Hamiltonian. In the case of a periodic system, we can analytically solve the model by transforming to momentum space and performing a Bogoliubov transformation \cite{gomez-santos_application_1987}, which yields
\begin{equation*}
    \mathcal{H}_{\text{PBC}}/N_{\text{s}}=E_{\text{GS}}+\frac{1}{N_{\text{s}}}\sum_\mathbf{k}\epsilon(\mathbf{k})b^\dagger_\mathbf{k}b_\mathbf{k}
\end{equation*}
with the ground state energy
\begin{align*}
    E_{\text{GS}} = &-2JS(S+1)+ \frac{1}{2N_{\text{s}}}\sum_\mathbf{k}\epsilon(\mathbf{k})
\end{align*}
where $\epsilon(\mathbf{k})$ is the magnon $b_\mathbf{k}$ dispersion, given by
\begin{align*}
    \epsilon(\mathbf{k})=&\sqrt{A_\mathbf{k}^2-B_\mathbf{k}^2} \\
    A_\mathbf{k}=&4JS\Big(1+\frac{\gamma_\mathbf{k}}{2}(1-\Delta)\Big) \\
    B_\mathbf{k}=&-2JS\gamma_\mathbf{k}(1+\Delta)
\end{align*}
with $\gamma_{\bf k}=\frac{1}{2}\big[\cos(k_x)+\cos(k_y)\big]$.
Since we want to describe a system with open boundaries, we work in real space coordinates. We use as basis the wave function $\psi^\dag=\left( a_1^\dag, a_2^\dag\dots a_{N_s}^\dag,a_1,a_2\dots a_{N_s} \right)$. By following the procedure detailed in Ref.~\cite{colpa_diagonalization_1978}, we get the Bogoliubov transformation matrices $U_{ik}$ and $V_{ik}$
\begin{gather}
\label{eq:a_to_b}
    a_i =\sum_{\mu=1}^{N_s} \left( U_{i\mu}b_{\mu}+ V_{i\mu}b^\dag_{\mu}   \right)    \\ \nonumber
    a^\dag_i =\sum_{\mu=1}^{N_s}\left( V_{i\mu}b_{\mu}+ U_{i\mu}b^\dag_{\mu} \right).
\end{gather}

\subsection{Creation of a coherent magnon state}
To create a coherent individual mode magnon state, we define the $\mu$-th magnon mode position operator as
\begin{gather}
    \hat{\mathcal{X}}_{\mu} = \frac{2}{\pi} \left(b_{\mu}+b_{\mu}^\dag \right) = \sum_i \hat{Z}_i \varphi_i^{\mu},
\label{eq:posop}
\end{gather}
a weighted sum of individual $\hat{Z}_i$ actions. We note that in the pure $XY$-model with antiferromagnetic order $\hat{Z}_j= \left( \hat{a}_j+\hat{a}^\dag_j \right) / \sqrt{2S}$; hence, using Eq.\,\eqref{eq:a_to_b}, we obtain
\begin{equation}
\label{eq:functions}
    \varphi_{i}^{\mu}=(-1)^{\#i}\frac{2}{\pi}\left(U_{i\mu}-V_{i\mu}\right),
\end{equation}
where the staggered spatial modulation $(-1)^{\#i}$ is needed to take into account the staggered quantization axis in the system. Therefore, we create such a state by performing $Z$-rotations on individual qubits
\begin{equation} \label{eq:magnon_creation}
    \vert \mu\rangle = e^{\pm i\frac{A}{2}\sum_j\varphi_{j}^{\mu}\hat{Z}_j}\vert 0\rangle = e^{\pm i \frac{A}{2} \hat{\mathcal{X}}_{\mu}} |0\rangle.
\end{equation}
with angles $\varphi_i^{\mu}$ defined in Eq.\,\eqref{eq:functions} and the average number of magnons $\langle \mu| b_{\mu}^\dag b_{\mu} |\mu\rangle = A^2/\pi^2$. We can also {\it measure} the magnon position operator  $\hat{\mathcal{X}}_{\mu}$by linearly superposing the individual $\hat{Z}_i$ qubit measurements.

\begin{figure}[t!]
    \centering
    \includegraphics[width=\columnwidth]{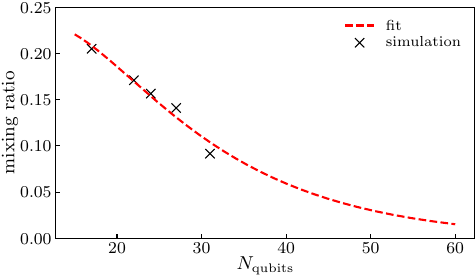}
    \caption{{\bf Robust magnon states}. Mixing between magnon modes (see definition in the text), obtained from exact simulation. The fit $\mathcal{M}(N) = B N e^{-a N} / (1 - B N e^{-a N})$ assumes leakage into $N$ possible modes decays exponentially with the system size. We observe clear reduction in the mixing ratio, indicating robust magnon modes.}
    \label{fig:mixing}
\end{figure}

To confirm that the functions in Eq.\,\eqref{eq:functions} create robust magnon modes, we evaluate the mixing between magnon modes defined as $\mathcal{M}(N)=\sum\limits_{m \neq n} \sum_t P(t, m, n) / \sum_{m} \sum_t P(t, m, m)$, where $P(t, m, n)$ is the population in mode $m$ at time $t$ after excitation in mode $n$. As shown in Fig.\,\ref{fig:mixing}, the degree of mixing quickly decays with system size, indicating that the functions are indeed robust. Notably, if we chose random functions instead, using the average projection on an $N$--dimensional sphere, we would expect mixing $\mathcal{M}(N) \propto N$. 

\subsection{Linear magnon response measurements}
Following excitation and evolution under the Hamiltonian for time $t$, our state is given by:
\begin{gather}
\ket{\psi_{\pm}(t)}=e^{-i\hat H t}e^{\pm i\frac{A}{2}\sum_i \varphi_i^{\mu}\hat{Z}_i}\ket{\psi_0},
\end{gather}
where $\ket{\psi_0}$ is the prepared low-temperature state. Hence, the expectation of $\hat{Z}_j$ after time $t$ yields:
\begin{gather}
\label{eq:protocolforresponsefunctions}
\langle Z_j(t)\rangle_{\pm}=\bra{\psi_0}e^{\mp i\frac{A}{2}\sum_i \varphi_i^{\mu}\hat{Z}_i} \hat{Z}_j(t) e^{\pm i\frac{A}{2}\sum_i \varphi_i^{\mu}\hat{Z}_i}\ket{\psi_0}.
\end{gather}

For simplicity, we start with the case of small $A$. To arrive at the response function, we isolate the terms that are even order in $\hat{Z}$ by subtracting the results from excitations with opposite signs:
\begin{gather}
\label{eq:protocolforresponsefunctions2}
\langle \hat{Z}_j(t)\rangle_--\langle \hat{Z}_j(t)\rangle_+=-iA\sum_i\varphi_i^{\mu}\langle [\hat{Z}_j(t),\hat{Z}_i(0)] \rangle + \mathcal{O}(A^3).
\end{gather}
Finally, computing the overlap with mode $\mu$ gives (using Eq.~\eqref{eq:posop})
\begin{gather}
\label{eq:greenf}
-iA\sum_{ij}\varphi_i^{\mu}\varphi_j^{\mu}\langle[\hat{Z}_j(t),\hat{Z}_i(0)] \rangle 
\approx \\ \nonumber
-i(4 A / \pi^2)\left(\langle[\hat{b}_{\mu}(t),\hat{b}_{\mu}^{\dagger}(0)]\rangle+\langle[\hat{b}^{\dagger}_{\mu}(t),\hat{b}_{\mu}(0)]\rangle\right)=\\ \nonumber (8A/\pi^2) \, \mathrm{Re}\,G^{\text{R}}_{\mu}(t),
\end{gather}
where $G^{\text{R}}_{\mu}(t)=-i\theta(t)[b_{\mu}(t),b_{\mu}^{\dagger}(0)]$ is the retarded Greens function of the $\mu$-th magnon mode, and terms quadratic in $b_{\mu}$ or $b^{\dagger}_{\mu}$ were neglected assuming weakly-interacting magnon modes. We note that while our experiment measures the real part of $G^{\text{R}}_{\mu}$ in the time domain, it contains both real and imaginary parts in Fourier space.

For arbitrary $A$ and a coherent state $\exp\left(\alpha \hat{b}_{\mu} - \alpha^* \hat{b}^{\dagger}_{\mu}\right) |0\rangle$ evolution without mode mixing rotates the coherent amplitudes $\alpha(t) = \alpha e^{-i \omega_{\mu} t}$ with $\omega_{\mu}$ being the mode energy. Hence, by the definition of a coherent state, $\langle \hat{\mathcal{X}}_{\mu} \rangle = 2 \, \text{Re} \, \alpha(t) = (2A / \pi) \sin \omega_{\mu} t,$
which gets modified by the exponential mode decay. In the limit of non-interacting magnon modes, this expression is equivalent to Eq.\,\eqref{eq:greenf}, since $\sum_i \varphi_i^{\mu} (\langle \hat{Z}_i(t)\rangle_--\langle \hat{Z}_i(t)\rangle_+) = \frac{2}{\pi} (\langle \hat{\mathcal{X}}_{\mu} \rangle_- - \langle \hat{\mathcal{X}}_{\mu} \rangle_+) = -\frac{8A}{\pi^2} \sin \omega_\mu t$.

\subsection{Measuring magnon pair processes}
Here, we discuss the measurements of magnon pairs. We simultaneously excite modes $\mu$ and $\nu$ with signs $s_\mu, s_\nu = \pm$, and the state evolves as: $\ket{\psi_{s_\mu s_\nu}(t)}=e^{-i\hat H t} \exp\left[ i s_\mu A \hat{\mathcal{X}}_\mu / 2 + i s_\nu A \hat{\mathcal{X}}_\nu / 2 \right]\ket{\psi_0}$. We measure the pair correlator $\langle \hat{\mathcal{X}}_\eta(t) \hat{\mathcal{X}}_\lambda(t) \rangle$. To extract the interaction, we superpose the experiments with $s_\mu, s_\nu = \pm$. First, assuming small $A$:
\begin{gather}
    \sum_{s_\mu, s_\nu = \pm} s_\mu s_\nu \langle \hat{\mathcal{X}}_\eta(t) \hat{\mathcal{X}}_\lambda(t) \rangle_{s_\mu, s_\nu} = \nonumber \\ 
    = -A^2 \langle \psi_0 | \left[ \left[ \hat{\mathcal{X}}_\eta(t)\hat{\mathcal{X}}_\lambda(t), \hat{\mathcal{X}}_\mu(0) \right], \hat{\mathcal{X}}_\nu(0) \right] | \psi_0 \rangle.
\end{gather}

Only cross terms $\propto s_{\mu} s_{\nu}$ survive, capturing non-trivial two-magnon processes. Subtracting the disconnected contribution $\langle \hat{\mathcal{X}}_\eta \rangle \langle \hat{\mathcal{X}}_\lambda \rangle$, the connected physical signal $S$ is the sum of the pure quantum scattering amplitude:
\begin{gather}
    S = A^2 (2 / \pi)^4 \langle \hat{b}_\eta(t) \left[ \left[ \hat{b}_\lambda(t), \hat{b}^\dagger_\mu(0) \right], \hat{b}^\dagger_\nu(0) \right] + \nonumber \\ 
    + \left[ \left[ \hat{b}_\eta(t), \hat{b}^\dagger_\mu(0) \right], \hat{b}^\dagger_\nu(0) \right] \hat{b}_\lambda(t) \rangle + \text{h.c.}.
\end{gather}

To time-evolve the bosonic operators, we write the effective interacting magnon Hamiltonian:
\begin{gather}
    \hat{H}_{\text{magnon}} = \sum_\kappa \omega_\kappa \hat{b}^{\dagger}_\kappa \hat{b}_\kappa + \frac{1}{2}\sum_{\kappa\rho\sigma\tau} \chi_{\kappa\rho\sigma\tau} \hat{b}^{\dagger}_\kappa \hat{b}^{\dagger}_\rho \hat{b}_\sigma \hat{b}_\tau,
\end{gather}
and in the first order in the interaction, 
\begin{gather}
    \hat{b}_\eta(t) \approx e^{-i\omega_\eta t} \left[ \hat{b}_\eta(0) - \right. \\ - \nonumber \left. i \sum_{\lambda,\mu,\nu} \chi_{\eta\lambda\mu\nu} \hat{b}^\dagger_\lambda(0) \hat{b}_\mu(0) \hat{b}_\nu(0) \mathcal{I}(\Delta \omega_{\text{pair}}, t) \right],
\end{gather}
where $\mathcal{I}(x, t) = \int_0^t e^{i x t'} dt' = t e^{i x t/2} \text{sinc}(x t/2)$ ensures energy conservation at large times, and $\Delta \omega_{\text{pair}} = \omega_\eta + \omega_\lambda - \omega_\mu - \omega_\nu$. Therefore, for the signal, we get
\begin{gather}    
    S = \left( -2i A^2 (2 / \pi)^4 \chi_{\eta\lambda\mu\nu} \mathcal{I}(\Delta \omega_{\text{pair}}, t) e^{-i(\omega_\eta + \omega_\lambda)t} \right) + \text{h.c.} = \nonumber \\
    = 4 A^2 (2 / \pi)^4 \, \text{Im} \left[ \chi_{\eta\lambda\mu\nu} \mathcal{I}(\Delta \omega_{\text{pair}}, t) e^{-i(\omega_\eta + \omega_\lambda)t} \right].
\end{gather}
This correlator captures the probability amplitude for the initial excitations to scatter into modes $\eta$ and $\lambda$ at time $t$. For arbitrary $A$, at weak interaction, the pumped modes still behave as coherent states with time-evolved amplitudes:
\begin{gather}
    \alpha_\gamma(t) \approx \left[ \alpha_\gamma(0)\right. - \\ - \nonumber \left.2i \sum_{\rho,\sigma,\tau} \chi_{\gamma\rho\sigma\tau} \alpha_\rho^*(0) \alpha_\sigma(0) \alpha_\tau(0) \mathcal{I}(\Delta \omega_\gamma, t) \right] e^{-i \omega_\gamma t}.
\end{gather}
Calculating the connected part of the phase-cycled measurement $\sum s_\mu s_\nu \langle \hat{\mathcal{X}}_\eta \hat{\mathcal{X}}_\lambda \rangle$ using these independent coherent amplitudes, we arrive at the exact same result for $S$. Thus, the method extracts the $2 \to 2$ scattering tensor $\chi_{\eta\lambda\mu\nu}$.

\subsection{Magnon interactions}
In order to explain theoretically the processes happening when exciting a particular magnon mode (Eq.~\eqref{eq:magnon_creation}), we consider the situation in which at temperature $T$, a non-thermal population of magnons in a specific mode $n$ is generated and is free to interact with the thermal environment. Given the small sizes of the lattices considered we expect momentum conservation to not restrict considerably the allowed processes, so we disregard it and consider only energy conservation. At lowest order the processes which will likely play a role are $1\leftrightarrow3$ and $2\leftrightarrow2$. For each of the processes, one needs to derive the corresponding vertex from the bosonic Hamiltonian. For symmetry reasons the $1\leftrightarrow2$ process is not allowed in the pure $XY$ phase, and we expect the $2\leftrightarrow2$ process to be the dominant source of decay since it is thermally activated by the thermal background inherently created in the state preparation.

For the $2\leftrightarrow2$ and $1\leftrightarrow3$ processes we derive the vertex associated with the four-boson part of the Hamiltonian. By inspection of Eq.~\eqref{eq:parametric_rotation}, we can determine that it derives from the terms $S^{\hat{z}}S^{\hat{z}}$ and from the higher order expansion of terms $S^{\hat{x}}S^{\hat{x}}$ and $S^{\hat{y}}S^{\hat{y}}$. We find the vertex to be given by three terms
\begin{equation}\label{eq:4bosonH}
\begin{split}
    \hat{V}^4=\sum_{\langle i,j\rangle}\bigg\{ &f_1\big[(a_ia^\dag_j+a^\dag_ia_j)a^\dag_ja_j + i\leftrightarrow j\big] \\
    +&f_2\big[(a_ia_j+a^\dag_ia^\dag_j)a^\dag_ja_j + i\leftrightarrow j\big] \\
    +&f_3\big[a^\dag_ia_ia^\dag_ja_j + i\leftrightarrow j\big]\bigg\}
\end{split}
\end{equation}
with parameters
\begin{equation*}
    f_1 = -\frac{J}{8}(1-\Delta), \quad
    f_2 = \frac{J}{8}(1+\Delta), \quad
    f_3 = -\frac{J}{2}
\end{equation*}
The matrix elements for the $2\leftrightarrow2$ process, which we define as
\begin{equation*}
    \hat{V}^{2\rightarrow2}= \sum_{nlmp} V_{n,l}(m,p) b^\dag_mb^\dag_pb_nb_l
\end{equation*}
are derived by expanding the Hamiltonian in Eq.~\eqref{eq:4bosonH} in terms of Bogoliubov bosons and keeping the terms with $2$ creation and $2$ annihilation operators. Note that we here use the notation $V_{n,l}(m,p)$ instead of $V_{nlmp}$ as used in the main text to differentiate from $1\leftrightarrow 3$ processes considered later on. We need to symmetrize the vertex over the $n\leftrightarrow l$, $m\leftrightarrow p$ and $nl\leftrightarrow mp$ indices. We then obtain
\begin{widetext}
\begin{equation*}
\begin{split}
V_{n,l}(m,p) = &\frac{f_1}{4} \sum_{\langle i,j\rangle} \Big[ U_{jn} V_{il} U_{jm} U_{jp} + 2V_{in} V_{jl} U_{jm} V_{jp} +2U_{jn}V_{jl}U_{im}U_{jp} 
+V_{jn}V_{jl}U_{im}V_{jp}+2U_{in}U_{jl}U_{jm}V_{jp} + U_{jn}V_{il}V_{jm}V_{jp} \\
&+ U_{jn}U_{jl}U_{jm}V_{ip} +2U_{jn}V_{jl}V_{im}V_{jp}  + (n\leftrightarrow l;m\leftrightarrow p;nl\leftrightarrow mp) \Big] \\
+ &\frac{f_2}{4} \sum_{\langle i,j\rangle} \Big[ U_{in} U_{jl} U_{jm} U_{jp} + 2U_{in} V_{jl} U_{jm} V_{jp} + 2U_{jn}V_{jl}U_{jm}V_{ip} 
+V_{jn}V_{jl}V_{im}V_{jp}+2U_{jn}V_{il}U_{jm}V_{jp} + V_{in}V_{jl}V_{jm}V_{jp} \\
&+U_{jn}U_{jl}U_{im}U_{jp} +2U_{jn}V_{jl}U_{im}V_{jp} +  (n\leftrightarrow l;m\leftrightarrow p;nl\leftrightarrow mp) \Big]\\
+&\frac{f_3}{4} \sum_{\langle i,j\rangle} \Big[ U_{in} V_{il} U_{jm} V_{jp} + U_{in} U_{jl} U_{im} U_{jp} + U_{in}V_{jl}U_{im}V_{jp} 
+U_{jn}V_{il}U_{jm}V_{ip} + V_{in}V_{jl}V_{im}V_{jp} + U_{jn}V_{jl}U_{im}V_{ip} \\
&+ (n\leftrightarrow l;m\leftrightarrow p;nl\leftrightarrow mp) \Big] + (i\leftrightarrow j)
\end{split}
\end{equation*}
\end{widetext}

Similarly, the matrix elements for the $1\leftrightarrow3$ vertex
\begin{equation*}
    \hat{V}^{1\rightarrow3}= \sum_{nlmp} V_n(l,m,p) \left(b^\dag_lb^\dag_mb^\dag_pb_n + H.c.\right)
\end{equation*}
are obtained by keeping the terms with $1$ creation (annihilation) and $3$ annihilation (creation) operators. We need here to symmetrize the vertex over the $l\leftrightarrow m\leftrightarrow p$ indices and obtain
\begin{widetext}
\begin{equation*}
\begin{split}
    V_n(l,m,p) =& \frac{f_1}{6} \sum_{\langle i,j\rangle}\Big[V_{in} V_{jl} U_{jm}U_{jp} + U_{jn}U_{il}U_{jm}U_{jp} +2V_{jn}U_{il}V_{jm}U_{jp}+U_{in}V_{jl}V_{jm}U_{jp}+2U_{jn}V_{il}V_{jm}U_{jp} + V_{jn}V_{il}V_{jm}V_{jp} \\ &+ (l\leftrightarrow m\leftrightarrow p)\Big] \\
    +& \frac{f_2}{6} \sum_{\langle i,j\rangle}\Big[U_{in} V_{jl} U_{jm}U_{jp} + U_{jn}V_{il}U_{jm}U_{jp} +2V_{jn}V_{il}V_{jm}U_{jp}+V_{in}V_{jl}V_{jm}U_{jp}+2U_{jn}U_{il}V_{jm}U_{jp} + V_{jn}U_{il}V_{jm}V_{jp} \\
    &+ (l\leftrightarrow m\leftrightarrow p)\Big] \\
    +& \frac{f_3}{6} \sum_{\langle i,j\rangle}\Big[U_{in} U_{il} V_{jm}U_{jp} + V_{in}V_{il}V_{jm}U_{jp} +U_{jn}U_{jl}V_{im}U_{ip}+V_{jn}V_{jl}V_{im}U_{ip} + (l\leftrightarrow m\leftrightarrow p)\Big] + (i\leftrightarrow j)
\end{split}
\end{equation*}
\end{widetext}

\subsection{Scattering and decay rates} \label{sec:theorydecay}
The $2\leftrightarrow2$ vertex describes a process in which an activated magnon $n$ scatters with a thermal mode into two others. This is thus a thermally activated process, whose decay rate can be computed by considering the magnon self-energy Feynman diagrams. The non-interacting magnon Greens function in imaginary time is 
\begin{equation*}
\begin{split}
    G_k(\tau) &= - \langle T_\tau b_k(\tau) b^\dagger_k\rangle \\
        &= - e^{-\epsilon_k \tau} \left[ \Theta(\tau) (n_B(\epsilon_k) + 1)
        + \Theta(-\tau) n_B(\epsilon_k)
        \right]
\end{split}
\end{equation*}
with $n_B(\epsilon_k)$ the Bose occupation factor. The decay rate of the coherent magnon amplitude, which is what we measure (half of the population decay rate) is related to the imaginary part of the self-energy by the standard relation 
\begin{equation*}
    \Gamma_k=-\text{Im}\;\Sigma_k(\epsilon_k).
\end{equation*}
We compute the self-energy in the interaction picture by expanding to second order (factor $1/2!$) and performing the Wick contractions ($8$ possibilities)
\begin{equation*}
    \Sigma_k(\tau) =4 \sum_{lmn} |V_{k,l}(m,n)|^2
        G_l (-\tau)
        G_m (\tau)
        G_n (\tau)
\end{equation*}
We then Fourier transform to Matsubara frequency, make the analytic continuation and take the imaginary part to obtain 
\begin{equation*}
\begin{split}
    \text{Im}\:\Sigma_k(\epsilon_k) & = 
        - 4 \pi \sum_{lmn} |V_{k,l}(m,n)|^2 
    n_B(\epsilon_l)(n_B (\epsilon_m)+1) \times \\ 
    &\times(n_B (\epsilon_n)+1)
    (e^{-\beta \epsilon_k}-1) \delta(\epsilon_k + \epsilon_l - \epsilon_m - \epsilon_n)
\end{split}
\end{equation*}
Finally, by inserting the Bose occupation terms $n_B(\epsilon) = \frac{1}{e^{\beta \epsilon} -1 }$ we get the scattering rate of the process
\begin{align*}
    \Gamma^{2\leftrightarrow2}_k=4\pi &\sum_{l,m,n} \vert V_{k,l}(m,n)\vert^2 \delta_\gamma(\epsilon_k+\epsilon_l-\epsilon_m-\epsilon_n)\\
    &\times\frac{(1-e^{-\beta\epsilon_k})e^{\beta(\epsilon_m+\epsilon_n)}}{(e^{\beta\epsilon_l}-1)(e^{\beta\epsilon_m}-1)(e^{\beta\epsilon_n}-1)}
\end{align*}

We proceed in the same way to compute the decay rate of the $1\leftrightarrow3$ process. In this case we have to distinguish between two relevant processes: either the activated mode decays into three thermal modes ($1\rightarrow3$) or it scatters with two thermal modes into another thermal mode ($3\rightarrow1$). We compute the rates as above, by keeping count of all the possible combinations to get
\begin{align*}
    \Gamma^{1\rightarrow3}_k=6\pi &\sum_{l,m,n}\vert V_k(l,m,n)\vert^2\delta_\gamma(\epsilon_k-\epsilon_l-\epsilon_m-\epsilon_n) \\
    &\times\frac{(1-e^{-\beta\epsilon_k})e^{\beta(\epsilon_l+\epsilon_m+\epsilon_n)}}{(e^{\beta\epsilon_l}-1)(e^{\beta\epsilon_m}-1)(e^{\beta\epsilon_n}-1)}
\end{align*}
for the first process, and
\begin{align*}
    \Gamma^{3\rightarrow1}_k=18\pi &\sum_{l,m,n}\vert V_l(k,m,n)\vert^2\delta_\gamma(\epsilon_l-\epsilon_k-\epsilon_m-\epsilon_n)\\
    &\times\frac{(1-e^{-\beta\epsilon_k})e^{\beta\epsilon_l}}{(e^{\beta\epsilon_l}-1)(e^{\beta\epsilon_m}-1)(e^{\beta\epsilon_n}-1)}
\end{align*}
for the second. Let us note that the first one is a pure decay, while the second one requires a thermal population of magnons to happen. We sum them to get the full rate of the vertex
\begin{equation*}
    \Gamma^{1\leftrightarrow3}_k=\Gamma^{1\rightarrow3}_k + \Gamma^{3\rightarrow1}_k.
\end{equation*} 
The total linear decay rate used in Figs. 3 and 4 of the main text is given by $\alpha_1=\Gamma^{2\leftrightarrow 2}+\Gamma^{1\leftrightarrow 3}$, where the first term dominates.

The finite decay time of the magnon modes leads to a finite energy broadening in a Lorentzian shape. The delta function in the equations above is therefore of the form $\delta_\gamma(x)=\frac{1}{\pi}\frac{\gamma}{x^2+\gamma^2}$. Note that this broadening allows decay processes to happen within the discrete energy spectrum caused by the finite size of our sample.

We further solve self-consistently for the decay rates in the Born approximation. When solving self-consistently, we set $\gamma=\sum_i \Gamma_i$ where $i$ runs over the other modes involved in the collision, where we remind that $\Gamma_i$ is the decay rate of the magnon amplitude.

\subsection{Frequency shifts from Hartree-Fock}
In Fig. 3 in the main text, we consider temperature-induced shifts in the magnon frequency from Hartree-Fock. Specifically, we compute the shifted frequency as:
\begin{gather}
\delta\omega_i=\sum_j (V_{i,j}(i,j)+V_{i,j}(j,i))n_j
\end{gather}

\subsection{Theoretical predictions of $\omega$ and $\Gamma$}
\begin{figure}
    \centering
    \includegraphics[width=\columnwidth]{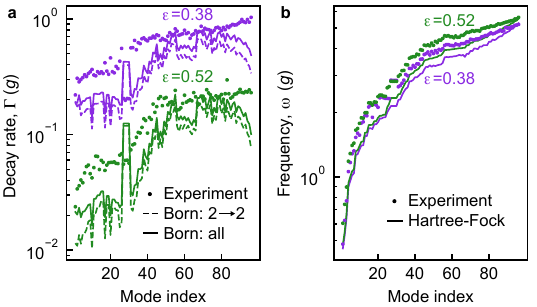}
    \caption{\textbf{Theoretical predictions of $\Gamma$ and $\omega$. a,} Mode index dependence of the decay rate at two energy densities, $\epsilon=-0.38$ (purple) and $\epsilon=-0.52$ (green), for experiment (circles), self-consistent Born calculation including only $2\rightarrow 2$ magnon scattering (dashed line), and including also $1\leftrightarrow 3$ magnon processes ("all"; solid line). The theoretical predictions are dominated by $2\rightarrow 2$ processes and qualitatively agree with the experimental data, but generally underestimates the decay rate, and also shows an increased decay rate for the corner modes (near mode index 30) instead of the suppressed rates observed experimentally. \textbf{b,} Hartree-Fock modeling (solid line) underestimates the magnon frequency compared to the experiment data (circles), but captures the temperature dependence relatively well.}
    \label{fig:mf_modeling}
\end{figure}

Fig.~\ref{fig:mf_modeling} shows the magnon decay rates and frequencies predicted theoretically using the self-consistent Born approximation and Hartree-Fock, respectively, at the two temperatures considered in Fig. 2 of the main text. While the theoretical model qualitatively agrees with the experimental data, including its temperature dependence, we find quantitative discrepancies: in particular, both $\Gamma$ and $\omega$ are underestimated, and the former shows larger increase with mode index than observed in the experiment. Comparison of the decay rates modeled with and without $3\leftrightarrow 1$ processes (solid vs dashed) indicates that $2\rightarrow 2$ scattering dominates. 

\subsection{Non-linear decay}
The decay rates derived above describe the change in amplitude of the activated magnon mode through the standard decay rate equation. In order to describe deviations from the linear decay, one needs a process in which multiple non-thermal magnons interact with each other (see Eq. 2 in the main text). Involving the $2\leftrightarrow2$ vertex, the process in which two non-thermal magnons scatter into other two will have the scattering rate
\begin{equation}
\begin{split}
    \tilde{\Gamma}^{2\leftrightarrow2}_{k}=4\pi&\sum_{l,m}\vert V_{k,k}(l,m)\vert^2\delta_\gamma(2\epsilon_k-\epsilon_l-\epsilon_m)\\
    &\times \frac{(1-e^{-2\beta\epsilon_k})e^{\beta(\epsilon_l+\epsilon_m)}}{(e^{\beta\epsilon_l}-1)(e^{\beta\epsilon_m}-1)}
\end{split}
\end{equation}
Instead, in the $1\leftrightarrow3$ vertex we can have two non-thermal magnons scattering with a third thermal one giving the rate
\begin{align*}
    \tilde{\Gamma}_k^{1\leftrightarrow3} = 18\pi&\sum_{lm}\vert V_{l}(k,k,m)\vert^2\delta_\gamma(2\epsilon_k+\epsilon_m-\epsilon_l)\\
    &\times\frac{(1-e^{-2\beta\epsilon_k})e^{\beta\epsilon_l}}{(e^{\beta\epsilon_l}-1)(e^{\beta\epsilon_m}-1)}
\end{align*}
The total nonlinear decay rate modeled in Fig. 4 in the main text is thus given by $\alpha_2=\tilde{\Gamma}^{2\leftrightarrow 2}+\tilde{\Gamma}^{1\leftrightarrow 3}$, where the former term dominates.
\subsection{Modeling temperature}
\begin{figure}
    \centering
    \includegraphics[width=\columnwidth]{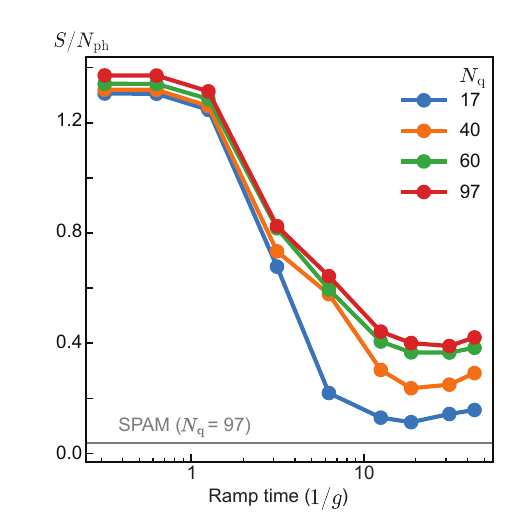}
    \caption{\textbf{Entropy from ramp-recovery measurements.} Entropy created by our ramp preparation procedure, as a function of system size. For each size $N_{\mathrm{q}}$, the number of photons initialized in the system $N_\text{ph}$ is $\lceil N_{\mathrm{q}}/2 \rceil$. The gray line indicates the baseline entropy measured without any ramps due to SPAM in the largest system.}
    \label{fig:sm_sn}
\end{figure}

To model the temperature of the system, we compare the energy measured in the experimental system with the energy of a thermal state of magnons in the mean-field calculation. What is measured experimentally is an average over all the bonds of
\begin{equation}
    E_{\text{exp}}(T) = \frac{1}{N_{\text{b}}}\sum_{\langle i,j\rangle}\frac{\langle X_iX_j+ Y_iY_j\rangle_T}{2}
\end{equation}
with $N_b$ the number of bonds. We can derive then
\begin{equation}
    E_{\text{exp}}(T)=-(S+1) + \frac{1}{J N_{\text{b}}}\sum_\mu\epsilon(\mu) + \frac{2}{J N_{\text{b}}}\sum_\mu\frac{\epsilon(\mu)}{e^{\beta\epsilon(\mu)}-1}
\end{equation}
where the first two terms are the ground state energy for the finite-size system and the last term is due to thermally occupied magnons at temperature $T=1/\beta$. We then derive the temperature dependence of the energy which gives a temperature of $\sim 6.4$\,MHz for an experimentally measured $E_{\text{exp}} = -0.524$ in a 97 qubit system. This temperature falls between the second and third excited mode, and corresponds to an average number of excitations in the system of $5.2$. As a complementary method, we also perform ramp-recovery tests in which we measure the round-trip entropy after preparing a correlated state with a linear ramp and returning to a product state with a reverse ramp of the same length (Fig. \ref{fig:sm_sn}). This provides an upper bound on the entropy created during our preparation, which is calculated from the site occupancies as $S/k_{\text{B}} = -\sum_i \left(p_{0i}\ln p_{0i} + p_{1i}\ln p_{1i} \right)$. For the full 97 qubit system at half-filling ($N_{\text{ph}} = 49$), our slowest ramps result in an entropy per photon of approximately $0.4\,k_{\text{B}}$. This corresponds to an average of 4.8 excitations, in reasonable agreement with the mean-field calculation.

\section{Accuracy of magnon observables}
\label{sec:sm_accuracy}

\subsection{Overview of small-system analysis}

We validate the end-to-end accuracy of our Hamiltonian learning procedure, as well as the coherence and stability of our quantum simulator, with comparisons between the experimental observables and those determined from exact numerical simulation for sizes up to $N_{\mathrm{q}}=31$. We use three observables as benchmarks: magnon frequency, magnon decay rate, and normalized cross-correlation between measured and target magnon time dynamics. Magnon frequency and decay rate are the key characteristics of interest for the physics we explore in the main text. The analysis in the main text presents frequency and decay as extracted from fits of the magnon time dynamics to exponentially decaying sinusoids, for simplicity and robustness to shot noise. However, for smaller sizes we see in both experiment and simulation substantial deviations from this functional form due to the discrete density of states (see e.\,g. Fig.\,\ref{fig:GGPbench}), making these fits less relevant. Therefore, for the purposes of precision characterization at small sizes we use modified definitions of these observables. We characterize magnon frequency by the mean value of the Fourier transform of the time dynamics, and the decay rate by its participation ratio:
\begin{align}
&\bar\omega = \frac{\int_0^{\infty}  \chi_{\mu}''(\omega)^2 \omega d\omega }{\int_0^{\infty}  \chi_{\mu}''(\omega)^2 d\omega} \label{eq:omegabar}\\
&\bar{\Gamma} = \frac{1}{2\pi}\frac{\left(\int_0^{\infty}  \chi_{\mu}''(\omega) d\omega\right)^2}{\int_0^{\infty}  \chi_{\mu}''(\omega)^2 d\omega}. \label{eq:gammabar}
\end{align}
with $\chi_{\mu}''(\omega)$ the imaginary part of the Fourier transform of $\chi_{\mu}(t)$. Note that we use barred symbols here to distinguish these from the $\omega$ and $\Gamma$ extracted from fits (as shown in the main text). In cases where our signals are well described by a single narrow Lorentzian peak, these measures of frequency and linewidth agree with those extracted from fits. However, for signals not well described by such a fit  -- which we see particularly in small size systems --- these measures provide more stable estimates than those obtained by fitting. Particularly, these measures are empirically seen to converge more quickly with the length of time of the signal than either the best fit frequency or the peak frequency of the Fourier transform, which aids classical simulation by requiring less evolution time.
In each case, we quantify accuracy using the magnitude of the fractional error relative to the simulation.

As an additional, model-independent comparison we also define a normalized cross-correlation between the experimental and numerical signals:
\begin{align}
&F_{\mathrm{XC}}= \nonumber\\
&\frac{\text{max}_t|\mathrm{XC}(\chi_{\text{exp}},\chi_{\text{sim}})(t)|}{\sqrt{\text{max}_t|\mathrm{XC}(\chi_{\text{exp}},\chi_{\text{exp}})(t)| \text{max}_t|\mathrm{XC}(\chi_{\text{sim}},\chi_{\text{sim}})(t)|}} \label{eq:xc}
\end{align}
with $\mathrm{XC}(a,b)(t) = \sum_{t'}a(t'+t) b^{*}(t')$. $F_{\mathrm{XC}}$ quantifies the correlation between two sets of time-series data, up to an overall time shift, with 1 indicating maximum correlation and 0 indicating no correlation.

When performing these comparisons, we observe that the experimental magnon signals have overall time delays of around 0.5 to 2.0 ns, depending on the device and preparation procedure, which we attribute to the finite turn-on time of the couplings. Therefore, we resample and shift the experimental datasets before computing the Fourier transforms, using a time shift value set by the state preparation. We find consistent values when comparing (1) the time shift that maximizes experiment and simulation agreement for $\bar\omega$, $\bar \gamma$, (2) the time-shift corresponding to the optimum in $F_{\mathrm{XC}}$, and (3) the time shift from direct fits (examples of which are shown in Fig. \ref{fig:sm_fits} of Sec. \ref{sec:sm-nonlinear}). Importantly, this time shift is held the same for all $N_{\mathrm{q}}$, to avoid any contamination of the scaling trends.

To simulate magnon dynamics in ramp experiments, we first optimize the simulated ramp time to achieve the target energy density, $\epsilon$. We find that the corresponding ramp times in simulation are a few percent away from those used in experiment for relatively fast ramps that prepare high-temperature states. For the slowest ramps, the energy density varies much more slowly with ramp time (and eventually converges to the ground state value in simulations), thus giving rise to a larger difference (up to 50\%). In numerical simulations of GGP, we similarly adjust the simulated period of $t_{\text{GGP}}$ by no more than 1 ns, to account for small variations in the effective interaction time due to the response of the control lines. 

\subsection{Additional magnon data with GGP preparation}
\begin{figure*}[ht!]
    \centering
    \includegraphics[width=\textwidth]{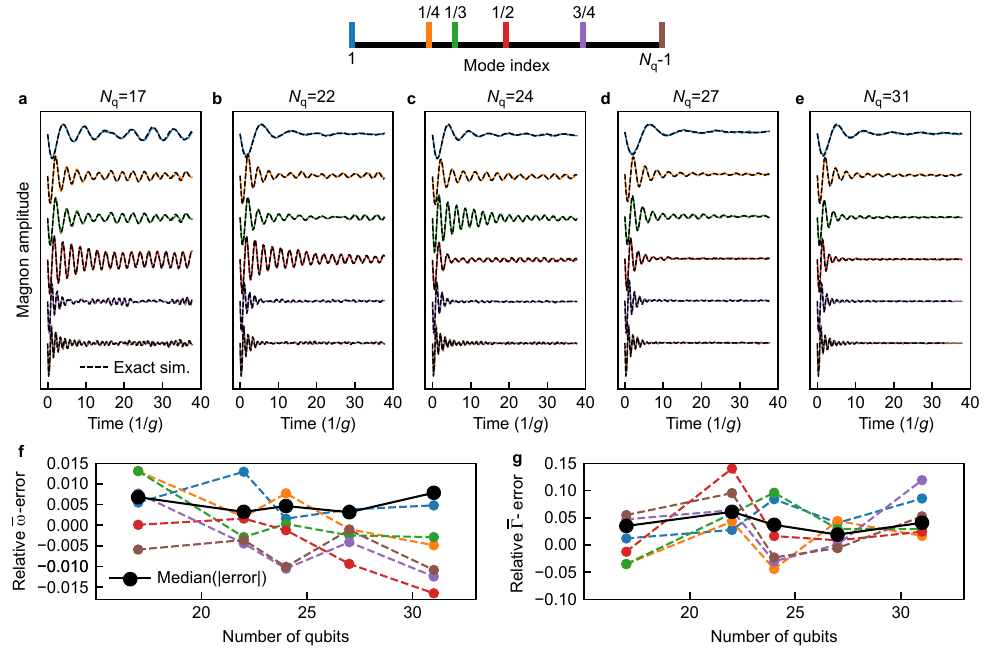}
    \caption{\textbf{Magnon data from GGP measurements in additional system sizes. a-e,} Magnon response in system sizes ranging from $N_{\mathrm{q}}$=17 to 31 qubits, for 5 modes uniformly distributed from mode index 1 to $N_{\mathrm{q}}$-1, as well as $\lfloor N_{\mathrm{q}}/3\rfloor$ to also include a corner mode. The data shows very good agreement with exact numerical simulations (black dashed curves). We note that $\lfloor N_{\mathrm{q}}/3\rfloor=5$ (green) for $N_{\mathrm{q}}=17$ overlaps with the 1/4 mode (yellow) and is thus repeated. \textbf{f,g,} Relative error in $\bar{\omega}$ (\textbf{f}) and $\bar{\Gamma}$ (\textbf{g}). The black curve shows median absolute error, indicating values below 0.01 and 0.1, respectively.}
    \label{fig:GGPbench}
\end{figure*}

In Fig. 1c in the main text, we present magnon data for 24 and 31 qubits following GGP preparation. Figs.~\ref{fig:GGPbench}a-e display similar data from additional system sizes for six magnon modes across the spectrum. We find very good agreement with exact simulations in all system sizes, as also indicated by the low relative errors in extracted values of magnon frequency and decay rate (Figs.~\ref{fig:GGPbench}f,g), with median errors (black curve) below 0.01 and 0.1, respectively.

\subsection{Small-system accuracy benchmarks}

\label{sec:accuracy_benchmark_small_size}

\begin{figure}
    \centering
    \includegraphics[width=\columnwidth]{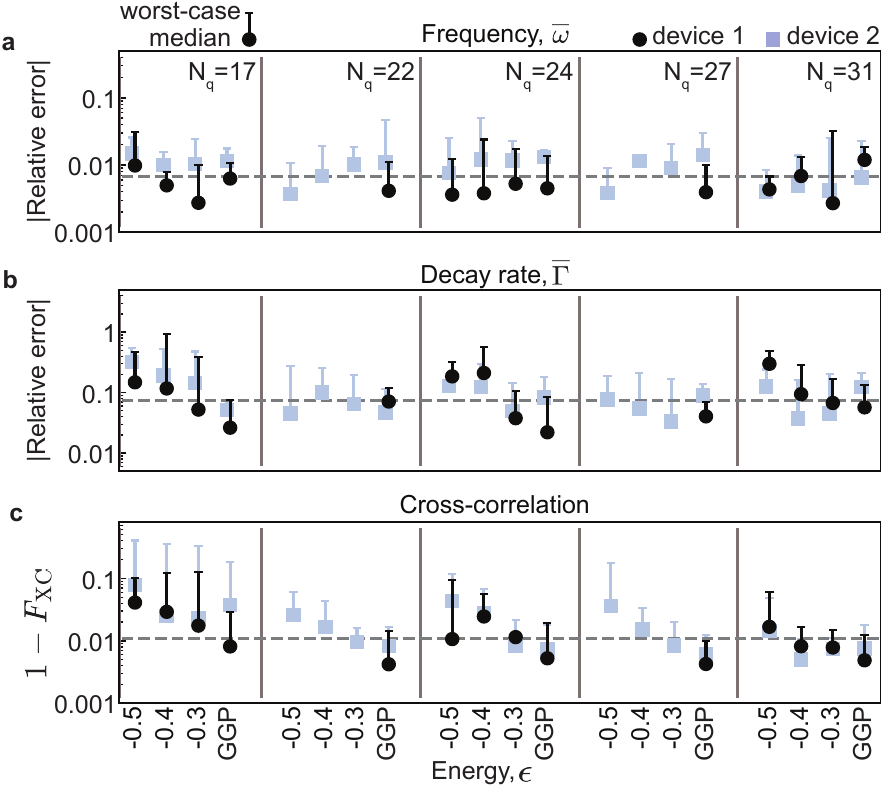}
    \caption{\textbf{Accuracy of extracted magnon observables vs exact simulation.} We benchmark using the magnon center frequency (\textbf{a}), the magnon decay rate (\textbf{b}), and the normalized cross-correlation between the signals (\textbf{c}; see text for details). Each point indicates the median and worst-case measurement over a set of six modes spanning the magnon spectrum, for the indicated system size and energy density. In addition to benchmarks on the experimental device used in the main text (device 1; black points), we include additional measurements on a secondary device (device 2; blue points) with similar performance. The errors remain low over all system sizes, with no visible degradation as size is increased. Dashed lines show the overall median of medians for each observable (0.007 for $\bar{\omega}$, 0.074 for $\bar{\Gamma}$, and 0.011 for $1-F_{\mathrm{XC}})$).}
    \label{fig:sm_accuracy}
\end{figure}

Fig. \ref{fig:sm_accuracy} presents accuracy benchmarks for a range of system sizes and a set of ramp and GGP preparations (the same GGP data shown in Fig. \ref{fig:GGPbench}) spanning the regime of interest.

We also include data from two separate physical devices, to demonstrate the reliability with which we can achieve this degree of accuracy. Each point is drawn from measurements of six magnon modes, containing five that are evenly spaced from 1 to $N_{\mathrm{q}}-1$, and one additional mode that is one of the long-lived corner modes near $N_{\mathrm{q}}/3$. In general, our observations do not indicate degradation in precision with system size, consistent with the observation that the disagreement between our two devices also does not increase in larger patches (see Section~\ref{sec:sm-nonlinear}).
In some cases the errors appear to be decreasing with increasing system size, which we attribute to reduced edge effects during Hamiltonian patching.
Overall median values for the relative error of magnon frequency and magnon decay rate are 0.007 and 0.078, respectively. In Sec. \ref{sec:complexity}, we present complexity scaling analyses that estimate the difficulty of numerically determining magnon properties to this precision in a large system.

\subsection{Coherent error model}

\begin{figure}
    \centering
    \includegraphics[width=\columnwidth]{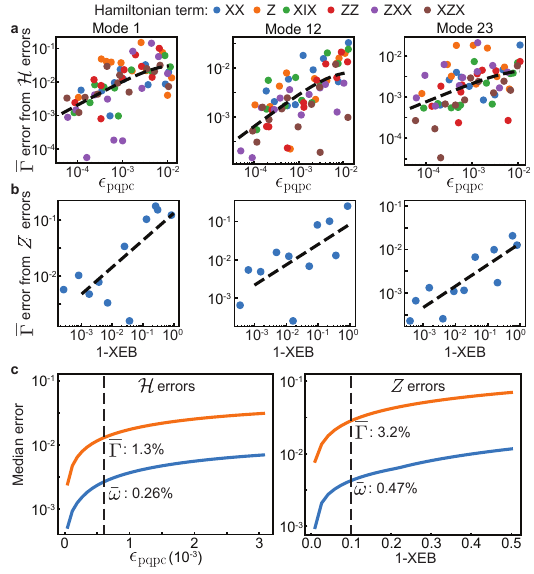}
    \caption{\textbf{Model of coherent errors. a,b,} We numerically simulate, for $N_{\mathrm{q}}=24$ and GGP preparation, the effect of random coherent errors as additional Hamiltonian terms (\textbf{a}) and random coherent $Z$ rotation errors in the GGP state preparation (\textbf{b}). In each case, we compute, for a given error perturbation, the degradation in XEB fidelity and in the observables $\bar{\omega}$ and $\bar{\Gamma}$, and fit the average correlation between these to a power-law functional form to extract the typical observable sensitivity (dashed lines). Hamiltonian errors are quantified in error per qubit per evolution time cycle ($\epsilon_{\mathrm{pqpc}}$), while $Z$ errors are shown as 1-XEB between the system with and without the perturbation. We use six magnon modes for our analyses (the same six used in the preceding section Sec \ref{sec:accuracy_benchmark_small_size}), and show three examples. \textbf{c}, Median error over modes, as determined from the fits. Dashed lines and values indicate the predicted contribution at our experimentally observed XEB errors. Overall levels of errors predicted from XEB fidelity are comparable to observable errors measured in experiment.}
    \label{fig:sm_coherent_error_obs}
\end{figure}

Our experimental platform unlocks the powerful ability to measure both exact state fidelities, quantified by XEB, and standard local observables. Here we use these capabilities to deepen our understanding of the errors that limit the accuracy of our observables. For a small system ($N_{\mathrm{q}}=24$), we numerically calculate evolution using an optimized Hamiltonian, and compare to one that has been perturbed with random terms of varied strengths and forms. For each result, we measure both the effect on the XEB fidelity and on the magnon observables. We also simulate the errors of imperfect calibrations of the phases in the $Z$ pulse, used to create the magnons and in GGP state preparation.

The results are summarized in Fig. \ref{fig:sm_coherent_error_obs}. While we see substantial scatter in the sensitivity of observables to various Hamiltonian terms and $Z$-phases, we see overall correlations that are reasonably captured by a power-law ansatz: $-\log(\text{err}) = A\log(B(1-\mathrm{XEB}))$. Furthermore, using the typical levels of measured fidelity errors (0.06\% per qubit at $gt=1$ for Hamiltonian errors, as shown in Sec. \ref{sec:hamlearning}, and 0.5\% per qubit for the $Z$-pulse, as shown in Sec. \ref{sec:zerr}), and adding the separate error contributions, we recover overall errors in magnon frequency and decay rate of 0.73\% and 4.5\%, respectively, comparable to the typical values observed in experiment. While this simple model neglects important effects, such as the structured Hamiltonian errors that might result from systematic errors or neglected terms, as well as incoherent errors, we conclude that, consistent with expectations from XEB purity measurements, our observable accuracies are still limited primarily by coherent errors, giving a strong motivation for continued development of calibration and learning procedures.

\section{Computational complexity of magnon signals}
\label{sec:complexity}
Response functions probe features of quantum dynamics across many scales --- from the microscopic, like the scattering of individual quasi-particles, to emergent long-wavelength phenomena, such as hydrodynamic transport. A wide variety of theoretical approaches for computing response functions have been introduced, resulting in qualitative understanding of many of the ingredients that appear in the setting of this paper.
However, these theoretical approaches either do not aim for microscopic accuracy or introduce uncontrolled approximations to become analytically tractable.
Accurate classical or quantum simulations are necessary to motivate which physical phenomena should be theoretically modeled and to validate the approximations in these calculations. 
This leads to the key question of whether our quantum simulations can serve as a more accurate benchmark than can be obtained with classical numerical methods. In this supplement, we argue that classical simulation benchmarks with controlled errors that reach comparable accuracy to the experiment are infeasible with methods known to us.

For the purposes of this section, we define the system of interest as the 2D spin-$\frac12$ $XY$-model on the square lattice with uniform nearest-neighbor couplings:
\begin{equation}
    \label{eq:HXY}
    H = \sum_{\langle ij\rangle} \frac{g}{2} \left( X_i X_j + Y_i Y_j\right),
\end{equation}
realized on finite grids with $N_{\mathrm{q}}$ qubits and open boundary conditions.
As in the main text, we consider quench dynamics starting with finite energy-density states, particularly those prepared with the GGP protocol outlined in Sec.~\ref{sec:ggp}, and we measure response functions 
\begin{equation}
    \label{eq:response}
    \chi_{\mu \nu}(t) = -i \theta(t) \sum_{i,j} \varphi^{\mu}_i \varphi^{\nu}_j \bra{\psi_0}\left[Z_i(t), Z_j\right] \ket{\psi_0}.
\end{equation}
The coefficients $\varphi^{\mu}_i$ are chosen to ensure coupling to the long-lived oscillatory magnon modes, as in Sec.~\ref{sec:magnons_construction}; below, we refer to $\chi_{\mu}(t)\equiv\chi_{\mu \mu}(t)$ as the signal from the $\mu^{\rm th}$ magnon mode. This response function is probed numerically with the same protocol that we use in the experiments and described in Eq~\eqref{eq:protocolforresponsefunctions}: one simulation is run with small amplitude $Z$-rotations of strength $A \varphi^{\mu}_i/2$ on each site $i$, and another is run with the opposite sign of rotations. As in Eq.~\eqref{eq:protocolforresponsefunctions2}, the measured response function is the difference of the magnon amplitude between these two simulations. Throughout this section, $A$ is chosen small enough to remain in the linear response regime. 

In comparison to the dynamics discussed in the rest of this text, in this section we use several simplifications which are favorable to classical simulation. First, we use the nearest-neighbor $XY$-model rather than the effective Hamiltonian describing the dynamics of the quantum device. The full device Hamiltonian, learned via the techniques described in  Sec.~\ref{sec:hamlearning}, has many additional correction terms which are small in magnitude but quite long-range. Incorporating the effect of such terms requires significant overhead for classical numerical methods but introduces minimal qualitative changes to the nature of the resulting signals. In addition, our analysis focuses on the global gate protocol (GGP) state preparation technique with no additional thermalization time included. We see in Fig.~\ref{fig:sm_bam_compare} that including the thermalization has only a small quantitative effect on magnon frequencies and linewidths; however, including it significantly lengthens the total evolution time and increases the difficulty of classical simulation. Finally, we use as our classical simulation targets the magnon frequency measure $\overline{\omega}$ and linewidth measure $\overline{\Gamma}$ defined in Eq.~\eqref{eq:omegabar}-\eqref{eq:gammabar}. We find that these measures converge with the least total evolution time among several tested frequency and linewidth measures, including identifying the peak frequency and using fits to Lorentzian mode shapes. Our empirical analysis of the dynamics with these simplifications below suggests that these changes do not alter the fundamental difficulty of the problem. 

In the remainder of this section, we present empirical evidence relating to the classical complexity of the magnon experiment using extensive state vector and tensor network computations. 
First, using exact state vector calculations, we analyze method-agnostic measures of signal complexity, namely the sensitivity of the magnon frequency and lifetime to changes in the initial state and the length of time of the simulation. We compare these exact signals to first-order perturbative predictions as well as predictions incorporating self-consistent renormalization of the magnon mode frequencies and lifetimes, which can be interpreted as a resummation of an infinite subset of magnon interaction diagrams. Neither of these methods captures the exact data quantitatively.
We follow this with extensive evidence from matrix product state (MPS) simulations, the most competitive classical method that we tested. We provide estimates for the resources needed to compete with the accuracy of the quantum simulation, which we argue follow an exponential scaling consistent with the volume-law growth of entanglement entropy when the energy is sufficiently high. Finally, we detail evidence from $2$D tensor network simulations, which use belief propagation to truncate projected entangled pair states (PEPS). We find that these methods are less performative than MPS for the problem at hand except in very high temperature regimes with limited spatial and temporal correlations. 

\subsection{Complexity of magnon signals from exact calculations}

The properties of elementary excitations in the $XY$-model in the zero-temperature context are readily studied with tensor network methods which take advantage of the area-law scaling of entanglement entropy to compress states. While quantum systems with frustrated spin-spin interactions or itinerant fermions can challenge tensor network methods, neither are present in the nearest neighbor square lattice $XY$-model. The increase in entanglement from populating a single excitation is constant, thus we cannot expect to find large classical complexity in the behavior of a finite number of magnon excitations at zero temperature.

By contrast, for finite-energy density quenches such as those considered in this paper, entanglement entropy grows linearly in time and saturates at a volume-law scaling value~\cite{2018-Jonay-Nahum-Coarse-grainedDynamicsStateEntanglement, 2019-Rakovszky-Pollmann-EntanglementGrowthInhomogenousQuenches}. Tensor network time-evolution algorithms typically require resources exponential in the entanglement entropy to reproduce the evolution with high fidelity~\cite{2008-Schuch-Cirac-EntropyScalingProductStates}, and we shall see that this is also the case here. The question at hand is whether this exponential complexity is also imbued on estimating the physical observables of interest, magnon frequency and lifetimes, to accuracies comparable to those achievable in the quantum simulation. 

\begin{figure*}[htb]
    \centering
    \includegraphics[width = 0.96\textwidth]{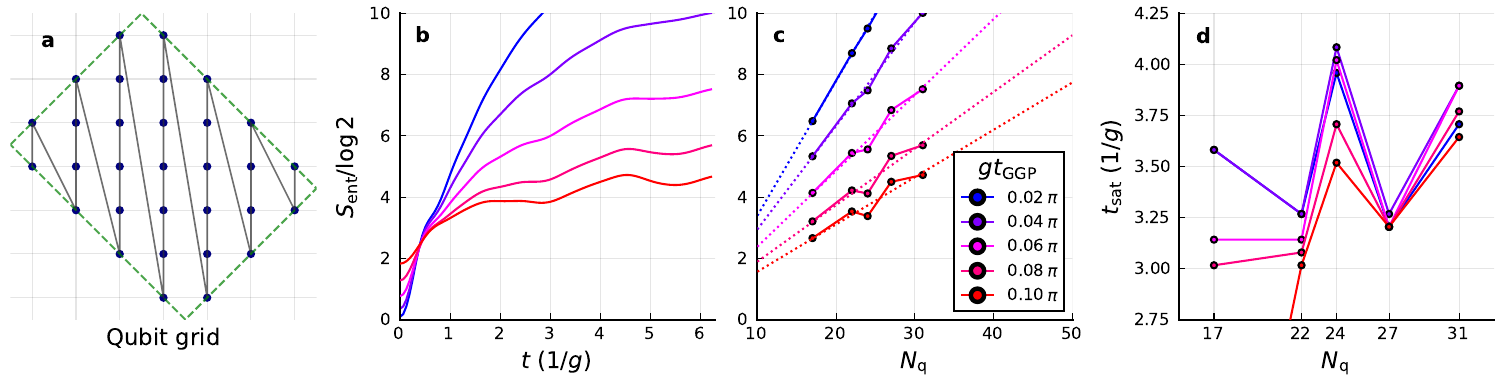}
    \caption{\textbf{Entanglement entropy in GGP quenches. a,} An example of the geometries probed in these calculations, here showing the $31$ qubit system on a rotated rectangular grid. The qubit ordering used for MPS calculations sorts qubits by row and then by column. \textbf{b,} Second-Renyi entanglement entropy versus time in quenches of varying energy densities, as controlled by the pulse time in the GGP protocol, as shown here for the $31$ qubit system. The entanglement cut separates the first and second halves of the qubits (rounded for odd $N_{\mathrm{q}}$) using the order shown in \textbf{a}. \textbf{c,} The maximum value of entanglement entropy reached in these quenches for several rotated rectangular grid geometries. This saturation entanglement scales linearly with system size, with a volume-law coefficient that increases with energy density. \textbf{d,} The saturation time $t_{\mathrm{sat}}$ to reach $90\%$ of the maximum entanglement entropy grows only slowly with system size. \label{fig:entropy}}
\end{figure*}

While classical complexity that grows exponentially in time is the generic situation for quantum dynamics from non-equilibrium quenches, there are several necessary criteria that we should check to ensure that the dynamics and observables of interest are not much easier to simulate. First, we will check that the entanglement entropy actually grows for initial states in the energy regimes of interest; states which are too close to the ground state in energy will have much slower entanglement growth. The final saturation entanglement entropy should show volume-law scaling with a sufficiently large coefficient in order to not be simple for MPS methods. Additionally, our physical observables must also require a sufficiently long time evolution in order for them to have high classical complexity. As our quantum simulation starts from a state with low entanglement, tensor network methods can capture an early time regime while the entanglement is sufficiently small. Moreover, signals that occur in spectral functions can sometimes be extrapolated from early times to late times, exploiting fits to simple functional forms or with signal processing techniques such as linear prediction \cite{2009-Barthel-White-SpectralFunctionsRenormalizationGroup}. 
Finally, as our case for complexity relies on the finite temperature background to generate entanglement, we must ensure that the observables we measure could not be learned from dynamics near the ground state. That is, we must impose the criterion that the magnon frequencies and lifetimes are sensitive to the finite temperature background at the precision of the experiment.
Similarly, we impose the criterion that our observables are sensitive to replacing our initial state with one of much higher energy density. This is necessary due to the existence of operator evolution methods that perform well near infinite temperature, such as those in Refs.~\cite{2023-Yi-Thomas-White-ComparingNumericalSpinModel, 2023-White-White-EffectiveDissipationQuantumHydrodynamics, 2022-vonKeyserlingk-Rakovszky-OperatorBackflowQuantumTransport, 2022-Rakovszky-Pollmann-Dissipation-assistedOperatorHydrodynamicTransport, 2019-Parker-Altman-UniversalOperatorGrowthHypothesis01,2024-Artiaco-Bardarson-EfficientLarge-scaleTimeEvolution,2025-Bauer-Artiaco-LocalInformationQuenchDynamics}.
We will see that these criteria also provide constraints on which parameter regimes and observables can exhibit high complexity, with short-wavelength magnon modes and very-high temperature signals being amenable to MPS simulations accurate over early times.

To probe these criteria for complexity, we use exact state-vector evolutions of the nearest neighbor $XY$-model. These simulations follow the same procedure detailed in the main text; specifically, we use initial states prepared with the nearest-neighbor $XY$ analog of the GGP protocol described in Sec.~\ref{sec:ggp}  and we use the same Holstein-Primakoff mode shapes derived in Sec.~\ref{sec:magnons_construction} with amplitude $A=1$.
The calculations are implemented using circuit simulators, for which a second-order Trotterization with time step size of $g \delta t = 0.04 \pi$ is used to convert the evolutions into circuits. This step size is chosen so that the effect of Trotter error is negligible compared to the size of truncation errors for most simulations. During the evolution used in the state preparation stage, we use a step size that is half as big due to the short duration of the evolution.
We use the high performance \texttt{qsim} state-vector simulator to perform exact simulations on up to $36$ qubits \cite{quantum_ai_team_and_collaborators_2025_4067237}.

\begin{figure*}[htb]
    \centering
    \includegraphics[width=\textwidth]{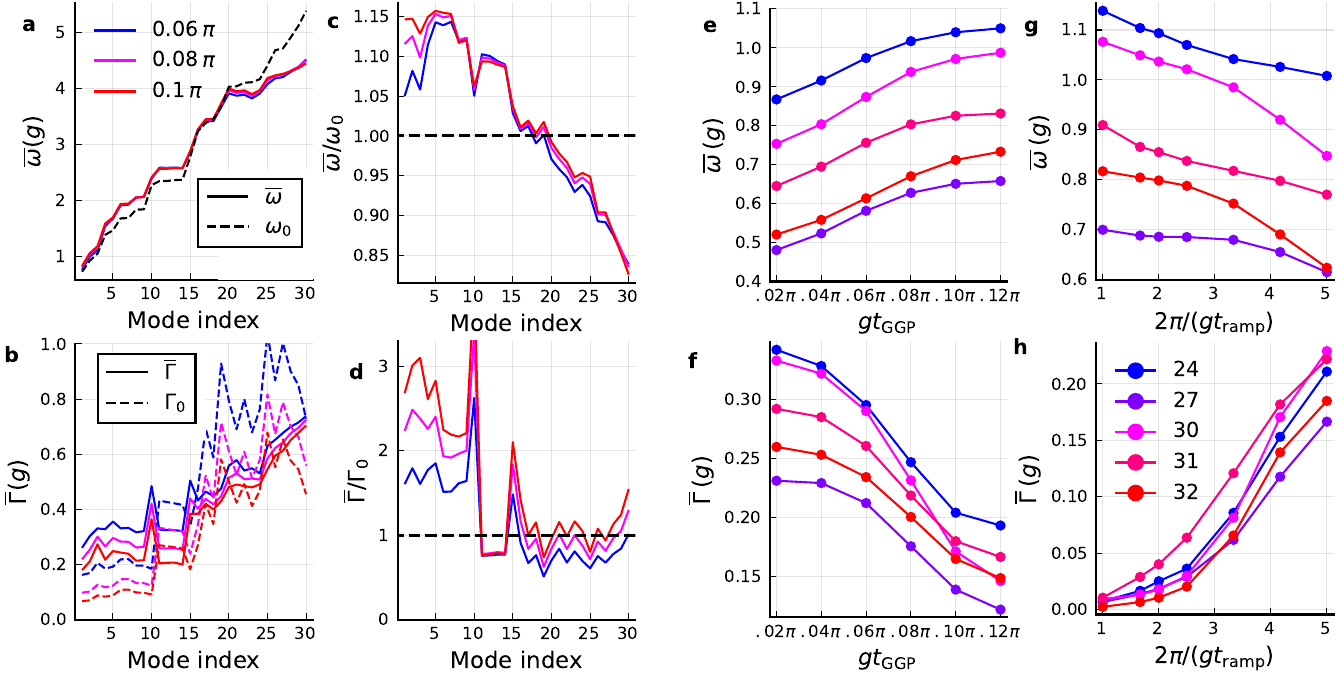}
    \caption{\textbf{Sensitivity of magnon frequency and lifetime. a,} The extracted mode frequency $\overline{\omega}$ in a 31 qubit geometry for states of varying energy densities, as controlled by the pulse time in the GGP protocol. The mean-field mode frequency $\omega_0$ is shown in comparison. \textbf{b}, The extracted mode linewidth $\overline{\Gamma}$ in the same geometry and initial states, compared to the self-consistent linewidths computed using the method of Sec.~\ref{sec:theorydecay}. \textbf{c,} Using $\omega_0$ to normalize mode frequencies, we see that these frequencies differ from the mean-field prediction. Moreover, the lowest modes have the highest relative sensitivity to $t_{\mathrm{GGP}}$, with shifts of $\sim 10\%$ which is easily detectable within experimental accuracy of $\sim 1\%$. \textbf{d}, The linewidth of the modes differ from the perturbative prediction, with the largest deviation and the largest sensitivity to shifts in $t_{\mathrm{GGP}}$ occuring for the lowest index modes. \textbf{e,f}, The lowest index mode frequency and width in exact computations across various system sizes as $t_{\mathrm{GGP}}$ is varied. Sensitivity large enough to be detected persists as system size is increased. \textbf{g,h}, The sensitivity of the lowest index mode frequency and width with respect to varying the ramp speed for initial states prepared by ramps. The slowest ramps achieve the lowest energies, where magnon modes have very long lifetimes and correspondingly narrow linewidths, which broaden significantly upon increasing the ramp speed and initial state energy. \label{fig:sensitivity}}
\end{figure*}

As shown in Fig.~\ref{fig:entropy}, the entanglement entropy reached during the quench dynamics shows clear volume-law scaling for initial states prepared by the global gate protocol. Specifically, we compute the dynamics of the second R\'{e}nyi entanglement entropy, as shown for $31$ qubit systems in Fig.~\ref{fig:entropy}b. The energy density of the initial states are tuned by changing the time of the global pulse $t_{\mathrm{GGP}}$; as shown in Fig.~\ref{fig:BAM_SM}, these pulse times result in initial states over a wide range of energy densities, from $\epsilon \sim 0.15$ which is close to infinite temperature to $\epsilon \sim 0.45$ which is the lowest temperature reachable using this protocol. At very early times, the entanglement entropy is upper bounded by an area-law scaling with qubit number, as the dynamics including the state preparation is well approximated by a short depth circuit acting on a product state. However, the area-law coefficient grows linearly with the depth of the circuit and thus the evolution time. This growth regime lasts until the entanglement entropy nears saturation. As shown in Fig.~\ref{fig:entropy}c, the maximum value of entanglement entropy reached scales linearly with qubit number. We measure the saturation time $t_{\mathrm{sat}}$, which we define as the time to reach $90\%$ of the maximum entanglement entropy, to set a rough boundary between the area-law scaling entanglement growth regime and the volume-law scaling regime. We see in Fig.~\ref{fig:entropy}d that $g t_{\mathrm{sat}} \sim 3.5$ and grows slowly with system size. The theoretical expectation is that entanglement saturation is reached in time $g t_{\mathrm{sat}} \sim L/v$, where $L$ is the longest linear dimension of the system and $v$ is a velocity of entanglement. Previous work on the $XY$-model identified a velocity $v \sim 2$, which is consistent with what we see here~\cite{Andersen2025}.

In Fig.~\ref{fig:sensitivity}, we show that at finite temperature these magnon frequencies and linewidths differ significantly from mean-field predictions and from their values near zero or infinite temperature. For magnon frequencies, we compare to the lowest-order theoretical prediction, which corresponds to the mode energies from the Holstein-Primakoff calculation detailed in Sec.~\ref{sec:magnons_construction}. While both the mean-field calculation and the exact results show ballistic magnon spectra across the temperature ranges probed (Fig.~\ref{fig:sensitivity}a), the measured frequencies for a system of $31$ qubits shows significant deviations of $\pm 15\%$ from the mean-field prediction (Fig.~\ref{fig:sensitivity}b). We also see, however, that high index modes do not appear to be very sensitive to changes in the background temperature; only the lowest index modes are. For magnon linewidths, we compare exact results to the linewidths predicted from perturbative magnon scattering with self-consistent broadening, as detailed in Sec.~\ref{sec:theorydecay}. In Fig.~\ref{fig:sensitivity}d, we see that once again the lowest index magnon modes differ significantly from this theoretical prediction and their linewidths are very sensitive to the background energy density of the initial state.
As both the temperature sensitivity and the deviation from the theoretically predicted linewidths suggest that the low index magnon modes are the most complex, we focus the remainder of our complexity analysis on the lowest index mode.

To provide more detail on how the lowest index mode frequency and linewidth changes with background energy density, we exactly compute $\overline{\omega}$ and $\overline{\Gamma}$ for initial states prepared by GGP and also for initial states prepared by diabatic ramps. These calculations use several rectangular and rotated rectangular (as depicted in Fig.~\ref{fig:entropy}) geometries of distinct sizes.
In both setups, we see that there is sensitivity to changes in the initial state for both frequency and linewidth of the lowest index mode that is large compared to the experimental accuracy. The mode frequency at intermediate energy densities as measured by $\overline{\omega}$ shifts by $\sim 10\%$ from either the smallest or largest energy densities measured, while the experimental accuracy on frequency is roughly $1\%$. Similarly, the mode lifetime shifts by $\sim 50\%$ from either the smallest or largest energy densities accessed with the GGP, and even more from the lowest energies measured in ramps, while the experimental accuracy is roughly $10\%$. Moreover, these shifts do not shrink significantly as the systems are increased in size from $24$ to $32$ qubits. This sensitivity suggests that classical computations need to deal with the finite temperature background in order to correctly model the magnon frequency shifts and linewidths. For methods based on time evolution, such as the tensor network state evolutions used in the next section, this  finite temperature background generates complexity through the growth of entanglement over time.

\begin{figure*}[htb]
    \centering
    \includegraphics[width = \textwidth]{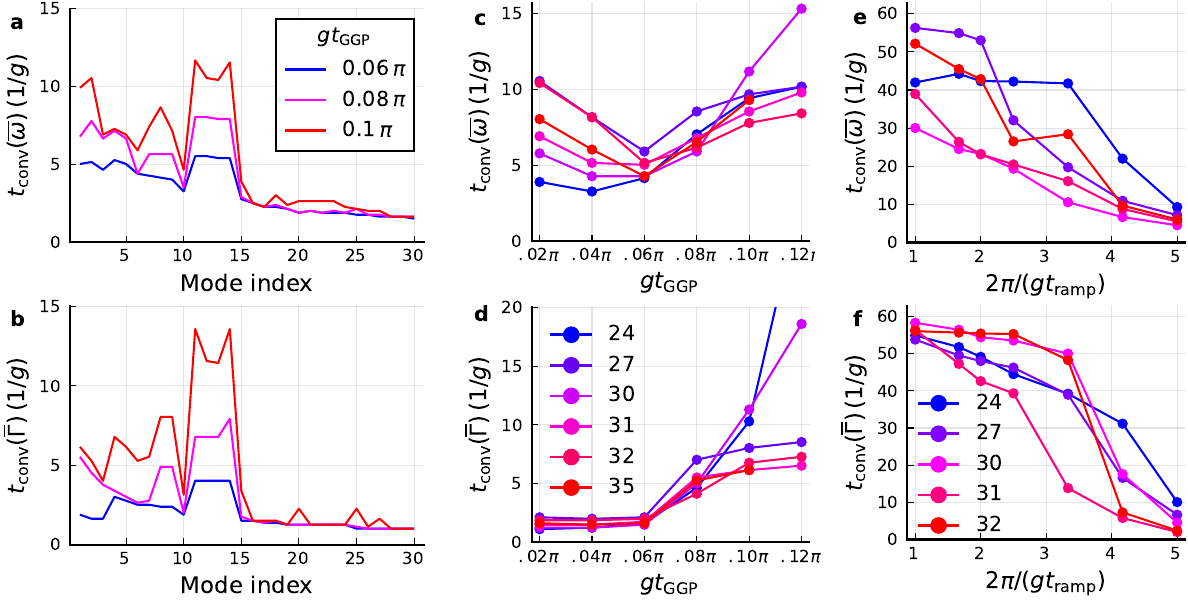}
    \caption{\textbf{Temporal complexity of parameter estimates.} The length of evolution time needed to extract estimates of $\overline{\omega}$ to an accuracy of $1\%$ (\textbf{a}, \textbf{c}, \textbf{e}) and $\overline{\Gamma}$ to an accuracy of $10\%$ (\textbf{b}, \textbf{d}, \textbf{f}), the accuracy targets used throughout this section. In \textbf{a} and \textbf{b}, we consider magnon modes across the magnon spectrum in a $31$ qubit geometry, for initial states prepared with the GGP protocal.  Higher index modes have much less temporal complexity for estimating magnon parameters. In \textbf{c} and \textbf{d}, we consider the same the lowest index mode for rectangular and rotated rectangular grids with various sizes $N_{\mathrm{q}}$. In \textbf{e} and \textbf{f}, we consider the same for ramp initial states of various speeds. In all cases, low energy states (where magnon signals live longest) require the most time evolution for accurate measurements of the parameters, which we attribute to the long period of oscillations. Notably, the time needed $t_{\mathrm{conv}}$ is larger than the time it takes entanglement entropy to approach saturation $t_{\mathrm{sat}}$. \label{fig:temporal}}
\end{figure*}

To show that the magnon signals require a sufficiently long time evolution for convergence of the magnon parameters, we repeat the extraction of frequency and linewidth described above using only the time evolution data before a cutoff time $t_{\mathrm{cut}}$. First, to show that we have obtained sufficiently long exact time evolutions, we verify that the frequency and linewidth estimates converge as $t_{\mathrm{cut}}$ is increased. Then, to find the time $t_{\mathrm{conv}}$ it takes to converge to $1\%$ accuracy for frequencies and $10\%$ accuracy for linewidths, we scan $t_{\mathrm{cut}}$ to find the last times the respective errors due to the time cutoff are larger than those values. The results are shown in Fig.~\ref{fig:temporal}. 
For the lowest index modes, $t_{\mathrm{conv}}$ is consistently longer than the entanglement saturation time $t_{\mathrm{sat}}$. Additionally, we see that the temporal complexity does not shrink dramatically as the system size is increased from $N_{\mathrm{q}}=24$ to $N_{\mathrm{q}}=35$.
However, we see that the necessary evolution time is much lower for high index modes as compared to low index modes, and at high temperatures as compared to low temperatures, which suggests that much shorter MPS evolutions can be used to predict magnon parameters in these regimes.

Thus, these temperature-dependent shifts of magnon frequency and magnon lifetime can serve as physically relevant targets for a quantum advantage experiment. In particular, these calculations suggest that a tensor network evolution must model long time dynamics on top of a finite temperature background in a $2$D quantum system; more specifically, the entanglement entropy in the evolution will need to approach a finite temperature saturation value which is a large fraction of the maximum possible entropy in these systems. For this reason, we expect exponential scaling of resources for tensor network methods; we investigate this expectation further in the next section.

\subsection{Matrix product state calculations}
In this section we present the results of matrix product state simulations of quench dynamics in the $2$D nearest neighbor $XY$-model, along with quantitative analysis of the accuracy of these calculation. We employ two simulation methods, each using a representation of the state as an MPS with fixed bond dimension $\chi$ but with distinct procedures for performing time-step updates. One method, a time-evolving block decimation (TEBD) algorithm utilizing a Trotter decomposition into gates, experiences truncation error which can be monitored as a violation of unitarity and energy-conservation. The other, an implementation of the time-dependent variational principle (TDVP)~\cite{Haegemann2011} with single site updates, explicitly conserves norm and energy while experiencing projection errors that are difficult to explicitly monitor~\cite{2019-Paeckel-Hubig-Time-evolutionMethodsMatrix-productStates}. We analyze the truncation errors of TEBD through the failure of norm- and energy-conservation across wide ranges of system sizes and bond dimension, showing that they behave predictably through scaling collapses, and empirically relate this behavior to the errors in estimated magnon frequencies and lifetimes using comparisons to exactly computed values on small systems. Extrapolating this behavior to large system size, we provide estimates of the resources needed for a hypothetical TEBD calculation to approach the estimated accuracy of the quantum simulations, which appear to be beyond the reach of reasonable classical simulations. Finally, we compare the behavior of our TDVP and TEBD calculations in the nearest neighbor $XY$-model, which indicate that TDVP with similar resources also struggles to capture the behaviors of the most challenging large scale magnon signals.

\begin{figure*}[htb]
    \centering
    \includegraphics[width = 1.05\textwidth]{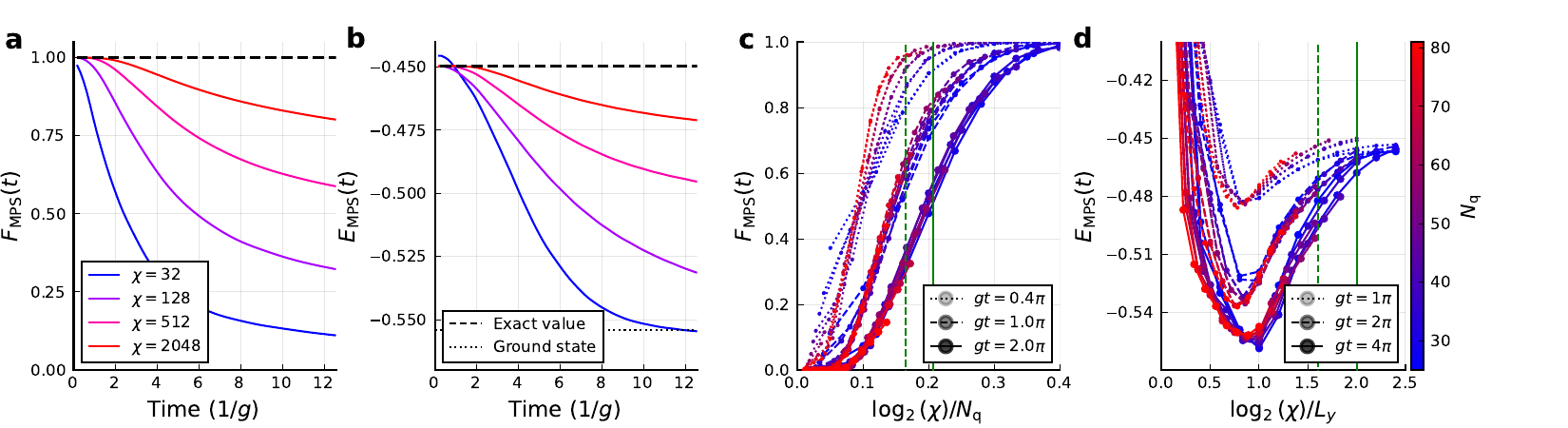}
    \caption{\textbf{Fidelity and energy error in TEBD simulations a,} The typical behavior of simulation fidelity $F_{\rm MPS}$ and \textbf{b}, measured energy density per bond of the MPS, as seen here for simulations on a $N_{\mathrm{q}}=36$ square grid after a GGP pulse of duration $g t_{\mathrm{GGP}} = 0.1 \pi$. \textbf{c,} The dependence of $F_{\rm MPS}(t)$ on bond dimension and system size is well described by a single variable, the scaled bond dimension $\log_2(\chi)/N_{\mathrm{q}}$, indicating a volume-law scaling of bond dimension to reach a fixed fidelity. Here, data is shown for three distinct evolution times for simulations after a GGP pulse of duration $g t_{\mathrm{GGP}} = 0.1 \pi$ on rectangular grids from $5 \times 4$ to $9 \times 9$ in size. The data collapses once entanglement entropy saturation is reached. Vertical green lines mark the values $\chi=2^{16}$ (dashed) and $\chi=2^{20}$ (solid) for the geometry with $N_{\mathrm{q}}=97$. \textbf{d,} The MPS energy is better described by bond dimension that scales exponentially in system width instead, as shown here by the rough collapse of energy for the same evolutions when plotted against $\log_2(\chi)/L_y$. For bond dimensions $\chi \lesssim 2^{L_y}$, energy roughly approaches the ground state energy over time. \label{fig:mpsfidelityenergy} }
\end{figure*}

Both of our MPS methods use one rank-3 tensor per site in an order that `snakes' through the system qubits row by row, as depicted in Fig.~\ref{fig:entropy}. As a consequence of the site ordering, these methods effectively map a $2$D system with only local interactions into a $1$D system with non-local interaction terms whose range scales with the system width. Alternate approaches for the site ordering (e.\,g. with the space-filling Hilbert curve) or for the network connectivity (e.\,g. hierarchical trees) could be used with either method, altering the distribution and range of the non-local connections that arise from the $2$D mapping \cite{2021-Cataldi-Montangero-HilbertCurveNetworkEfficiency, 2020-Kloss-BarLev-StudyingDynamicsNetworkStates}. Due to the volume-law scaling of entanglement entropy in the dynamics that we target, we expect that these methods would result in similar conclusions and we do not study them here; direct comparisons between MPS and tree tensor network dynamics at finite energy support this \cite{2025-Vovrosh-Dauphin-SimulatingDynamicsClassicalNumerics}.

The time-dependent variational principle (TDVP)~\cite{Haegemann2011}, aims to update the MPS tensors by solving the Schr\"{o}dinger equation projected to the manifold of states with constant bond dimension. As a consequence, the effect of the non-local Hamiltonian interactions are approximately incorporated into local MPS updates without growing the bond dimension (single-site TDVP) or only growing the bond dimension locally (two-site TDVP)\cite{2019-Paeckel-Hubig-Time-evolutionMethodsMatrix-productStates}. During each local update, the effective basis of states in which the Hamiltonian is projected -- i.\,e. the span of the tensor products of a Schmidt vector to the  left of the update site, a Schmidt vector to the right of the update site, and a local basis state -- will only partially capture the action of non-local Hamiltonian interactions, introducing an error known as projection error. We employ a subspace expansion technique to enhance this basis with a random vector \cite{Yang2020}, partially alleviating the projection error. TDVP with single-site updates is designed to explicitly conserve the norm and energy; additionally, we explicitly enforce the $U(1)$ symmetry that arises from conservation of total $Z$ magnetization using block-sparse tensors~\cite{2011-Singh-Vidal-TensorNetworkU1Symmetry}.

We employ TDVP in its two-site update version, including subspace expansion with a random vector, in order to be able to sufficiently increase the bond dimension during the state preparation stage of our simulations, i.\,e. during ramps or during the GGP pulse. We have observed that both the two-site algorithm without subspace expansion as well as the one-site version with subspace expansion are not able to create enough new states in the MPS manifold during the evolution to faithfully follow the state during the stage in which the bond dimension grows from the initial value $1$. In all other stages of the time evolution, we employ the single-site TDVP with subspace expansion, with the advantage that energy is conserved. When evolving with the time-dependent ramp Hamiltonian, we use a reduced time step of $1/500$ of the ramp time, necessary to keep time discretization errors to negligible levels; otherwise, we use a time step of $dt = 0.5$ ns. In all computations, the Hamiltonian is created as a matrix product operator whose representation is compressed via the delinearization algorithm of Ref.~\cite{Hubig_2017}.

By contrast, the TEBD method we employ uses a Trotterization of the dynamics into two-site gates. Each gate is applied exactly using an exact contraction of the form of a matrix product operator acting on a matrix product state, temporarily increasing the bond dimension from $\chi$ to $4 \chi$ for all MPS sites between the support of the gate in the snake ordering. After each gate application, a single sweep of truncated SVD compression is applied, reducing the bond dimension back to $\chi$ across the affected area. The error produced by such a compression is easily monitored from the cumulative size of truncated singular values, or equivalently by tracking the shrinking of the square of the norm of the MPS. The latter quantity, which we refer to as the simulation fidelity $F_{\mathrm{MPS}}$, is in many cases strongly correlated with the actual fidelity of the simulation~\cite{QuantinuumNature2026Magnetism,2025-Thompson-Niesen-Non-zeroNoiseTensorNetworks,2020-Zhou-Waintal-WhatLimitsComputers,2023-Ayral-Waintal-Density-matrixRenormalizationFiniteFidelity}. Tracking this simulation fidelity across simulations with multiple bond dimensions can also allow for extrapolated predictions of observables that converge faster than the results from the largest bond dimension used~\cite{QuantinuumNature2026Magnetism, 2025-Mandra-Kechedzhi-HeuristicMatrixIsingModels}.

\begin{figure*}[htb]
    \centering
    \includegraphics[width = \textwidth]{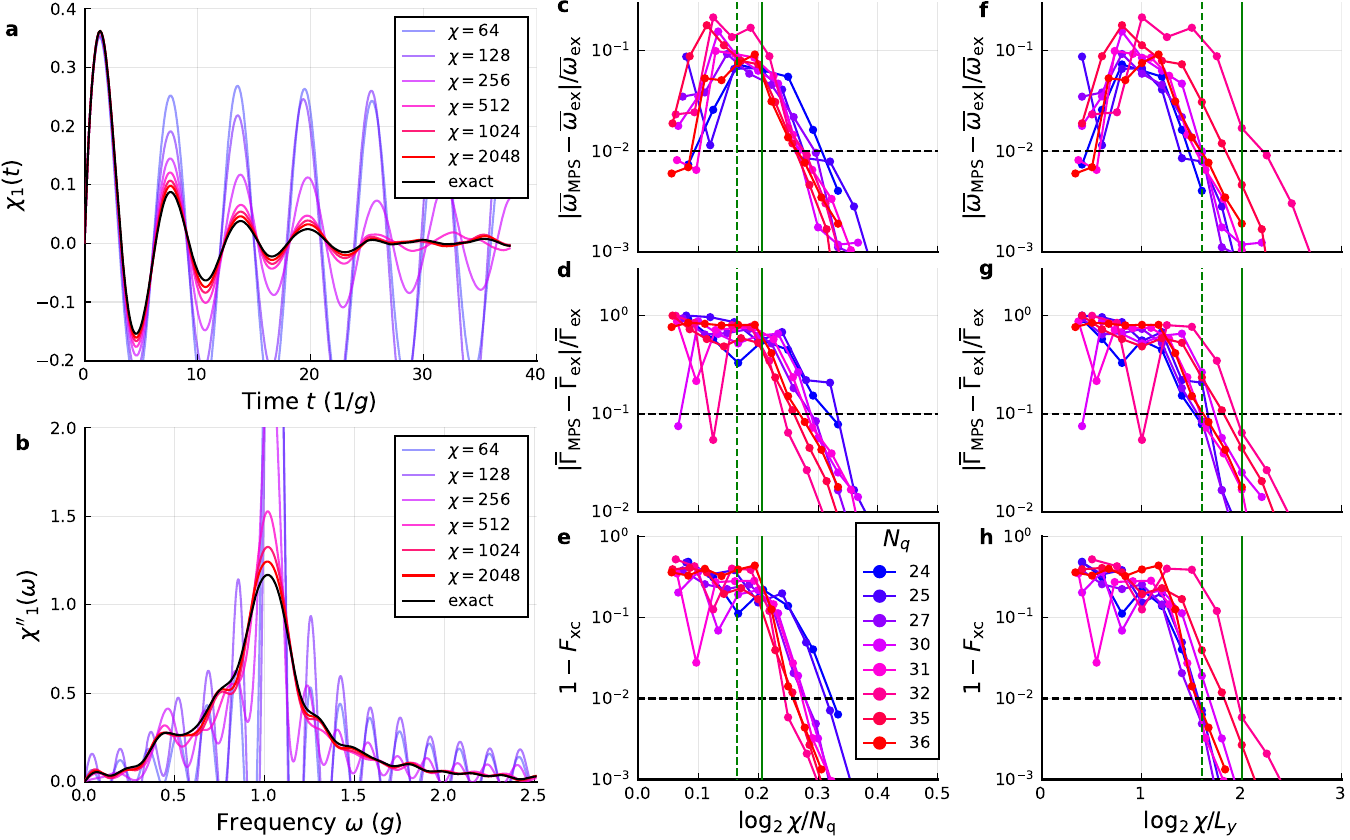}
    \caption{\textbf{Effect of TEBD truncations on magnon amplitudes. a,b,} The typical behavior of magnon amplitudes in TEBD simulations with insufficient bond dimension is a signal that oscillates with similar frequency and amplified magnitude compared to the exact result, as seen here for the lowest index magnon mode of a $N_{\mathrm{q}}=36$ square grid after a GGP pulse of duration $g t_{\mathrm{GGP}} = 0.1 \pi$ in the time domain (\textbf{a}) and the frequency domain (\textbf{b}). \textbf{c-e,}  Accuracy of magnon frequencies (\textbf{c}) and linewidths (\textbf{d}), and cross-correlation between the exact and MPS time-domain signals (\textbf{e}), via comparisons to exact calculations on $8$ rectangular and rotated rectangular grid geometries, also for initial states created by a GGP pulse of duration $g t_{\mathrm{GGP}} = 0.1 \pi$. The approximate collapse with scaled bond dimension $\log_2(\chi)/N_q$ implies exponential scaling of bond dimension to reach fixed accuracy values. Accuracy targets comparable to experimental accuracies ($1\%$ for frequencies, $10\%$ for linewidths, and $99\%$ cross-correlation) are shown as horizontal dashed lines, and scaled bond dimensions for $\chi=2^{16}$ and $\chi=2^{20}$ on a $N_{\rm q}=97$ qubit geometry are shown in vertical lines (dashed and solid, respectively). \textbf{f-h,} The same data when shown rescaled by the width of the system in the direction along the MPS snake. In rotated rectangular geometries, the width is not constant along this direction; an estimated effective width is used instead (see Fig.~\ref{fig:mpswidth}). An area-law scaling hypothesis would result in a smaller prediction for the necessary bond dimension. However, the volume-law collapses (\textbf{c-e}) better describe the accuracy results in the largest system sizes.\label{fig:mpsmagnon}}
\end{figure*}

The accuracy of the TEBD simulations can be gauged by monitoring the simulation fidelity (via the shrinking of the norm of the wave function) and the failure of energy conservation. The results are summarized in  Fig.~\ref{fig:mpsfidelityenergy}, for quench simulations starting from the initial state prepared by the GGP protocol with $g t_{\rm GGP}= 0.1 \pi$.
We see that the truncations of TEBD systematically reduce the energy of the time-evolved MPS, with the rate of energy loss increasing as the amount of truncation increases by decreasing the bond dimension. Specifically, we measure the energy with respect to the normalized MPS wave function, so that the energy is lower bounded by the ground state energy. When the bond dimension becomes sufficiently small, we see the surprising result that the energy of the state converges over time to roughly the ground state energy, the lowest possible by the variational property. In Fig.~\ref{fig:mpsfidelityenergy}d, we see that for the $XY$-model on rectangular grids of size $L_x \times L_y$, this occurs roughly when $\chi \sim 2^{L_y}$. For even smaller bond dimensions $\chi \lesssim 2^{L_y}$, the state does not lose as much energy --- however, this is because it is impossible to represent states with lower energies rather than an indication of increased accuracy. For larger bond dimensions $2^{L_y} \lesssim \chi \lesssim 2^{2L_y}$, the energy loss gradually becomes less. The approximate collapse of the curves in Fig.~\ref{fig:mpsfidelityenergy}d indicates that the energy can become converged with a bond dimension that scales exponentially with the width of the system $L_y$ rather than with the total qubit number $N_{\mathrm{q}}$. 

The simulation fidelity $F_{\rm MPS}$ shown in Fig.~\ref{fig:mpsfidelityenergy}a also undergoes an initial rapid decay followed by a saturation at long times, with the rate of decay of fidelity increasing with increasing amount of truncation (decreasing bond dimension). The collapse shown in Fig.~\ref{fig:mpsfidelityenergy}c shows that the fidelity is governed by the scaled bond dimension $\log_2(\chi)/N_{\mathrm{q}}$, which takes the value $0.5$ for a MPS state with maximal bond dimension on an $N_{\mathrm{q}}$-qubit system. This volume-law like scaling starts after an initial early time regime where $t \lesssim \pi/g$. This indicates that reaching a fixed value of the fidelity requires a bond dimension that scales exponentially with $N_{\mathrm{q}}$. This is consistent with the expectations from the volume-law entanglement entropy scaling illustrated in Fig.~\ref{fig:entropy}b-c, which show that the entanglement entropy has approximately reached saturation at time $t \approx \pi/g$. Unlike the measurements of Fig.~\ref{fig:entropy}, this volume-law scaling of bond dimensions to reach large simulation fidelity is measured on systems up to $N_{\mathrm{q}}=81$ in size in Fig.~\ref{fig:mpsfidelityenergy}, as an exact calculation is not needed as a reference to compute $F_{\rm MPS}.$

\begin{figure*}[htb]
    \centering
    \includegraphics[width = \textwidth]{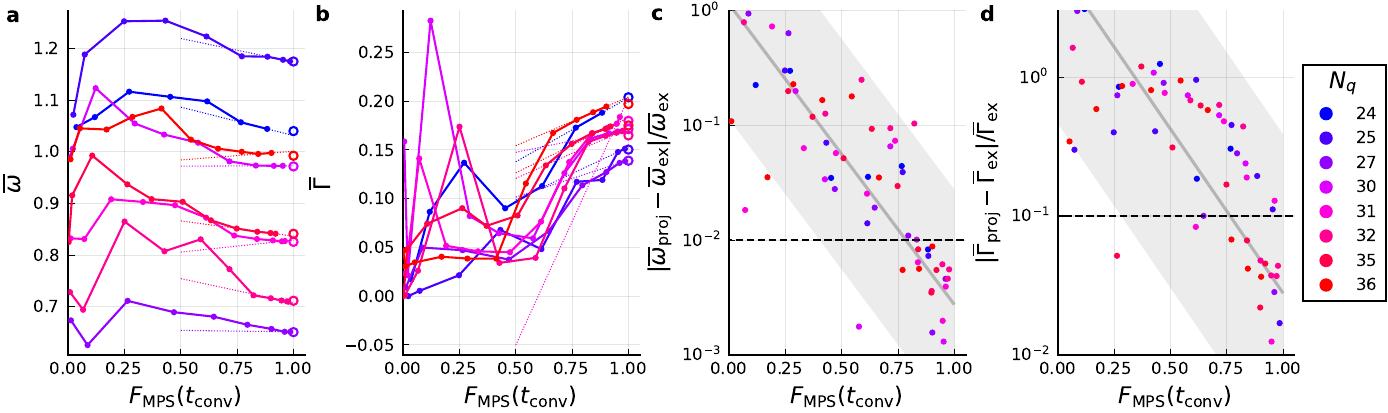}
    \caption{\textbf{Fidelity extrapolation of magnon frequency and linewidth in TEBD data. a,b,} Extracted frequency (\textbf{a}) and linewidth (\textbf{b}) as a function of simulation fidelity $F_{\rm MPS}$ for the data shown in Fig.~\ref{fig:mpsmagnon}. The evolution time over which $F_{\rm MPS}$ is measured is the minimum time necessary to converge the frequency and linewidth computed as in Fig.~\ref{fig:temporal}. Linear extrapolations of $F_{\rm MPS} \rightarrow 1$ using three consecutive data points are shown (dotted lines), and the extrapolated value is compared to the exact value (open circles). Due to the non-monotonic behavior with bond dimension, extrapolations from low fidelity are unlikely to be accurate. \textbf{c,d,} To make a quantitative estimate of the necessary fidelity for accuracies comparable with the experiment, extrapolation accuracy is plotted against maximum fidelity for sets of three consecutive data points with increasing bond dimensions for frequency (\textbf{c}) and linewidth (\textbf{d}). Linear fits (gray solid lines) to the errors suggest a target simulation fidelity $F_{\rm MPS} \sim 0.75$. \label{fig:mpsfidelityextrap} }
\end{figure*}

In these TEBD evolutions, a large simulation fidelity is sufficient for reaching accurate values of local observables in the evolved state, but apriori it may not be necessary. Fig.~\ref{fig:mpsmagnon} shows the effect of the MPS truncation on the magnon response functions, and as well on the extracted magnon frequency $\overline{\omega}$ and linewidth $\overline{\Gamma}$. The empirically observed effect is that even in heavily truncated evolutions, magnon signals oscillate at frequencies that are somewhat close to correct, but with amplitudes that decay far too slowly. A typical example of these dynamics and the resulting Fourier-domain response functions are shown in Fig.~\ref{fig:mpsmagnon}a and b, respectively. At small bond dimensions, the Fourier-domain signal shows a greatly amplified peak with large side-lobes, which are an artifact of the discontinuity at the finite-time cutoff. Both the approximate accuracy of the magnon frequency and the long lifetime of the magnon signal at low bond dimensions are consistent with the fact that the MPS state has evolved to have much lower energies.

In Fig.~\ref{fig:mpsmagnon}c-e, we quantify the accuracy of magnon frequency $\overline{\omega}$ and linewidth $\overline{\Gamma}$ as a function of bond dimension by comparing with exact (state-vector) results for $8$ different geometries from $24$ to $36$ qubits in size. These geometries are both rectangular grids ($N_q = 25, 30, 32, 35, 36$) and rotated rectangular grids ($N_q = 24, 27, 31$). The accuracy targets of $1 \%$ error for the frequency estimate and $10 \%$ error in the lifetime estimate, which are comparable to experimental accuracy, are shown as horizontal lines as a guide to the eye. While even low bond dimension TEBD calculations appear to have error $\lesssim 10\%$ in frequency, achieving the $1\%$ accuracy target requires bond dimensions that are much larger. The approximate collapse of the error curves in Fig.~\ref{fig:mpsmagnon}c appears to indicate that the relevant bond dimension scaling is exponential in the number of qubits $N_{\mathrm{q}}$. For the linewidth estimate $\overline{\Gamma}$, the errors are close to $100 \%$ at low bond dimension; the MPS signal does not decay at all, giving a linewidth estimate that is very close to $0$. In Fig.~\ref{fig:mpsmagnon}d, we see that comparably large bond dimensions are necessary to reach the accuracy target $10\%$ in the linewidth measure $\overline{\Gamma}$. Assuming the exponential scaling implied by these figures continues to hold in larger systems, the implied bond dimension to reach experimental accuracy for a system with $N_q = 97$ qubits would be in the range of $\chi \sim 2^{25}-2^{30}$. A similar result is implied by analyzing the cross-correlation  $F_{\rm XC}$, as defined in Eq.~\eqref{eq:xc}, between the exact and MPS simulated curves. As in Sec.~\ref{sec:sm_accuracy}, we use the time interval from $gt = 0$ to $gt = 12 \pi$ for this comparison.

The collapse of the curves in Fig.~\ref{fig:mpsmagnon}c-e is only approximate; moreover, the range of system sizes available in exact calculations is limited. To probe how the argument above might fail, we consider an alternate area-law scaling of bond dimension by plotting the accuracy of magnon frequency and linewidth versus $\log_2(\chi)/L_y$ in Fig.~\ref{fig:mpsmagnon}f-h, where $L_y$ is the width of the system in a direction along the MPS snake. While a significant part of the dependence of the error could be explained with either the area-law scaling hypothesis or the volume-law one, the drift of the curves at larger sizes in the area-law case suggests that the volume-law scaling hypothesis is more accurate.

\begin{figure}[htb]
    \centering
    \includegraphics[width = 0.9 \columnwidth]{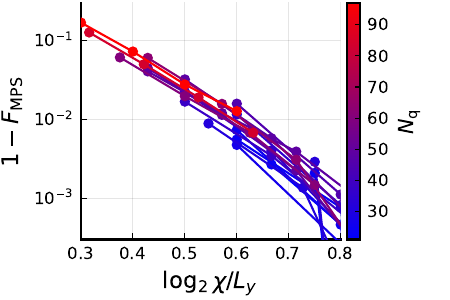}
    \caption{\textbf{Determining effective width for MPS calculations with non-uniform width.} The simulation fidelity $F_{\rm MPS}$ of the initial state prepared with the GGP pulse shows area-law scaling: in rectangular geometries, scaling bond dimension by the width $L_y$ causes the curves to approximately collapse. For non-rectangular geometries, we use this collapse to measure an effective $L_y$ for each geometry considered. The plot shows the resulting $F_{\rm MPS}$ for both rectangular geometries and rotated rectangular geometries up to $97$ qubits in size. \label{fig:mpswidth}}
\end{figure}

For the rotated rectangular geometries, the width is non-uniform, raising the question of how to compare to rectangular geometries in these area-law collapses. For this purpose, we use the simulation fidelity of the initial state prepared after the GGP protocol at various bond dimensions, which shows an area-law collapse when the bond dimension is scaled by $L_y$ in rectangular geometries, as shown in Fig.~\ref{fig:mpswidth}. For the rotated geometries, we choose an effective $L_y$ that collapses the same fidelity curves onto the result for rectangular geometries. In particular, the width of the $97$ qubit geometry has $L_y=10$.

\begin{figure*}[htb]
    \centering
    \includegraphics[width = 1.0\textwidth]{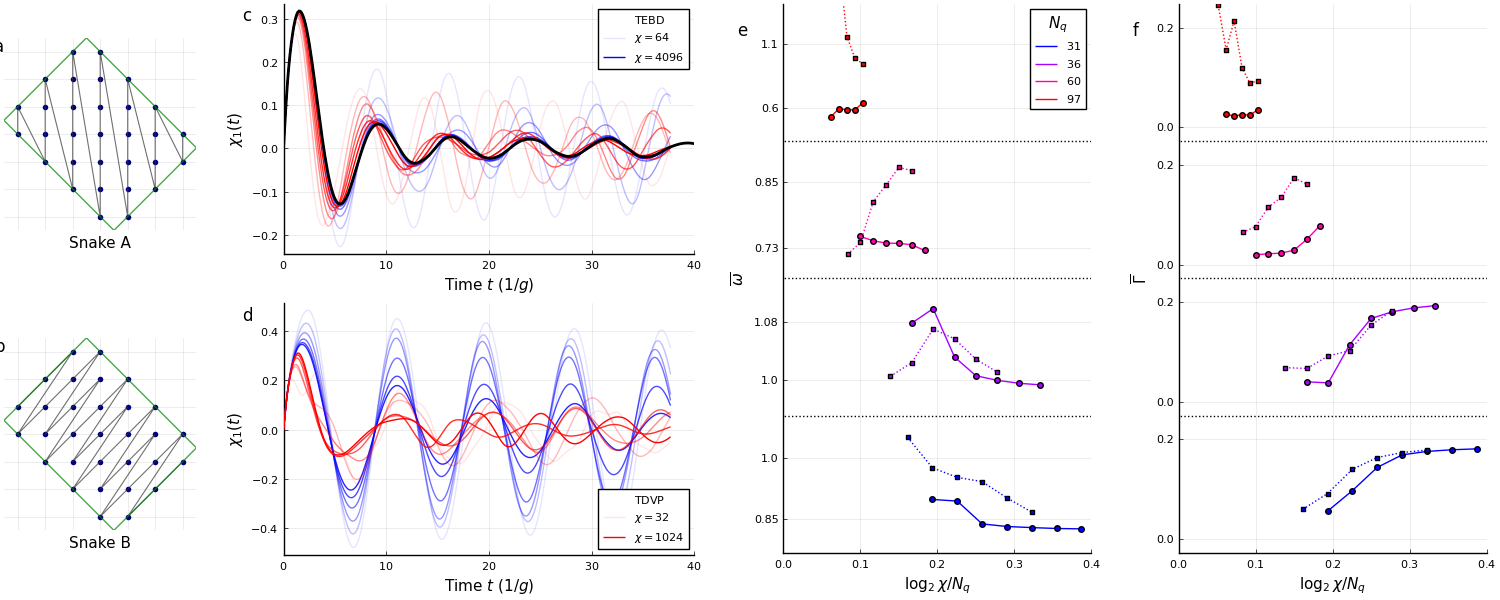}
    \caption{\textbf{Comparison of TEBD and TDVP calculations for the lowest-index magnon mode}. \textbf{a,} The qubit ordering used in TEBD calculations on rotated rectangular geometries and \textbf{b}, the ordering used in TDVP calculations. Direct comparison of the MPS convergence of TEBD and TDVP is shown for systems with $N_{\mathrm{q}} = 31$ qubits (\textbf{c}) and $N_{\mathrm{q}} = 60$ qubits (\textbf{d}). In each, blue curves of increasing opacity represent TEBD data with increasing bond dimension (from $\chi=32$ to $\chi=2048$) and red curves represent TDVP data of increasing bond dimension (from $\chi=32$ to $\chi=1024$). For $31$ qubit data, the corresponding exact data is shown in black. \textbf{e,f,} The extracted magnon frequency $\overline{\omega}$ and linewidth $\overline{\Gamma}$ are shown for $4$ system geometries, with TEBD results shown as solid curves and TDVP results shown as dotted lines. Horizontal gray bars show the exact result for $\overline{\omega}$ with an interval of $1\%$ and $\overline{\Gamma}$ with an interval of $10\%$ where known. Vertical marks denote location of $\chi=2^{16}$ (dashed green line) and $\chi=2^{20}$ (solid green line) for the corresponding system sizes $N_{\mathrm{q}}$. Calculations with bond dimensions $\chi \sim 2^{16}$ are extremely challenging but potentially possible with large classical compute resources, while those with $\chi \sim 2^{20}$ are unfeasible on classical machines. In all cases, the disagreement between TDVP and TEBD supports our finding that convergence does not happen until large values of the scaled bond dimension $\log_2 \chi / N_{\rm q}$. \label{fig:tdvpcomparison} }
\end{figure*}

Extrapolating trends from data taken at several different bond dimensions has proven effective in accelerating the convergence of MPS dynamics simulations~\cite{QuantinuumNature2026Magnetism,2025-Mandra-Kechedzhi-HeuristicMatrixIsingModels}. In Fig.~\ref{fig:mpsfidelityextrap}, we illustrate one possible method for extrapolations which treats magnon frequency and linewidth as a function of the simulation fidelity $F_{\rm MPS}$ and extrapolates to the exact limit $F_{\rm MPS} \rightarrow 1$. Specifically, we directly apply this extrapolation to the extracted magnon frequency and linewidth, rather than extrapolating $\chi_\mu(t)$ for each time $t$ independently. The latter approach (not shown) produces poor results in this setting with oscillating signals, as the signals with various bond dimensions become out-of-phase with each other, resulting in non-monotonic dependence for $\chi_\mu(t)$ on bond dimension. The fidelity of the MPS computation for this extrapolation is taken to be the fidelity at the earliest time the frequency has converged to a level of $1 \%$, or linewidth has converged to a level of $10 \%$. These convergence times are determined individually for each magnon signal by computing frequency and linewidth as a function of total evolution time and finding the last time it deviates above those levels, as shown in Fig.~\ref{fig:temporal}. This allows for using as short as possible time evolutions to learn the magnon parameters. For each set of $3$ consecutive increasing bond dimensions in the data set, we linearly extrapolate $\overline{\omega}$ and $\overline{\Gamma}$, as illustrated in Fig.~\ref{fig:mpsfidelityextrap}a,b. The accuracy of the results is plotted against the largest fidelity of the data points used in each extrapolation in Fig.~\ref{fig:mpsfidelityextrap}c-d. We see non-smooth convergence of $\overline{\omega}$ and $\overline{\Gamma}$ to the corresponding exact values. A linear regression of the logarithm of accuracy against $F_{\rm MPS}$ estimates the necessary fidelity to compete with experimental accuracy as roughly $F_{\rm MPS} \sim 0.75$, reflecting the fact that convergence to exact results is seen in our data sets only with high fidelity MPS calculations.

\begin{figure}[htb]
    \centering
    \includegraphics[width = 0.9 \columnwidth]{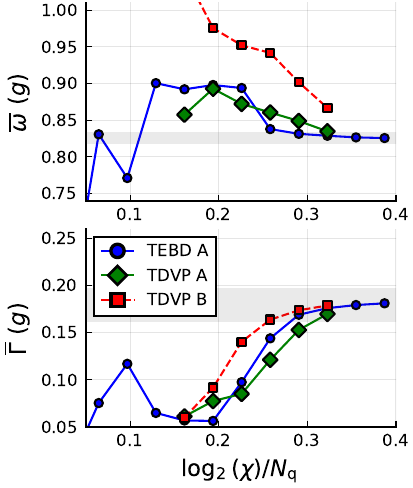}
    \caption{\textbf{Comparison between MPS qubit orderings.} Extracted frequency and lifetime of the lowest index mode on the rotated rectangular geometry with $N_{\mathrm{q}}=31$ using TDVP calculations with two different qubit orderings. The orderings are labeled `A' and `B' as pictured in Fig.~\ref{fig:tdvpcomparison}a-b. Additionally, the TEBD data computed using the qubit ordering `A' is shown. Snake ordering `A' consistently shows faster frequency convergence, while snake ordering `B' shows faster linewidth convergence. The distinction between the performance of the snake orderings is much less pronounced for calculations with higher index modes (not shown).\label{fig:snake}}
\end{figure}

The quantitative analyses above use TEBD data as they rely on knowledge of the simulation fidelity computed from truncation errors, which could apriori be quite different in effect from the projection errors in the TDVP method. In Fig.~\ref{fig:tdvpcomparison}, we directly compare the results of TEBD and TDVP calculations for the lowest index magnon mode across $4$ rotated rectangular geometries with increasing qubit numbers. We note that these calculations use distinct qubit orderings, shown in Fig.~\ref{fig:tdvpcomparison}a,b. For the $31$ qubit system, where we have exact data for comparison, we see convergence of frequency and linewidth to the correct values at similar sizes of the scaled bond dimension $\log_2(\chi)/N_{\mathrm{q}}$, with frequency converging a bit quicker for TEBD calculations and linewidth converging quicker for TDVP calculations with rotated rectangular geometries. 
In part, this variation can be explained by the different choices of qubit orderings for the two calculations. In Fig.~\ref{fig:snake}, we show the results of the TDVP calculations with both qubit orderings, which shows that TDVP has similar behavior to TEBD when the qubit orderings are identical.

For large systems, such as the $60$ qubit geometry shown in Fig.~\ref{fig:tdvpcomparison}d, we see a more stark distinction between the two methods. While neither method has reached convergence, the TEBD signal has a much more consistent frequency whereas the TDVP signal varies more wildly. Evidence from smaller systems suggest that the stable frequency of the TEBD calculation is more likely to be close to the correct answer, as Fig.~\ref{fig:mpsmagnon} shows that even with insufficient bond dimension, TEBD computes frequencies that are within $\sim 10\%$ of the right value. On the other hand, TEBD signals are greatly amplified and linewidths greatly suppressed compared to expected physical values, suggesting that the smaller amplitude of the TDVP signals is closer to the truth. The data from smaller systems and the distinctions between these methods both support the conclusion that both methods are far from convergence at the computed bond dimensions, which are up to $\chi=2048$ for TEBD and $\chi=1024$ for TDVP.

\subsection{Belief Propagation PEPS}
The quantum states we consider in this work are the result of evolution under Hamiltonians on $2$d geometries. This begs the question of whether a simulation approach based on $2$D tensor networks could be more efficient than ones based on $1$D MPS. Two-dimensional projected entangled pair states (PEPS) can represent states with area-law entanglement using a bond dimension $\chi$ that may grow with the correlation length of the state but that is independent of system size. Such a representation uses a memory footprint that is linear in the number of sites; more precisely, a PEPS on the square lattice (with coordination number $4$) has $\mathcal{O}(N_{\mathrm{q}}\chi^4)$ parameters. In contrast, an MPS representing an area-law entangled state in a $2$D geometry  typically requires a bond dimension $\chi$ that scales exponentially in the smallest linear dimension of that geometry. In practice, however, PEPS are notoriously challenging to work with and have historically been inefficient for simulating quantum dynamics due to prohibitive complexity beyond $\chi \sim 10$.

The main difficulty in working with PEPS is to \textit{gauge} the network. Gauging here refers to the computation of effective environments for each site, which allow physically local updates and contractions to remain local computations. During the evolution, the quality of the environments determines how well the state can be compressed into a finite bond dimension $\chi$. At the time of computing expectation values, the environments determine the accuracy of the estimates even if the underlying state is exact. 

Belief propagation (BP) has recently been introduced as an efficient method for gauging PEPS using the BP approximation to the environments~\cite{tindall2023gauging}. This is, in effect, a tree approximation to the structure of the state correlations, with runtime scalings mirroring those of tree tensor networks. Single- and two-qubit gates can be applied in time $\mathcal{O}(\chi^{z+1})$, i.\,e., $\mathcal{O}(\chi^{5})$ on 2D grids, allowing the bond dimension, and thus the ability to capture entanglement, to grow into the dozens or beyond. The BP approximation is particularly well-suited to compress states with large loops or ones with weak ``loop correlations''. 

When it comes to estimating local expectation values, PEPS require significantly more accurate contractions than BP can provide. While there exist so-called loop series~\cite{evenbly2026loop} and cluster expansions~\cite{gray2025tensor}, planar geometries are currently most efficiently contracted via \textit{boundary MPS} environments. Given a boundary MPS bond dimension of $R$, local expectation value contractions require $\mathcal{O}(N_{\mathrm{q}}R^3\chi^5)$ time~\cite{rudolph2025simulating}, where $R=1$ resembles the BP limit.

To rule out BP-PEPS as a method for classically simulating our experiments, we need to demonstrate that
\begin{enumerate}
    \item the PEPS bond dimension $\chi$ must be unfeasibly large to capture the evolved states to sufficient fidelity,
    \item expectation values from low-fidelity PEPS are indeed inaccurate.
\end{enumerate}

Let us first consider the fidelity. Ref.~\cite{rudolph2025simulating} describes a method to approximate the running fidelity of a PEPS during simulation of a quantum circuit as a product of the per-gate fidelities. These fidelities are computed from the magnitude of singular values thrown away, which is precisely how fidelities are typically estimated for MPS. In our case, the singular values are conditioned on the BP approximation. We call this the BP fidelity $F_{\text{BP}}$ and the exact fidelity is $F_{\text{ex}}$. Ideally, the BP fidelity would be representative to gauge the quality of the simulation at scale. 

\begin{figure}
    \centering
    \includegraphics[width=1\linewidth]{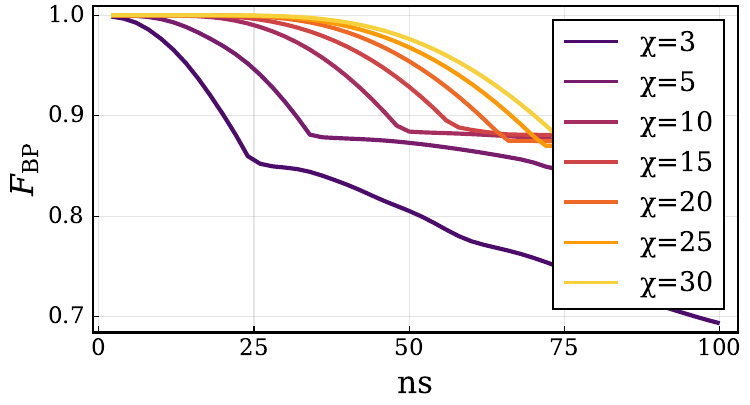}
    \includegraphics[width=1\linewidth]{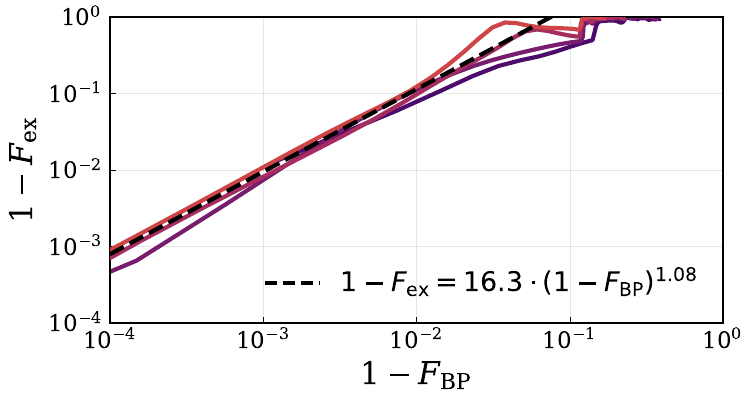}
    \includegraphics[width=1\linewidth]{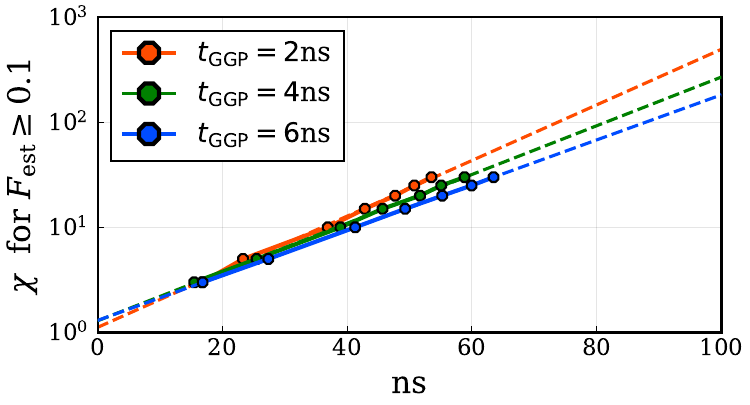}
    \caption{\textbf{Extrapolating BP-PEPS fidelity on a 17-qubit rotated square geometry}. Top: BP fidelity $F_{\text{BP}}$ from truncating the BP-PEPS during evolution under the $XY$ Hamiltonian and $t_{\mathrm{GGP}}=6$ ns. Middle: For $\chi=3,5,10,15$, we compute the exact fidelity $F_\text{ex}$ via exact contraction, and calibrate a fidelity rescaling. Here $t_{\mathrm{GGP}}=6$ ns is shown. Bottom: Extrapolating the bond dimension necessary for the BP-PEPS to achieve estimated fidelity $F_{\text{est}} \geq 0.1$ at different temperatures. These values underestimate the bond dimension required in larger systems.}
    \label{fig:PEPS_fidelity}
\end{figure}

In Fig.~\ref{fig:PEPS_fidelity}, we see results of BP-PEPS $XY$-model evolution on a rotated square lattice with 17 qubits using the open-source TensorNetworkQuantumSimulator.jl library. It is important to highlight that the decay of fidelity is not expected to differ qualitatively when scaling up to the $\sim 100$ qubit regime due to the nature of PEPS simulation. That being said, fidelities for larger lattices will be lower, and our results should be viewed as underestimating the resources required for accurate PEPS simulation. At first glance, the top panel depicts high $F_{\text{BP}}$ throughout. The exact fidelities $F_{\text{ex}}$ we can compute until $\chi=15$, on the other hand, are significantly lower than indicated by the BP truncation error. Fortunately, we can extract a robust, temperature-dependent relation between $F_{\text{ex}}$ and $F_{\text{BP}}$, allowing us to calculate a rescaled fidelity estimate $F_{\text{est}}$ that corresponds to approximately $\times10$ to $\times20$ larger errors than the BP truncation error indicated. Note that the BP fidelity is expected to be representative for less loop-correlated states, or ones on geometries with larger loops. This itself hints at the difficulty of accurately contracting the resulting PEPS.

We can now use the estimated fidelity $F_{\text{est}}$ with simulations up to $\chi=30$, to extrapolate the bond dimension required to attain PEPS with fidelity above $0.1$. The bottom panel of Fig.~\ref{fig:PEPS_fidelity} depicts our results. At $100$ ns, all temperatures require a bond dimension significantly above $100$, which is currently outside reach for contracting with boundary MPS -- even with only 17 qubits. We also observe a temperature dependence, where higher temperatures require a higher bond dimension for similar $F_{\text{est}}$. This aligns with general tendencies in state-based Schrödinger-picture simulation via MPS or other tensor networks. On the flipside, we find that higher temperatures exhibit lower loop correlations, indicating that one may achieve accurate expectation values from the PEPS with a lower boundary MPS rank $R$ than lower-temperature PEPS at the same bond dimension $\chi$.

\begin{figure}
    \centering
    \includegraphics[width=1\linewidth]{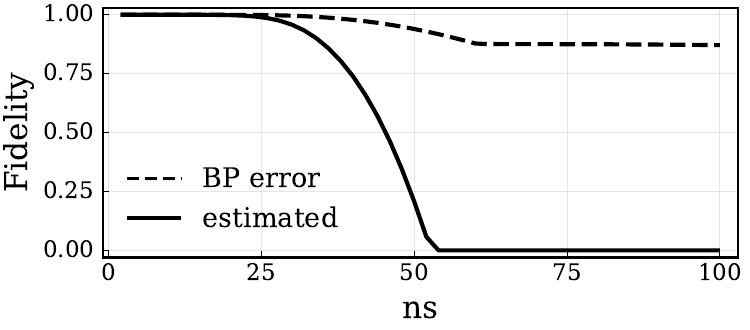}
    \includegraphics[width=1\linewidth]{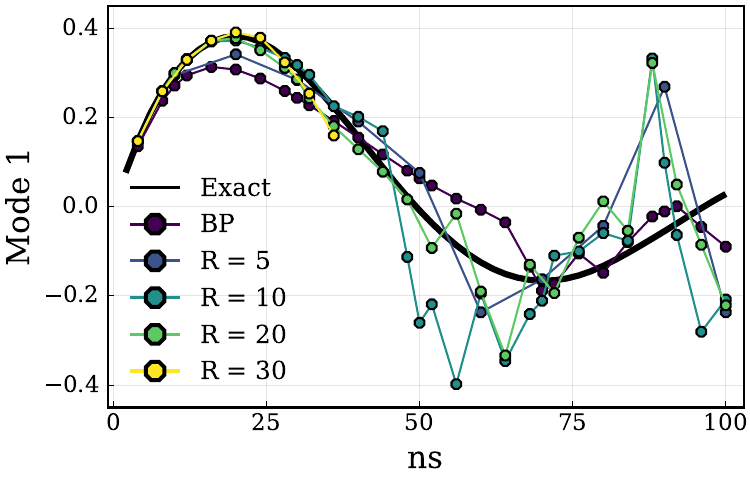}
    \caption{\textbf{Example mode 1 estimates for BP-PEPS on a 24-qubit rotated square geometry.} Top: BP truncation estimated and rescaled fidelities for a $\chi=25$ BP-PEPS $t_\text{GGP}=6$ ns. Bottom: Mode 1 computation for contraction via BP or boundary MPS with rank $R$. The drop of the rescaled fidelity coincides with where more and more faithful contraction remains inaccurate. By the time the estimated fidelity exhibits the ``kink'', all expectation values are highly accurate. We note that higher temperatures, i.\,e., shorter $t_\text{GGP}$, are more accurately contracted with low $R$ or even with BP until the kink in the fidelity.}
    \label{fig:PEPS_expectation}
\end{figure}

One could argue that high global fidelity is not required for accurately estimating local expectation values. Before testing the absolute quality of magnon modes for different boundary MPS ranks $R$, we return to the top panel of Fig.~\ref{fig:PEPS_fidelity}. It shows that around $F_{\text{BP}} \approx 0.85$ there is a ``kink'' in the estimated fidelity, after which its decay is slowed down. We have found that BP truncation systematically breaks the PEPS quantum state, after which the $XY$-model dynamics break down and $F_{\text{ex}}$ drops sharply. 

Finally, in Fig.~\ref{fig:PEPS_expectation} we depict a 24-qubit BP-PEPS simulation, $F_{\text{BP}}$ and $F_{\text{est}}$ for $\chi=25$, as well as mode 1 computations with BP and various $R$ at $t_\text{GGP}=6$ ns. As the estimated fidelity drops after approximately $30$ ns, we see the quality of even $R=30$ deteriorate. Higher temperature modes are more accurately estimated with BP (not shown), which suggests that our low- to intermediate temperature states primarily require larger $R$ for accurate estimates at $50$ ns. Beyond this evolution time, particularly after the kink in $F_{\text{BP}}$ at $\approx60$ ns, BP-PEPS simulations require both drastically larger $\chi$ and $R$ to reproduce the magnon mode. As we increase the system size, the cost of contracting BP-PEPS will scale at least linearly in the number of sites. The bond dimensions of the PEPS and boundary MPS required for the same accuracy also tend to increase, resulting in PEPS that cannot be accurately contracted with current techniques. 

\subsection{Discussion}\label{sec:complexitysummary}

The magnon dynamics that are the focus of this work vary to a large extent in their complexity. The sensitivity analysis indicates that high-index magnon frequencies are only barely affected by fairly large changes to the background energy density of the state, while the low-index magnon signals are more sensitive. Additionally, high-index magnon modes --- which have short periods of oscillation --- can have their frequency and linewidth estimated using much shorter time evolutions. These short times necessary for estimating these physical parameters allows for lower classical resources (e.\,g. lower MPS bond dimensions). One possible explanation for this simplification is that these physical signals probe smaller length scales and time scales. By contrast, the lowest index magnon modes probe length scales that are on par with the linear dimensions of the device.

For this reason, the bulk of our complexity analysis focused on computations related to the lowest index magnon mode. Empirical evidence suggests that accurately computing the frequency of this mode requires large simulation fidelity $F_{\rm MPS} \sim 0.75$ in an MPS calculation that runs over the necessary length of time, as illustrated in Fig.~\ref{fig:mpsfidelityextrap}. The time necessary for an accurate frequency determination at the level of experimental accuracy for this slowest mode is $t_{\rm conv} \sim 8/g$, which is larger than the time scale $t_{\rm sat} \approx 4/g$ in which the entanglement entropy saturates to a volume-law behavior, as detailed respectively in Fig.~\ref{fig:temporal} and Fig.~\ref{fig:entropy}. The consequence of this large entanglement entropy on simulation accuracy is shown clearly in the collapse of the system size and bond dimension dependence of $F_{\rm MPS}$ into a single function of $\log_2(\chi)/N_{\mathrm{q}}$, as shown in Fig.~\ref{fig:entropy}. This implies that reaching $F_{\rm MPS} \sim 0.75$ requires bond dimensions that scale exponentially in system size $N_{\rm q}$.

A natural question to ask is whether the exponential growth in complexity of this observable will continue as $N_{\mathrm{q}}$ is scaled to large values, or if the complexity growth will slow down. The logic of the previous paragraph implies that as long as the time scales probed by the observable are long enough for entanglement to scale as a volume-law, and as long as the observable cannot be extrapolated to sufficient accuracy from low-fidelity simulations, then MPS methods will see this exponential scaling of complexity.
The lowest mode corresponds roughly to a mode with momentum $k \sim \pi/L$, where $L$ is the largest linear dimension of the system; thus, as we increase the system size, larger and larger length scales are probed. 
For the ballistic magnon signals of this work, the oscillation periods will continue to grow $T \sim 1/\omega \propto L$. We expect the time scale needed to measure the magnon frequency also grows with the oscillation period. The entanglement saturation time will also grow with the linear dimension of the system. As these times scale similarly, we expect that volume-law entanglement entropy continues to hold at the relevant time scale as the system size grows.

However, the theoretical expectation is that the response function at the largest length scales --- in the hydrodynamic regime where wavelengths are long compared to scattering lengths ---  shows diffusive behavior instead of the ballistic magnon oscillation. As the system is probed at smaller momenta, the time-domain response function should eventually experience a crossover from showing a damped oscillating signal with frequency $\omega \sim v k$ to an overdamped exponential decay with width $\Gamma \sim D k^2$. 
As the nature of the signal changes, the quantitative analyses applied here may not directly apply beyond that scale. 
However, the experimental evidence is that underdamped, ballistic oscillations appear at even the longest wavelengths accessed in the largest systems ($97$ qubits) considered in our experiment; the crossover to diffusive behavior, assuming it occurs, requires still longer wavelengths. This suggests that the empirically measured exponential growth of complexity for MPS methods extends at least to that size.

\end{document}